\documentclass[apjl]{emulateapj}

\shorttitle{{\it HST}/NICMOS OBSERVATIONS OF THE TOOMRE SEQUENCE NUCLEI}
\shortauthors{Rossa et al.}

\newcommand{\lta}{\lesssim}
\newcommand{\gta}{\gtrsim}

\newcommand{\Mpc}{\>{\rm Mpc}}
\newcommand{\pc}{\>{\rm pc}}

\begin{document}

\title{THE TOOMRE SEQUENCE REVISITED WITH {\it HST}/NICMOS: NUCLEAR
BRIGHTNESS PROFILES AND COLORS OF INTERACTING AND MERGING
GALAXIES\altaffilmark{1}}

\author{J\"{o}rn Rossa\altaffilmark{2,3},
Seppo Laine\altaffilmark{4},
Roeland P. van der Marel\altaffilmark{2},
J. Christopher Mihos\altaffilmark{5},\\
John E. Hibbard\altaffilmark{6},
Torsten B\"{o}ker\altaffilmark{7}, and
Ann I. Zabludoff\altaffilmark{8}}

\altaffiltext{1}{Based on observations made with the NASA/ESA Hubble
Space Telescope, obtained at the Space Telescope Science Institute,
which is operated by the Association of Universities for Research in
Astronomy, Inc., under NASA contract NAS 5-26555. These observations
are associated with proposal \#9402.}

\altaffiltext{2}{Space Telescope Science Institute, 3700 San Martin
Drive, Baltimore, MD 21218; marel@stsci.edu}

\altaffiltext{3}{Present address: Department of Astronomy, University
of Florida, 211 Bryant Space Science Center, P.O.Box 112055,
Gainesville, FL 32611-2055; jrossa@astro.ufl.edu}

\altaffiltext{4}{Spitzer Science Center, Mail Code 220-6, California
Institute of Technology, Pasadena, CA 91125; seppo@ipac.caltech.edu}

\altaffiltext{5}{Department of Astronomy, Case Western Reserve
University, 10900 Euclid Avenue, Cleveland, OH 44106; mihos@case.edu}

\altaffiltext{6}{National Radio Astronomy Observatory, 520 Edgemont
Road, Charlottesville, VA 22903-2475; jhibbard@nrao.edu}

\altaffiltext{7}{Astrophysics Division, RSSD, European Space Research
and Technology Centre (ESTEC), NL-2200 AG Noordwijk, The Netherlands;
tboeker@rssd.esa.int}

\altaffiltext{8}{Steward Observatory, University of Arizona, 933 North
Cherry Avenue, Tucson, AZ 85721-0065; azabludoff@as.arizona.edu}


\begin{abstract}
We discuss the near-infrared properties of the nuclei in the 11
merging galaxies of the Toomre sequence, based on high spatial
resolution $J$, $H$, and $K$ imaging data using NICMOS onboard the
{\it Hubble Space Telescope (HST)}. The observations are less affected
by dust extinction than our previous {\it HST}/WFPC2 observations and
offer higher spatial resolution than existing ground-based near-IR
data. Nuclear positions are generally found to be consistent with
those reported from data in other wavebands. In NGC\,7764A, we detect
for the first time two nuclei with a separation of about 260\,pc,
consistent with its placement in the middle of the merging sequence.
We see a marginal trend for the nuclei to become bluer with advancing
merger stage, which we attribute to a dispersal of dust at late times
in the merging process. Our data also indicate a statistically
significant trend for the nuclei in the sequence to become more
luminous, within an aperture of fixed physical size and after
correcting for dust extinction, with advancing merger stage.  We
derive $K$-band surface brightness profiles for those nuclei for which
the morphology allows a meaningful isophotal analysis, and fit the
profiles with a ``Nuker law'' for comparison with other samples of
galaxies observed with {\it HST}. The majority of the nuclei have
steep profiles that can be characterized as power-law type. In
general, the Toomre sequence galaxies tend to have steeper profiles
and higher central luminosity surface densities than E/S0's. Our
findings can be explained if the Toomre sequence galaxies have newly
formed stars that are concentrated toward their centers. We derive
$V-K$ color profiles for the nuclei to further address this
possibility, but find that the large amounts of dust extinction
complicate their interpretation. Overall, our results are consistent
with the generic predictions of $N$-body simulations of spiral galaxy
mergers. If left to evolve and fade for several Gyrs, it is possible
that the properties of the Toomre sequence nuclei would resemble the
properties of the nuclei of normal E/S0 galaxies. Our results
therefore support the view that mergers of spiral galaxies can lead to
the formation of early-type galaxies.
\end{abstract}





\section{INTRODUCTION}
\label{s:intro}

It is commonly assumed that the merging of two spiral galaxies
following a close interaction will eventually result in their
transformation into an elliptical galaxy. This outcome was first
suggested by the early simulations of \citet{too72} and later
confirmed by various other investigators \citep[e.g.,][]{bar88,bor91}.
Evidence gathered over the last few decades conclusively shows that at
least a fraction of elliptical galaxies were indeed formed by mergers
\citep[see, e.g., the review by\,][]{sch98}. Such evidence includes
the finding from ground-based imaging that on global scales many
merger remnants are well fit by the $R^{1/4}$ luminosity profiles that
are characteristic of elliptical galaxies \citep[e.g.,][]
{sch82,hib99,rot04}.

Previous studies of spiral-spiral mergers have focused extensively on
the tidal tails \citep[e.g.,][]{hib96,hib01}, globular clusters
\citep[e.g.,][]{whi93,whi95,sch96,sch04}, young star clusters
\citep[e.g.,][]{whi99,whi05} and dwarf galaxies that may have formed
during the merger \citep[e.g.,][]{brai01,wei03}. Less attention has
been paid to the nuclei, however \citep[although see][]{sch96a,sco00}.
This is in contrast with the case for, e.g., the nuclear regions of
spiral galaxies \citep{car97,boe02} and ellipticals \citep[e.g.,][]
{lau95,rav01,res01,lai03a,fer06}, which have been studied at high
resolution using the {\it Hubble Space Telescope (HST)}. In fact, one
of the most important physical processes that occurs during the
merging process is the inflow of gas into the nuclear region
\citep[e.g.,][]{nog88,hos94,hos96,bar96}. To study the question of
whether and how mergers convert the structure of the nuclear regions
from disk galaxy bulges or pseudo-bulges into the density profiles
seen in the centers of elliptical galaxies, it is therefore important
to investigate the nuclear regions at the highest achievable angular
resolution.

\citet{too77} assembled a small sample of prospective merger systems
in various stages of merging that later became known as the {\em
Toomre sequence}. We investigated various properties of this optically
selected merger sample with {\it HST}/WFPC2, in particular the
response of the nuclei to the merging process \citep[][hereafter
Paper~I]{lai03b}. However, in the optical several of the nuclear
regions are hidden behind dust, which makes identification of the
nuclei difficult, if not impossible. Near-infrared (NIR) observations
are much less affected by dust, and thus are better suited to
identifying the location of the nuclei within the often complex
circumnuclear environments of the host galaxies.

Previous investigations have studied a range of basic properties of
various merger samples \citep[e.g.,][]{jos84,lon84,sta91,bus92,rot04}.
Most of the Toomre sequence galaxies have been studied previously from
the ground at NIR wavelengths: NGC\,4038/39 \citep{bus90,hib01,kas03};
NGC\,4676 (J.~Hibbard, unpublished); NGC\,7592 \citep{bus90};
NGC\,6621/22 \citep{bus90}; NGC\,3509 \citep{bus92}; NGC\,520
\citep{bus92,kot01}; NGC\,3256 \citep{kot96}, and more recently, all
the four late-stage mergers \citep{rot04}. However, these observations
either lacked adequate spatial resolution to reveal the detailed
structure of the nuclei, or were aimed at investigating properties on
larger scales. In order to explore the nuclear regions themselves, the
high spatial resolution of the {\it HST} (5--50 pc at the distance of
the Toomre sequence galaxies) is imperative.

We present here a study of the entire Toomre sequence with {\it HST}
/NICMOS. This work is part of a high spatial resolution multi-instrument 
{\it HST} survey of the nuclear regions in the Toomre sequence 
(cf.~Table~\ref{t:ts}), including WFPC2, NICMOS and STIS
observations. Earlier (see Paper~I) we reported on the nuclear
morphologies and luminosity densities based on WFPC2 broad- and
narrowband observations. In a future paper \citep[][hereafter
Paper~III]{ros07} we will present stellar population studies based on
STIS G430L spectra. In this paper we focus on the nuclear morphology
and photometric properties as revealed by the NICMOS observations. We
make a detailed comparison with our previous WFPC2 observations and
other existing multi-wavelength observations, and discuss the results
in the context of the evolution of merging galaxies and the formation
of ellipticals.

Our investigation is the first comprehensive high spatial resolution
study of an optically selected sample of mergers. Merging galaxies
often have high infrared luminosities \citep{san88}, sometimes in
excess of $L_{\rm{IR}}\geq10^{11}\,\rm{L_{\sun}}$, and samples of
LIRGS (Luminous InfraRed Galaxies) and ULIRGS (Ultra Luminous InfraRed
Galaxies) are dominated by merging systems \citep{san96}. The present
optically selected sample thus provides a useful comparison to {\it
HST}/NICMOS studies of merger remnants selected only on the basis of
their infrared luminosity \citep[e.g.,][]{sco00,bor00}.

This paper is structured in the following way: in \S~\ref{s:sampleobs}
we present the observations and the data reduction procedures. In
\S~\ref{s:imaging} we show results on the nuclear morphology and color
mapping for each individual system separately (grouped in three merger
stages), and compare them with other multi-wavelength studies. In
\S~\ref{s:colors} we present and discuss photometry, colors, and
radial color profiles for the Toomre nuclei. In \S~\ref{s:sfbprof} we
present $K$-band surface brightness profiles, and discuss the results
in the context of the evolution of merging galaxies and the formation
of early-type galaxies. In \S~\ref{s:summary} we summarize the main
results of the paper.

\section{SAMPLE AND OBSERVATIONS}
\label{s:sampleobs}

\subsection{Sample}
\label{ss:sample}

Our sample is the Toomre sequence of relatively nearby ($D \leq$
122\,Mpc) strongly interacting and merging galaxies, ordered according
to the estimated degree of completeness of the merging process
\citep{too77}. The basic properties of the eleven systems are listed
in Table~\ref{t:ts}. Distances in the table and throughout this paper
are based on a Hubble constant of $H_{0}=75\,\rm{km\,s^{-1}\,Mpc^{-1}}$, 
save for the distance to the Antennae which is based on observations of 
red giant branch stars \citep{sav04}. The systems span a wide range of 
dynamical phases, from galaxies early in the merging process (e.g., 
NGC\,4038/39) to late stage systems (merger remnants; e.g., NGC\,7252). 
A detailed description of the optical morphology of the nuclear regions is 
presented in Paper~I.  We emphasize that the Toomre sequence was chosen 
originally based on the visual inspection of photographic plates, mostly 
those presented in the {\em Atlas of Peculiar Galaxies} \citep{arp66}, and 
the rough merger stage has been derived from the degree of coalescence 
\citep{too77}. Although the very early and very late stages can be 
distinguished quite unambiguously, an exact placement in a sequence may be 
possible only after deriving the dynamical history of the merger through 
detailed simulations.

\subsection{Observations and Data Reduction}
\label{ss:obsreduc}

All NIR images discussed in this paper were obtained with the Near
Infrared Camera and Multi-Object Spectrometer (NICMOS) aboard
{\it HST}, which uses a $256\times256$ pixel HgCdTe array. We used the
NIC2 camera which has a field of view of $19\farcs2\times19\farcs2$
and a pixel scale of $0\farcs075$\,pix$^{-1}$. We observed seven of
the eleven Toomre systems in three filter passbands (F110W, F160W and
F205W), which resemble but do not quite match the ground-based $J$,
$H$, and $K$ bands. Additionally, we made use of archival NICMOS
images of the remaining four late-stage Toomre galaxies (NGC\,2623,
NGC\,3256, NGC\,3921 and NGC\,7252), which originate from the
following GO programs: NGC\,2623: \#7219 (P.I.: N.~Scoville);
NGC\,3256: \#7251 (P.I.: J.~Black); NGC\,3921 and NGC\,7252: \#7268
(P.I.: R.~van der Marel). All NICMOS observations were performed in
the {\em MULTIACCUM} observing mode. We note that for NGC\,2623 and
NGC\,3256 no F205W observations have been carried out, so we used
available F222M observations instead. Furthermore, for NGC\,3256 no
F110W observations were available. A complete listing of all the
observations is given in Table~\ref{t:obs}.

Our observations were obtained shortly after NICMOS was reactivated
through the installation of a cryocooler (installed on March 7, 2002).
After this reactivation, the NICMOS aperture locations shifted
slightly with respect to their prior locations in the focal plane.
However, the aperture location table used by the telescope pointing
system was not updated until later in the year. As a result, some of
our observations showed an offset of a few arcseconds between the
target and the intended location on the detector. This generally was
not a problem, since in almost all cases the target still fell on the
detector. Only our observations for NGC\,7592 were somewhat adversely
affected, as discussed in \S~\ref{ss:earlystage} below.

The data obtained in the F110W and F160W passbands were pipeline-reduced, 
including the basic reduction steps such as dark and flatfield correction 
using the STSDAS\footnote{The Space Telescope Science Data Analysis System 
is a product of the Space Telescope Science Institute, which is operated 
by AURA for NASA.} tasks {\em calnica} and {\em calnicb}. We used a dither 
pattern for the observations in order to allow for good cosmic ray removal 
and correction of instrumental blemishes such as bad pixels and the 
``grot'' (areas of reduced sensitivity due to paint flecks that have 
fallen on the detector). For the F205W observations a slightly 
different observing strategy was used, in order to subtract the 
thermal background emission from the telescope. This is only an issue 
for observations at or longward of the $K$-band. We used a small scale 
dither pattern in combination with a chop between the object and 
background positions. The chop positions were separated by either 
60\arcsec\,or 90\arcsec\,(depending on the extent of the galaxy) from 
the galaxy and contain little or no emission from the galaxy. As this 
combined strategy is currently not supported in the pipeline reduction, 
we had to manually adjust a few steps, and edit the respective table-files 
in {\em calnica}, before we could go on with processing the images with 
{\em calnicb}. For further details concerning the general reduction steps 
we refer to the NICMOS Data Handbook \citep{dic02}. Throughout the remainder 
of the paper, we refer for simplicity to the F110W, F160W and F205W/F222M 
filterbands as the $J$, $H$, and $K$-bands.

For the photometric comparison with our WFPC2 data and in order to
create color maps, we had to register and resample the images so that
the WFPC2 and NIC2 images are on the same spatial scale. In most cases
we used only the PC chip of the WFPC2 (field of view: 35\arcsec\
$\times$ 35\arcsec, as compared to 19\farcs2 $\times$ 19\farcs2 for
NIC2) on which all the nuclei save for the Antennae were centered. We
used the STSDAS task {\em reflresamp} to perform the necessary steps,
aligning all images to the NIC2 images. Rotations, translations and
scale differences were taken into account using a combination of
header information about the telescope pointing and orientation, and
matching object positions in the field. We measured the positions of
at least two reference points (e.g., stars, compact clusters or
\ion{H}{2} regions) in both the WFPC2 and NIC2 images and checked the
final registering by blinking the images. In the case of NGC\,520 S
and NGC4676 A we note that due to the lack of any reference points in
the NIR that had visible counterparts in the optical, the final
registering between the WFPC2 and NICMOS images may be off by 2--3
pixels. We did not correct for differences in the FWHM of the point-spread 
function (PSF) between the optical and infrared datasets; this may cause 
small artifacts in the central $\sim 0\farcs2$ of the color maps presented 
below. However, none of the main conclusions of our paper are affected by 
these artifacts.

Magnitudes were obtained from count rates (CR) using $m_{\rm Filter} =
-2.5\log({\rm CR}) + {\rm ZP}_{\rm Filter}$, where ZP$_{\rm Filter}$
is the magnitude zeropoint of the individual filter. Because the
NICMOS filters do not correspond exactly to the $J$, $H$ and $K$-bands, 
there are color terms in the zeropoints of the photometric transformation 
equations. These color terms can vary throughout each galaxy, due to stellar 
population gradients and the generally very inhomogeneous distributions of 
dust. To calculate and correct for the color terms on a pixel by pixel basis 
is not straightforward. We have therefore taken a simplified approach. First, 
we adopted an initial guess zeropoint that depends only on the filter. These 
zeropoints were based on the color terms appropriate for a $10^9$ year old 
population \citep[as given by][]{bru03} without additional reddening. The 
color terms were calculated using the {\em calcphot} task in the STSDAS
package {\em synphot}. We then refined the zeropoints on a galaxy-by-galaxy 
basis. For each galaxy we calculated (as discussed in \S~\ref{s:colors}) 
the count rates in several apertures of fixed sizes. These were combined 
with the initial-guess zeropoints to obtain initial guesses for the $J$, 
$H$ and $K$ magnitudes. We then used these magnitudes to estimate the colors 
in the aperture, and subsequently we used these colors to improve the 
zeropoint estimates.

From a variety of tests we infer that this iterative scheme yields
magnitudes that are accurate to $\sim 0.05$ mag in each of the $J$,
$H$ and $K$-bands. This is generally much smaller than the variations
in color either within a given galaxy, or between different galaxies.
When quoting aperture magnitudes in \S~\ref{s:colors} we use the
zeropoints calculated specifically for each given combination of
filter, galaxy and aperture. When displaying the images, constructing
surface brightness profiles, and constructing color maps, we adopt a
single zeropoint for each combination of filter and galaxy, by
averaging the zeropoints calculated for 100\,pc and 1\,kpc apertures.
Our calibrated magnitudes are in a system defined by the $J$, $H$ and
$K$-band filter transmission curves of the Bessell system
\citep{bes88}.

\section{NUCLEAR MORPHOLOGY}
\label{s:imaging}

\subsection{Introduction}
\label{ss:nucmorph}

\begin{figure}
\includegraphics[angle=270,scale=0.7,clip=t]{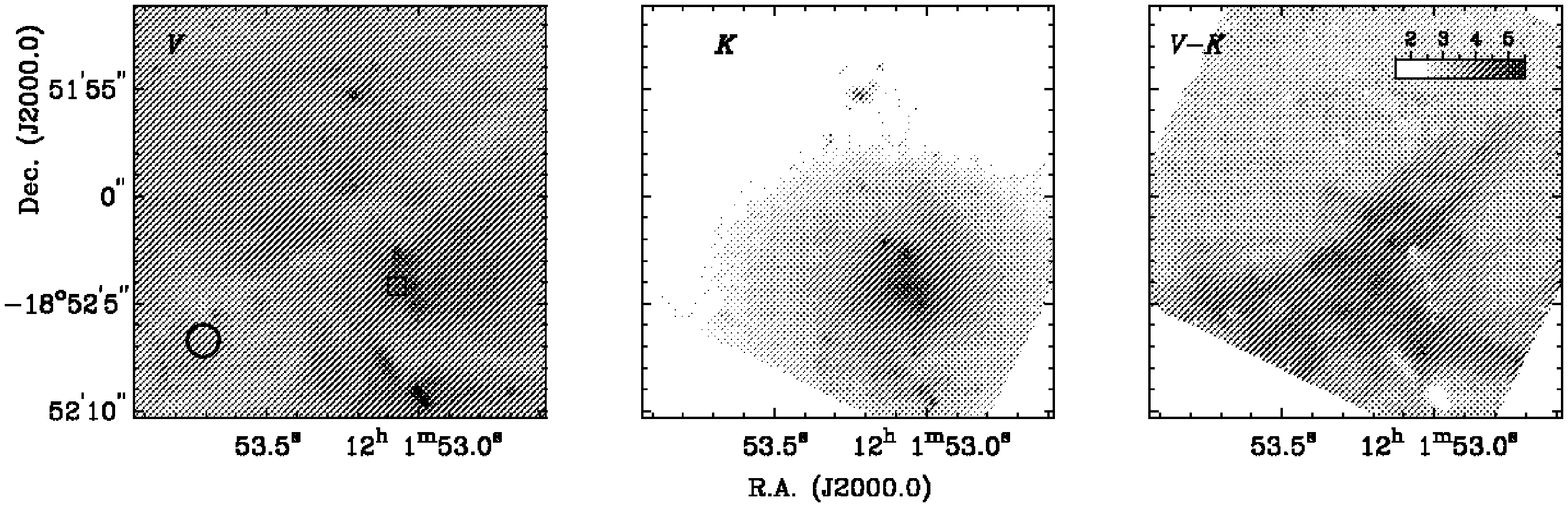}
\caption{$V$-band (left), $K$-band (center), and $V-K$ (right) images
of the circumnuclear region in NGC\,4038. The nucleus is marked by a
black square in the $V$-band image, as in this particular case several
nuclear components are visible. The displayed field of view (fov)
translates to a physical size of 1.3\,kpc $\times$ 1.3\,kpc. To
visually indicate the scale we show an aperture of diameter 100\,pc
(solid circle) and, for all nuclei other than the Antennae, also a
1\,kpc (dashed circle) at the bottom left of the $V$-band image.
\label{f:fortyfortyeightvk}}
\end{figure}

Figures~\ref{f:fortyfortyeightvk} (NGC\,4038) through
\ref{f:seventytwofiftytwovk} (NGC\,7252) show images of the nuclear
regions of the Toomre sequence galaxies. Most figures show three
panels arranged in the following order: the left panel in each figure
shows the WFPC2 F555W (hereafter $V$-band) image from Paper~I; the
middle panel shows the NICMOS $K$-band image from the present study;
and the right panel shows the $V-K$ image. A comparison of the $V$ and
$K$ bands provides the longest baseline in color space available from
our {\it HST} data. We also created color images of $J-H$, $J-K$, and
$H-K$, composed entirely of the NICMOS data. However, these generally
show the same structure as the $V-K$ images, at lower contrast, so we
do not present them here. Also, most galaxies reveal very similar
morphology in the $J$, $H$, and $K$ band, so we only show the $K$
images.

In this section we describe the nuclear morphologies on a
galaxy-by-galaxy basis, focusing on the information revealed by the
$J$, $H$, and $K$ images. Of course, many of the Toomre sequence
galaxies have also been observed at other wavelengths. For example,
seven of the eleven systems have X-ray imaging from either Chandra
and/or XMM-Newton in the X-ray regime, while NGC\,6621/22 was studied
with ROSAT/PSPC; only NGC\,3509, NGC\,7592 and NGC\,7764A have not yet
been studied in X-rays. Many of the Toomre sequence galaxies have also
been studied with moderate angular resolution in \ion{H}{1} and in the
radio continuum and in CO lines
\citep[e.g.,][]{nor95,hib96,geo00,hib01,ion05}. All these observations
at other wavelengths have provided a wealth of information. However,
we highlight in the discussions below only those aspects that have a
direct impact on the interpretation of features observed in the NICMOS
images.

One advantage of our high spatial resolution NICMOS data is that it
allows better identification of the nuclei in the Toomre sequence
galaxies than do either ground-based or optical data. In this context
we define ``nucleus'' loosely as a light concentration that has
properties consistent with it being the center of the galaxy itself,
or one of its merging progenitor galaxies. Table~\ref{t:ts} lists the
nuclei identified in our data, including their right ascension and
declination. These positions are measured from the NICMOS data, and
have an absolute accuracy limited to $\sim 1$\farcs6 \citep{pta06},
due to guide star catalog uncertainties and uncertainties in the
relative positions of the {\it HST} instruments and Fine Guidance
Sensors (FGS) in the {\it HST} focal plane. For the NICMOS imaging of
NGC\,3921, we have some evidence that the astrometry is in fact off by
several arcseconds, and believe that the absolute positions inferred
from the WFPC2 data, as listed in Paper~I, are more accurate. While 
we typically have not marked in 
Figures~\ref{f:fortyfortyeightvk}--\ref{f:seventytwofiftytwovk} the 
positions of the nuclei we identify, their positions can always be easily 
located (if they are not already obvious from the images) by combining the 
positions in the table with the coordinates listed along the image axes.

In order to unambiguously identify the nuclei of the Toomre sequence, 
it is important to compare the derived infrared positions of the 
nuclei with measurements at other wavelengths that are relatively 
insensitive to dust, such as X-rays, radio continuum and CO emission. 
Table~\ref{t:pos} lists the offsets between the absolute positions 
inferred from our NICMOS data and those inferred from data in these 
other wavelengths (where available). In general the positions from 
different wavelengths agree within the uncertainties; in the 
discussions below we address the nuclear positions only in those cases 
where the agreement is not good.

Many of the Toomre galaxies have quite complex nuclear morphologies in
the NICMOS images. Some interesting features are obvious when blinking
or displaying images at various contrast levels on a high resolution
computer screen, but can be much harder to identify and interpret on
the figures reproduced here. The discussions presented below are
therefore intended to guide the reader through the most relevant
features. Descriptions such as ``compact'', ``concentrated'', and
``extended'' are qualitative in nature and refer only to the visual
impression obtained from the images. The term ``circumnuclear region''
is used loosely to denote the region in the immediate vicinity of the
nucleus. No particular scale is necessarily implied, but we generally
use this term to denote only a small fraction (say, 10\%--30\%) of the
size of the field-of-view. The NIC2 field of view corresponds to
$7.5$\,kpc at the median distance of the sample galaxies, so the
circumnuclear region does not generally extend beyond $\sim 1$\,kpc
from the nucleus. When referring to the entire NIC2 field of view
(which is generally much smaller than the full size of the sample
galaxies) we refer to the ``central region" of the galaxy. More
quantitative measures of the nuclear structure, such as detailed
aperture photometry, color index profiles and surface brightness
profiles of the nuclei, are presented and discussed in
\S\S~\ref{s:colors} and~\ref{s:sfbprof}.

\subsection{Early-stage systems: NGC\,4038/39 (The Antennae),
NGC\,4676 (The Mice), NGC7592, NGC\,7764A}
\label{ss:earlystage}

\begin{figure}
\includegraphics[angle=270,scale=0.7,clip=t]{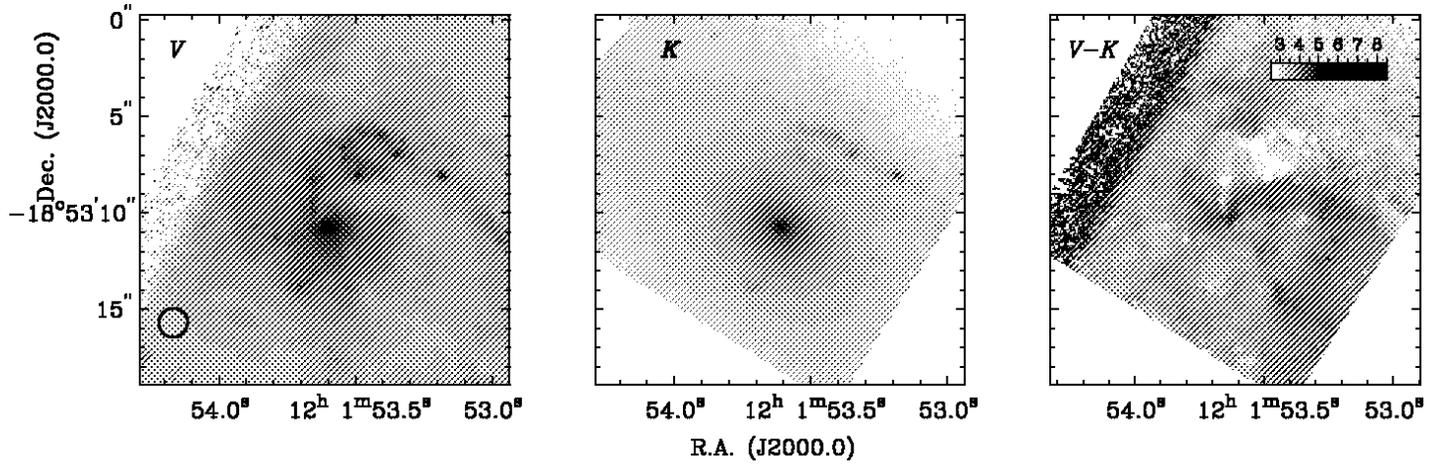}
\caption{As Figure~\ref{f:fortyfortyeightvk}, but for NGC\,4039 (fov:
1.3\,kpc
$\times$ 1.3\,kpc). \label{f:fortyfortyninevk}}
\end{figure}

\begin{figure}
\includegraphics[angle=270,scale=0.7,clip=t]{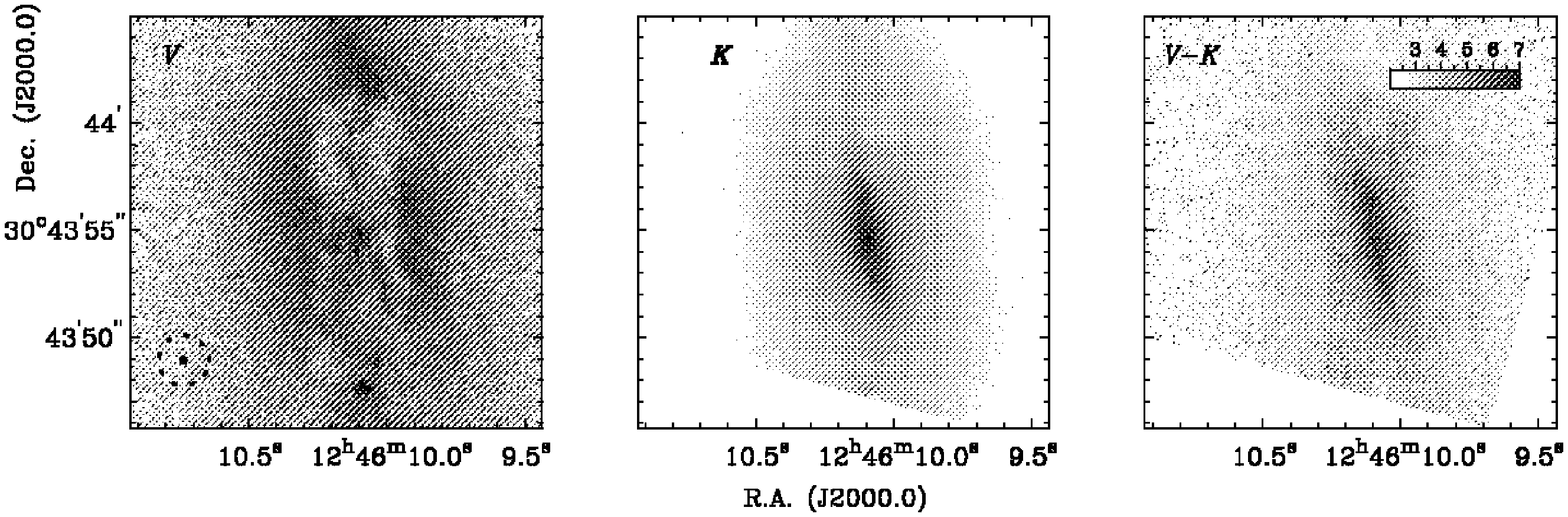}
\caption{As Figure~\ref{f:fortyfortyeightvk}, but for NGC\,4676 A
(fov: 8.1\,kpc $\times$ 8.1\,kpc). \label{f:fortysixseventysixavk}}
\end{figure}

\begin{figure}
\includegraphics[angle=270,scale=0.7,clip=t]{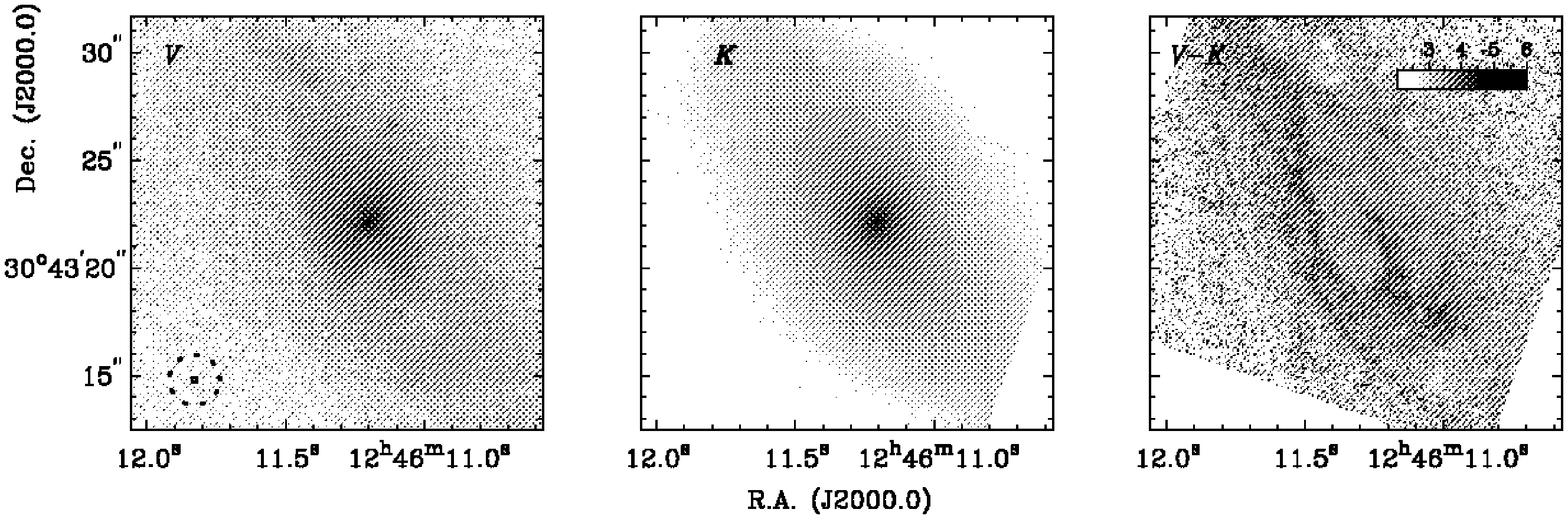}
\caption{As Figure~\ref{f:fortyfortyeightvk}, but for NGC\,4676 B
(fov: 8.1\,kpc $\times$ 8.1\,kpc). \label{f:fortysixseventysixbvk}}
\end{figure}

\begin{figure}
\includegraphics[angle=270,scale=0.7,clip=t]{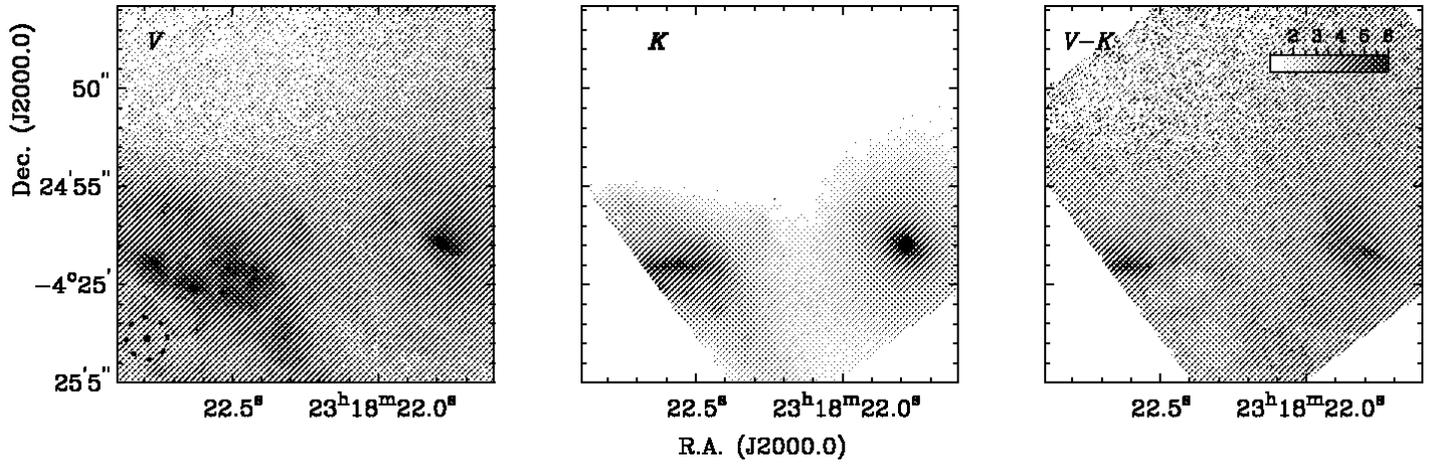}
\caption{As Figure~\ref{f:fortyfortyeightvk}, but for NGC\,7592 (fov:
8.9\,kpc $\times$ 8.9\,kpc). \label{f:seventyfiveninetytwovk}}
\end{figure}

\begin{figure}
\includegraphics[angle=270,scale=0.7,clip=t]{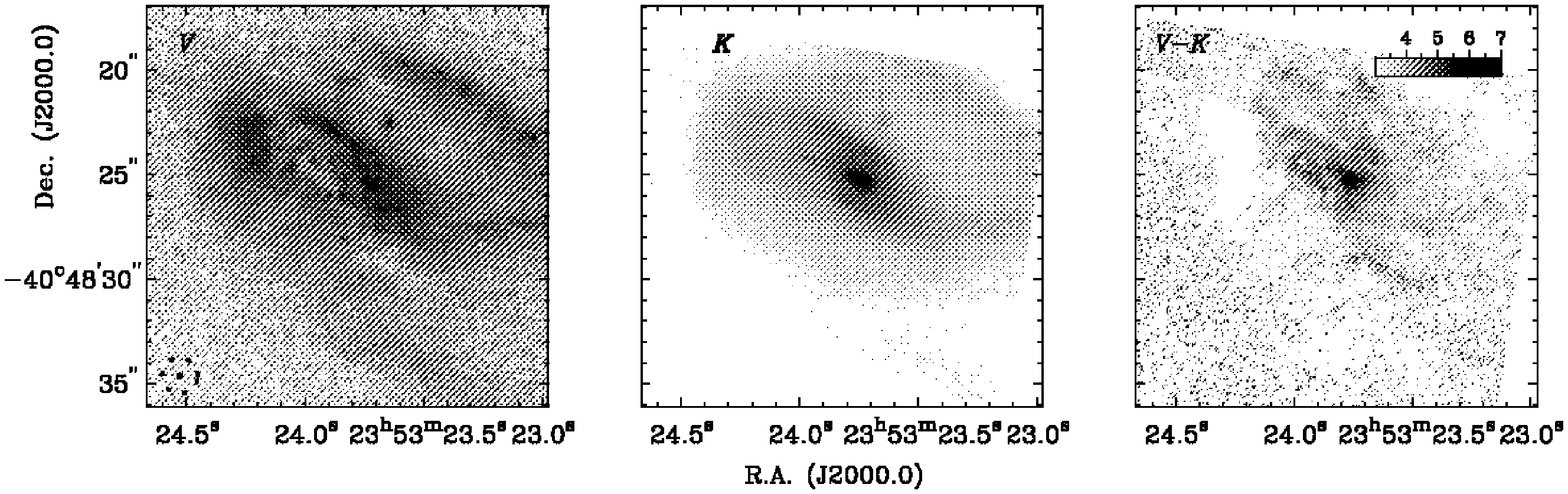}
\caption{As Figure~\ref{f:fortyfortyeightvk}, but for NGC\,7764A (fov:
11.3\,kpc $\times$ 11.3\,kpc). \label{f:seventysevensixtyfourvk}}
\end{figure}

\begin{figure}
\includegraphics[angle=270,scale=0.7,clip=t]{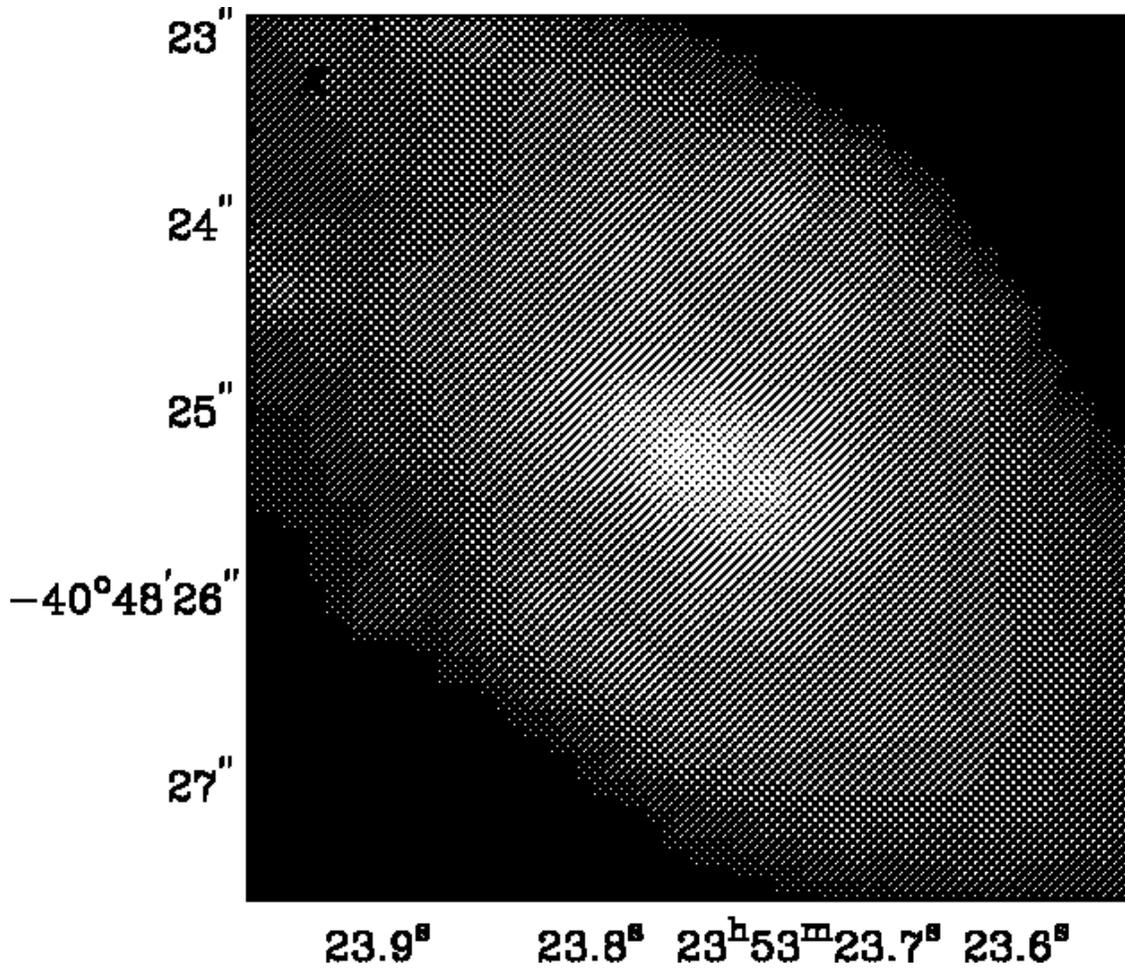}
\caption{Enlarged view of the $K$-band image of the circumnuclear
region in NGC\,7764A, shown in inverted grey-scale and revealing the
two nuclei. The displayed field of view translates to a physical size
of 2.7\,kpc $\times$ 2.7\,kpc. \label{f:seventysevenenlarged}}
\end{figure}

The Antennae\footnote{We use a different distance to the Antennae as
compared to Paper~I. In that paper we adopted the distance of $21.6
\Mpc$ implied by the observed redshift of the Antennae, neglecting any
potential peculiar velocity of the system. Instead, we adopt here the
value of 13.8\,Mpc, which was recently derived from measurements of
the tip of the red giant branch \citep[TRGB][]{sav04}. While this
distance puts the Antennae closer than has been assumed in most
previous work, the TRGB-method has been well-demonstrated and the
inferred distance should be more accurate than the canonical 
redshift-distance.} were imaged in two separate pointings, one centered 
on the nucleus of NGC\,4038 (Figure~\ref{f:fortyfortyeightvk}) and the other 
centered on the nucleus of NGC\,4039 
(Figure~\ref{f:fortyfortyninevk}). As a result of the NICMOS aperture 
location offset discussed in \S~\ref{ss:obsreduc}, the images were not 
centered exactly on the presumed nuclear positions (determined from 
previous WFPC2 imagery). The nucleus of NGC\,4039 is an easily 
identified concentrated source (see Figure~\ref{f:fortyfortyninevk}). 
By contrast, NGC\,4038 does not seem to possess a highly concentrated 
nuclear region; however, a probable nucleus can be identified in all 
three NIR filterbands (marked with a box in 
Figure~\ref{f:fortyfortyeightvk}). Given the patchy dust obscuring the 
view into the nucleus in the optical regime, optical identifications 
of the nuclei can be misleading. In the case of NGC 4038, the nucleus 
identified in Paper~I is most likely a circumnuclear \ion{H}{2} region 
\citep[see also][]{whi99}. It is very blue in $V-K$, and is likely a 
foreground star-forming knot located in front of a region of high 
extinction. The nucleus identified here using the NICMOS images is 
barely visible in the optical.

A comparison with the results obtained from data at other wavelengths
does not provide an unambiguous answer to the question of whether the
nucleus of NGC\,4038 has been correctly identified. This may not be
surprising, given the intrinsically complex nature of the nuclear
region. We have identified the clump closest to the center of the
overall light profile as the nucleus; this position is consistent with
that implied by radio continuum data, if one assumes that the nucleus
coincides with the brightest source observed at $\lambda$6\,cm (as
assumed for Table~\ref{t:pos}). However, there are several bright
sources in the nuclear area. If the radio continuum nucleus is
presumed to reside at the centroid of these sources (as do
\citet{nef00}, see their Figure~5a) then the radio continuum and
NICMOS positions are not in mutual agreement. A more accurate
alignment of the near-IR and radio data might be possible using the
technique of \citet{whi02}, but we do not pursue that here.

The Antennae have also been studied extensively in the X-ray regime
\citep[e.g.,][and follow-up papers]{fab97,fab01}. The NICMOS position
of NGC 4038's nucleus is consistent with source \#24 in the list of
\citet{zez02a,zez02b}, and may well be the nucleus (as assumed for
Table~\ref{t:pos}). However, \citet{zez02a,zez02b} actually identify
their source \#25 as the nucleus, offset in declination by about
6\arcsec\ from the NICMOS position. There is little evidence in the
NICMOS images to favor source \#25 as the nucleus, and in fact source
\#24 has a much larger X-ray flux.

We have observed NGC\,4676 in two pointings, and clearly detect the
nuclei of both galaxies. The northern galaxy, NGC\,4676\,A
(Figure~\ref{f:fortysixseventysixavk}), shows a highly concentrated
infrared source, completely obscured in the optical, which we identify
as the nucleus. The second nucleus (NGC\,4676\,B) shows a concentrated
nuclear region with a smooth appearance
(Figure~\ref{f:fortysixseventysixbvk}).

We acquired NICMOS images of NGC\,7592 in one pointing, as both nuclei
in principle fit into the field of view of the NIC2 aperture (see
Figure~\ref{f:seventyfiveninetytwovk}). However, this observation was
affected by the NICMOS aperture location offset, discussed in
\S~\ref{ss:obsreduc}, which placed the second circumnuclear region
right on the edge of the NIC2 aperture. Although the image covers most
of the circumnuclear region, this region shows a very complex
morphology and there is a chance that the true eastern nucleus
(NGC\,7592 E) is located outside the field of view. In contrast, the
western nucleus (NGC\,7592\,W) is clearly defined and compact, and was
previously identified by \citet{raf92} to be of Seyfert\,2 type.

The nuclear region of NGC\,7764A is shown in
Figure~\ref{f:seventysevensixtyfourvk}.  When enlarged, all three
NICMOS images ($J$, $H$ and $K$) reveal what appears to be a double
nucleus in this system (see Figure~\ref{f:seventysevenenlarged}). If
these truly are the two merging galaxy nuclei, it is consistent with
the placement of NGC\,7764A in the middle of the Toomre sequence. The
two putative nuclei are separated in a nearly east-west direction by
$0\farcs43$, which corresponds to 260\,pc at the distance to
NGC\,7764A. This is similar to the separation of the two nuclei in the
merging galaxy Arp\,220 \citep{sco98}.

\subsection{Mid-stage systems: NGC\,6621/22, NGC\,3509, NGC\,520}
\label{ss:midstage}

The nucleus of NGC\,6621 is detected in the infrared, and has a
fainter counterpart in the optical images (see
Figure~\ref{f:sixtysixtwentyonevk}). The circumnuclear region has a
complex morphology, even in the $K$-band. In contrast, the nuclear
region of NGC\,6622 appears much less complex, showing only a single
concentrated source which we identify as the nucleus
(Figure~\ref{f:sixtysixtwentytwovk}).

The nuclear region of NGC\,3509 looks rather inconspicuous, with only
a single, well-defined nucleus (Figure~\ref{f:thirtyfivezeroninevk}),
appearing as a concentrated but clearly resolved source. As noted
already in Paper~I, NGC\,3509 has possibly experienced a {\em minor}
merger in the past, as only one clear tidal tail is visible in deep
optical images \citep[cf.][]{arp66}. The relatively low value of the
IRAS flux density ratio at $60\mu$m over $100\mu$m ($f_{60}/f_{100}$
$\approx$ 0.33) also favors a scenario in which star formation has
been suppressed compared to the other mid-stage mergers in the Toomre
sequence, which show more vigorous signs of recent or ongoing star
formation (cf.~Figures 7 and 8 of Paper~I).

NGC\,520 was also observed in two pointings. The NIR observations of
the southern nucleus (NGC\,520 S) are quite striking
(Figure~\ref{f:fivetwentyavk}), revealing a nucleus completely hidden
in the optical behind an enormous dust complex. The $J$-band
observations show a considerable amount of dust present around the
nuclear region. At longer wavelengths ($H$ and $K$) we peer deeper
through the dust toward the nucleus, but even the $K$ image is
affected by patchy dust structure around the circumnuclear
region. However, the nucleus is clearly visible, appearing as an
extended disk composed of at least seven major knots. The brightest
knot, which we identify as the nucleus, is slightly extended toward
the north. Interestingly, the nuclear disk, which appears almost
edge-on in the E-W direction, is slightly skewed with respect to the
large scale dust lane. Such a misalignment is also seen in Centaurus~A
\citep{schr98}.

Previous sub-arcsecond resolution ground-based NIR imaging also
revealed an extended disk around NGC\,520 S \citep{kot01}. However,
considerable sub-structure was only seen in Br$\gamma$ observations,
not in their $K$-band image. Radio observations at $\lambda$\,2\,cm
\citep{car90} also showed an extended disk with five prominent knots,
very similar in morphology to our five brightest components. More
recently, high spatial resolution radio continuum observations at
$\lambda$\,21.4\,cm obtained with MERLIN revealed an extended disk
with at least nine bright components \citep{bes03}, also consistent
with the NIR morphology. The much less obscured northern nucleus
(NGC\,520 N) reveals a bright centrally confined, compact morphology,
embedded in an amorphous disk that is highly inclined to the 
line-of-sight (see Figure~\ref{f:fivetwentybvk}).

\subsection{Late-stage systems: NGC\,2623, NGC\,3256, NGC\,3921,
NGC\,7252}
\label{ss:latestage}

The nuclear region of NGC\,2623 shows a highly concentrated light
profile (Figure~\ref{f:twentysixtwentythreevk}). Consistent with our
previous findings (see Paper~I), only one nucleus is identified,
appearing slightly elongated in the E-W direction. There is no clear
nucleus discernible in the optical image.

Our analysis of the late-stage merger galaxy NGC\,3256 is based on
archival NICMOS images in $H$ and F222M (hereafter referred to as $K$;
see Figure~\ref{f:thirtytwofiftysixvk}); no $J$ band data exists. The
nuclear region (NGC\,3256\,N) appears as a concentrated and smooth
source. Radio imaging at $\lambda$\,3\,cm and $\lambda$\,6\,cm,
performed by \citet{nor95}, identified a second nucleus $5\arcsec$
south of the primary (NGC\,3256\,S). Studies of molecular gas
emission also hint at a second nucleus at the same location
\citep{sak06}. \citet{boe97} also imaged NGC\,3256 in the $N$-band and
detected a source at this same position. Based on radio and X-ray
power and their ratios \citet{nef03} argue that both nuclei may be
low-luminosity active galactic nuclei (LLAGNs). In our NICMOS data
there is a compact source at the position of the secondary nucleus,
but with fainter intensity than the primary nucleus. The source appears 
projected onto the middle of a spiral arm, lying in a highly reddened region 
corresponding to a local maximum in the $V-K$ image. It has an 
extinction-corrected $K$-band magnitude of $-18.2$, similar to the nuclei in 
the other Toomre sequence galaxies (see \S\ref{ss:mergerstage}). All this 
supports the view that there is in fact a highly extincted infrared 
counterpart to the secondary nucleus that was detected in radio, MIR, and 
X-ray data.

We also analyzed archival images of NGC\,3921. The nuclear region
shows a slightly elongated core in the NNE to SSW direction (see
Figure~\ref{f:thirtyninetwentyonevk}), with only one nucleus
detected. Our findings are consistent with earlier studies
\citep{sch96a} and with the fact that NGC\,3921 is generally regarded
as a late-stage merger remnant with properties resembling elliptical
galaxies. In a recent X-ray study by \citet{nol04}, however, the X-ray
luminosity of the hot diffuse gas was found to be two orders of
magnitude less than that typical of elliptical galaxies.

Our images of the latest-stage merger remnant of the Toomre sequence,
NGC\,7252, show a nuclear spiral (see
Figure~\ref{f:seventytwofiftytwovk}), as already detected in the
optical and described in our previous study (Paper~I). Only one
nucleus is identified in all three filter passbands, consistent with
previous ground-based studies \citep[e.g.,][]{sch82,hib94} and with
the general notion of NGC\,7252 being a prototype merger remnant, with
an estimated age of about 1\,Gyr \citep[e.g.,][]{sch98}.

\section{PHOTOMETRY AND COLORS}
\label{s:colors}

\subsection{Photometry}
\label{ss:photom}

We measured the $J$, $H$, and $K$ magnitudes for all identified Toomre
sequence nuclei within apertures 100\,pc and 1\,kpc in size.
Depending on the distance to each galaxy, the 100\,pc aperture
corresponds to 2--20 pixels and the 1\,kpc aperture corresponds to
23--200 pixels. Fluxes were measured with the IRAF\footnote{IRAF is
distributed by the National Optical Astronomy Observatory, which is
operated by the Association of Universities for Research in Astronomy,
Inc. (AURA) under cooperative agreement with the National Science
Foundation.} task {\it phot}, centered on the identified nuclei; the
resulting magnitudes are listed in Table~\ref{t:colors} and
Table~\ref{t:colorsonek}. The photometric random errors on the
magnitudes and colors are very small (with a median of 0.01 mag) and
are not listed in the tables. The actual errors are dominated by
systematic calibration uncertainties ($\lta 0.05$\,mag in each band,
see \S~\ref{ss:obsreduc}).

\subsection{Near-Infrared Colors}
\label{ss:colors}

Figure~\ref{f:jmhhmkonehpeg} shows the $J-H$ vs. $H-K$ color-color
diagram for the 100\,pc aperture measurements of the Toomre sequence
nuclei. Separate symbols are used for the early-stage, mid-stage, and
late-stage mergers, as indicated in Table~\ref{t:ts}. For a direct
comparison of the late-stage mergers with elliptical galaxies, we also
show in the figure the sample of elliptical galaxies studied by
\citet{sil98}. We transformed their data, which were calibrated to the
CIT photometric system, to the Bessell system in which our NICMOS data
were calibrated \citep{bes88}. We also used the respective $J-H$ and
$H-K$ values from the {\it GALAXEV} database to generate and overplot
the evolutionary track of an aging simple population \citep{bru03}. We
used the Bruzual-Charlot models with the Padua 1994 stellar evolution
library for a standard Chabrier initial mass function (IMF) from
$0.1\,\rm{M_{\sun}}$ to $100\,\rm{M_{\sun}}$ \citep{cha03} and solar
metallicity. Two of the three late-stage merger remnants (those which
had $J$, $H$, and $K$ measurements) fall close to the positions of the
elliptical galaxies studied by \citet{sil98}. All Toomre sequence
nuclei are closely aligned along the reddening vector direction, as
expected for dusty stellar populations.

Figure~\ref{f:jmhhmkonehpms} shows the same color-color data for the
Toomre nuclei as Figure~\ref{f:jmhhmkonehpeg}, but focuses on a
smaller region of the color-color diagram. We compare our data to
ground-based measurements from \citet{bus92} (also transformed to the
Bessell system) for a larger sample of interacting galaxies selected
from the Arp atlas \citep{arp66}. The average and median colors are
slightly redder for the Toomre sequence nuclei than for the
\citet{bus92} galaxies, but this result is only marginally significant
given the large scatter in both distributions. Moreover, a direct
comparison between the samples is not straightforward. \citet{bus92}
performed photometry in a fixed angular aperture of 5\farcs4, which is
generally much larger than the apertures of a fixed physical size that
we have considered for the NICMOS data. Also, Toomre's criteria for
characterizing a galaxy as interacting or merging were more stringent
than those of \citet{bus92}. In any case, Figure~\ref{f:jmhhmkonehpms}
shows that the majority of the Toomre sequence nuclei fall in the
region of the $J-H$ versus $H-K$ diagram that is populated by the
sample of \citet{bus92}. The Toomre sequence nuclei therefore have
colors that are comparable to those seen in other interacting and
merging galaxies.

In principle it should be possible to constrain the stellar
populations in the nuclei of the Toomre sequence galaxies using the
multi-band $V$, $I$, $J$, $H$ and $K$ photometry presented here and in
Paper~I. Such analyses have already been performed for other merging
galaxies \citep[e.g.,][]{pas04}. However, in the presence of
significant dust, as is the case here, the results are quite
uncertain. We therefore postpone a discussion of the stellar
populations to a subsequent paper (Paper~III), in which we will
present {\it HST}/STIS spectra obtained in the context of our
investigation. These spectra can be used to investigate the
populations of the nuclei with moderate obscuration. By contrast,
near-infrared spectroscopy would be required to study the populations
of the most heavily obscured nuclei.

\subsection{Dependence on Merger Stage}
\label{ss:mergerstage}

In Figures~\ref{f:jminuskonehpc} and~\ref{f:vminuskonehpc} we examine
color trends along the proposed merger sequence. The NIR $J-K$ colors
in apertures of 100\,pc and 1\,kpc are shown as a function of merger
stage in Figure~\ref{f:jminuskonehpc}. The $J-K$ colors vary between
1.0 and 2.1 for all but the two most heavily obscured nuclei
(NGC\,2623 and NGC\,520 S), which have much redder values within the
100\,pc and 1\,kpc apertures (see Tables~\ref{t:colors} and
\ref{t:colorsonek}, respectively). We have also plotted $V-K$ colors
(listed in Tables~\ref{t:colors} and~\ref{t:colorsonek}) as a function
of merger stage (see Figure~\ref{f:vminuskonehpc}), therefore
extending the spectral baseline to optical wavelengths. By and large,
these yield results consistent with those seen for $J-K$. At the
latest stages, the NIR colors tend to be less red and closer to those
of normal E/S0's. Out of 15 nuclei, the three bluest nuclei in a
$100\,\pc$ aperture are all found in the five latest galaxies of the
sequence. While this is suggestive of a trend for the nuclei to become
bluer with increasing merger stage, it is clear that there is large
scatter in color along the sequence, and that any trend is weak at
best and not monotonic.

Unfortunately, the near-IR colors measured here do not provide much
insight into variations in age along the Toomre sequence, or
variations in age between the Toomre sequence galaxies and elliptical
galaxies. The $J-K$ color of a stellar population does not vary much
as a function of age. For example, a solar-metallicity population with
a \citet{cha03} IMF has $J-K = 0.80$, 0.57, 0.82, and 0.89, at ages of
$10^{7}$, $10^{8}$, $10^{9}$, and $10^{10}$ years, respectively
\citep[using the Padua 1994 stellar library and an IMF that extends
from $0.1\,\rm{M_{\sun}}$ to $100\,\rm{M_{\sun}}$;][]{bru03}.
Moreover, realistic stellar populations for the galaxies in the
present sample cannot produce $J-K$ colors that are much redder than
$\sim 1.0$. Therefore, the redder $J-K$ colors seen in
Figure~\ref{f:jminuskonehpc} must be due to dust extinction. As a
corollary, the trend that the early-stage mergers are redder than the
late-stage mergers merely indicates that the former are dustier than
the latter.

Models of merging galaxies with black holes often invoke feedback from
an AGN to disperse obscuring dust and shut off star formation
\citep[e.g.,][]{hop05}. Combined with the hierarchical formation of
structure over a Hubble time, such models can successfully explain a
variety of observed properties related to galaxy formation and quasar
evolution \citep[e.g.,][]{hop06}. The existence of AGN-driven feedback
is supported by observations of individual galaxies such as NGC\,6240,
which has two AGN and a strong superwind \citep{kom03,ger04}. Our
finding that the dust content of the Toomre sequence nuclei has a
decreasing trend with merger stage is qualitatively consistent with
the predictions of feedback models. While the Toomre sequence does not
currently host many strong AGN (see Table~\ref{t:ts}), this is not
difficult to reconcile with the feedback models if bright AGN stages
are either very short-lived or highly obscured \citep{hop05}. It will
be valuable to use numerical models to make quantitative predictions
for the dust and gas content of interacting and merging galaxies as a
function of merger (st)age, and compare these to results such as those
obtained here.

As in our previous WFPC2 study, we also calculate the total luminosity
within the 100\,pc and 1\,kpc apertures for the sample galaxies, in
this case for the $K$-band. The $K$-band luminosity ($L_K$) is shown
in Figure~\ref{f:toomrelumikonehpc} as a function of merger stage,
where the luminosities have been corrected for dust extinction. To
perform this correction, we first assume that the stellar population
has an intrinsic $J-K \approx 0.85$. The $K$-band extinction for each
nucleus can then be estimated from Tables~\ref{t:colors} and
\ref{t:colorsonek} using the equation $A_K = 0.659 (J-K-0.85)$, based
on the extinction law of \citet{rie85}.\footnote{For NGC 3256 we used
a similar formalism for calculating $A_K$ based on the $H-K$ color,
because no $J$-band data are available for this galaxy.} Our results
are not significantly affected for other realistic choices for the
intrinsic $J-K$ color. Figure~\ref{f:toomrelumikonehpc} shows that
$L_K$ (within an aperture of fixed physical size) tends to increase
with advancing merger stage.  To statistically test this trend, we
calculate the Spearman rank-order correlation coefficient ($r_s$) for
each of the 100\,pc and 1\,kpc datasets. In both cases we find that
$r_s=0.61$, indicating a 2$\sigma$ correlation (99\% CL). We note that
the Antennae drive much of this correlation; removing the Antennae
nuclei from the analysis lowers the correlation coefficient to
$r_s=0.5$, indicating a 1.5$\sigma$ correlation (93\% CL). We caution,
however, that this analysis assumes that there are no systematic
errors which depend on the merger stage.  While not shown here, we
have done the same analysis for the {\it fractional} nuclear
luminosity (i.e., $L_{K,nuc}/L_{K,tot}$) as a function of merger stage
and find a similar trend. Because we have corrected for extinction,
the underlying cause of this trend must be either an increasing
stellar density with advancing merger stage, or a younger average
stellar age with advancing merger stage (or both). We cannot
discriminate between these two possibilities on the basis of the
near-IR imaging data alone.  Available far-infrared data also cannot
address this issue, since the global star formation rates measured by
IRAS \citep[e.g.,][] {jos85,cas91} are not necessarily a good
indicator of the star formation rates in the nuclei \citep{ken87}. As
a caveat we would like to note that the observed trend of increasing
$L_K$ as a function of advancing merger stage is based on the
initially assumed merger stage \citep{too77}, which may not
necessarily reflect the actual merger (st)age for all of the eleven
merger systems. Nonetheless, we use the pre-defined merger stage
(which is a fixed value) rather than the merger age. While the very
early- and very late-stages are well defined, an actual ranking
(i.e. an exact placement within this sequence) may be possible only
after deriving the dynamical history of the merger through detailed
simulations.

\subsection{Radial Color Gradients}
\label{ss:nucprof}

\begin{figure}
\includegraphics[angle=270,scale=0.7,clip=t]{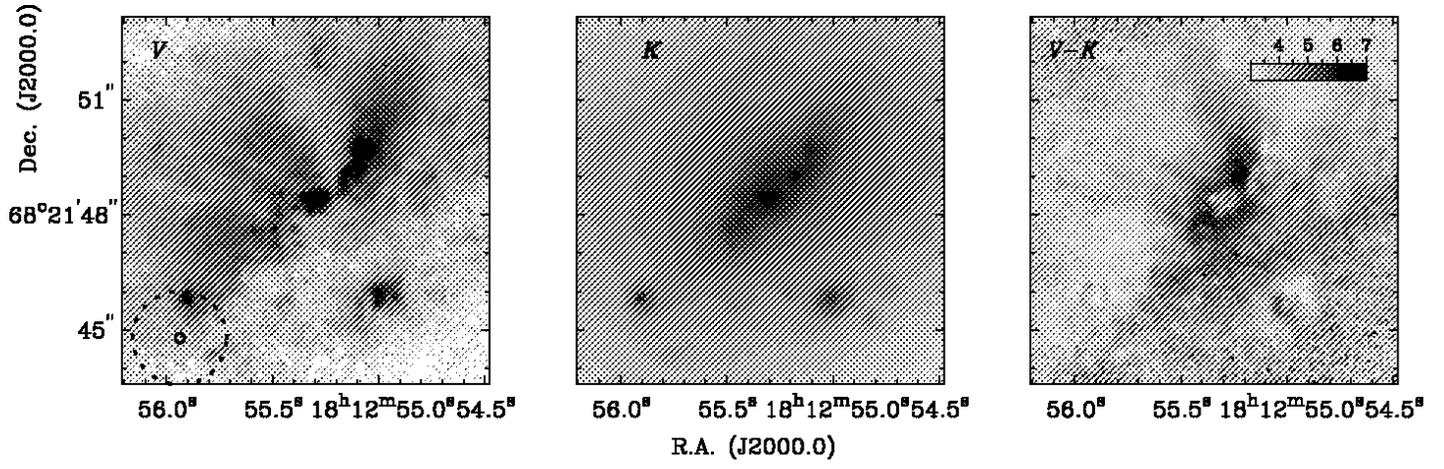}
\caption{As Figure~\ref{f:fortyfortyeightvk}, but for NGC\,6621 (fov:
3.9\,kpc $\times$ 3.9\,kpc). \label{f:sixtysixtwentyonevk}}
\end{figure}

\begin{figure}
\includegraphics[angle=270,scale=0.7,clip=t]{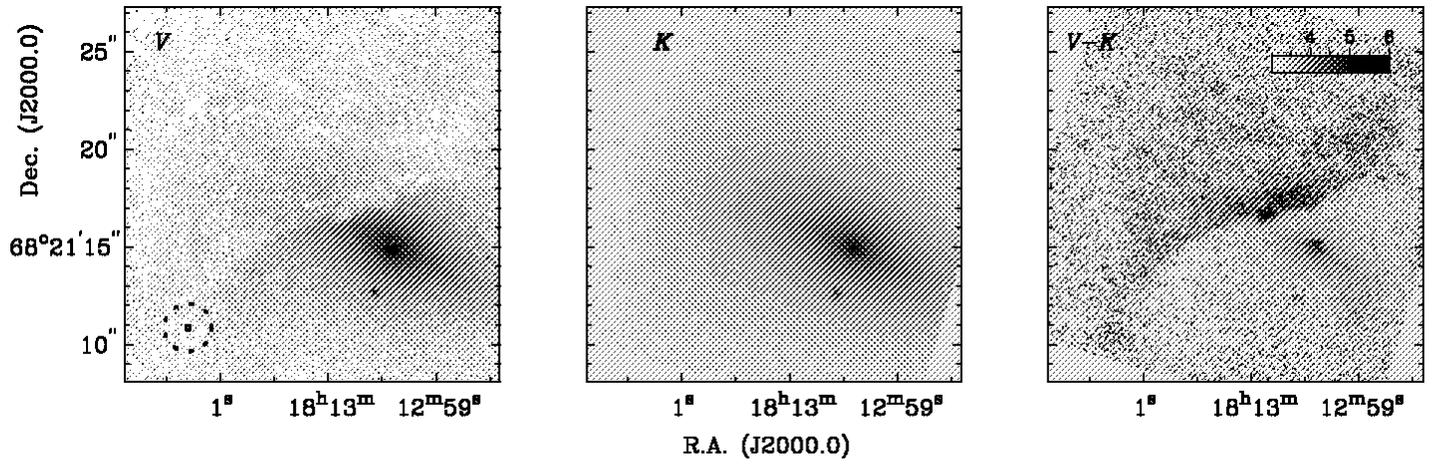}
\caption{As Figure~\ref{f:fortyfortyeightvk}, but for NGC\,6622 (fov:
7.8\,kpc $\times$ 7.8\,kpc). \label{f:sixtysixtwentytwovk}}
\end{figure}

\begin{figure}
\includegraphics[angle=270,scale=0.7,clip=t]{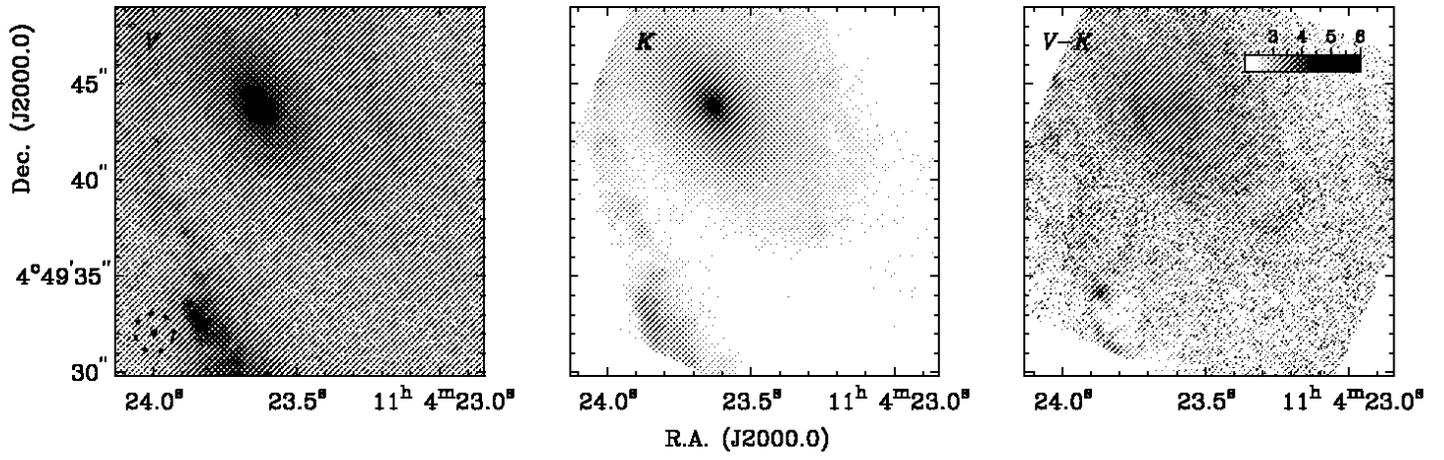}
\caption{As Figure~\ref{f:fortyfortyeightvk}, but for NGC\,3509 (fov:
9.5\,kpc $\times$ 9.5\,kpc). \label{f:thirtyfivezeroninevk}}
\end{figure}

\begin{figure}
\includegraphics[angle=270,scale=0.7,clip=t]{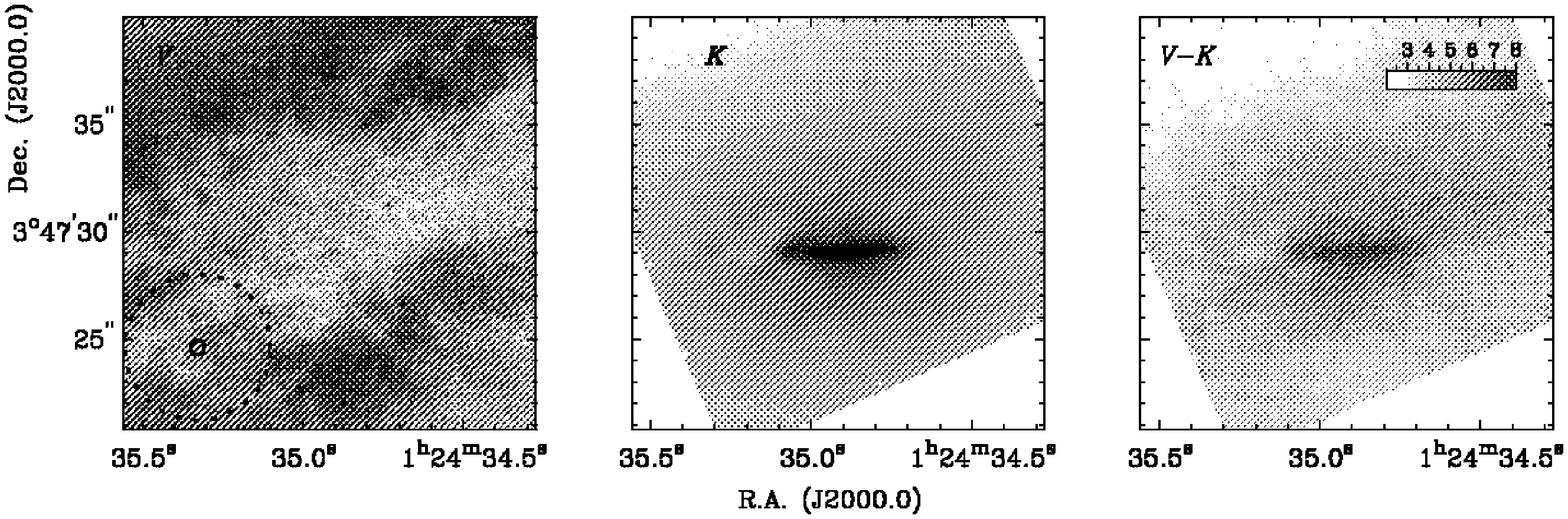}
\caption{As Figure~\ref{f:fortyfortyeightvk}, but for NGC\,520 S (fov:
2.8\,kpc $\times$ 2.8\,kpc). \label{f:fivetwentyavk}}
\end{figure}

\begin{figure}
\includegraphics[angle=270,scale=0.7,clip=t]{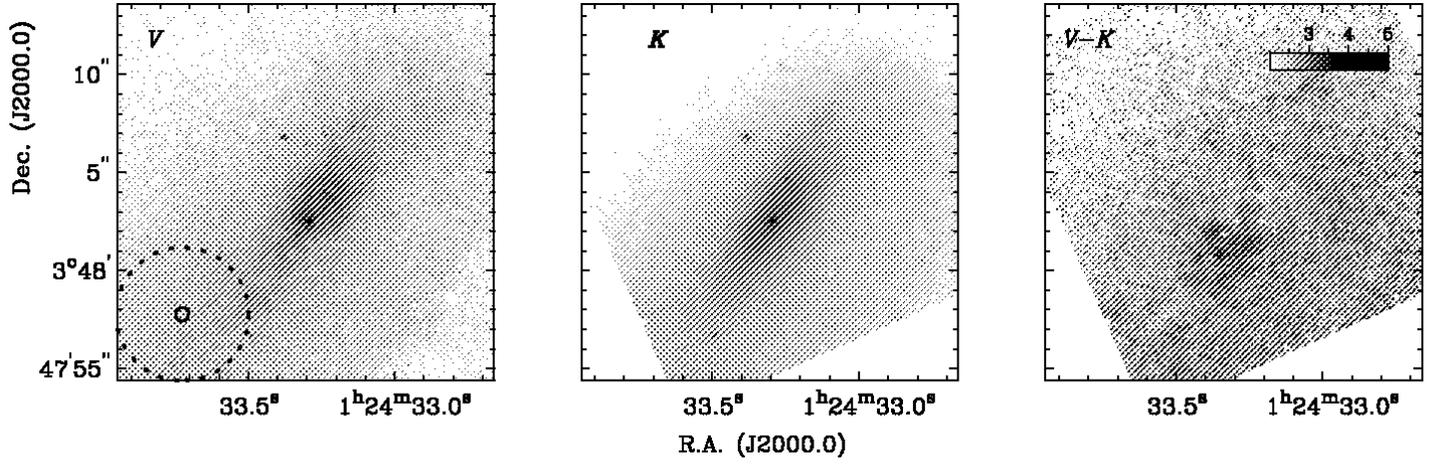}
\caption{As Figure~\ref{f:fortyfortyeightvk}, but for NGC\,520 N (fov:
2.8\,kpc $\times$ 2.8\,kpc). \label{f:fivetwentybvk}}
\end{figure}

\begin{figure}
\includegraphics[angle=270,scale=0.7,clip=t]{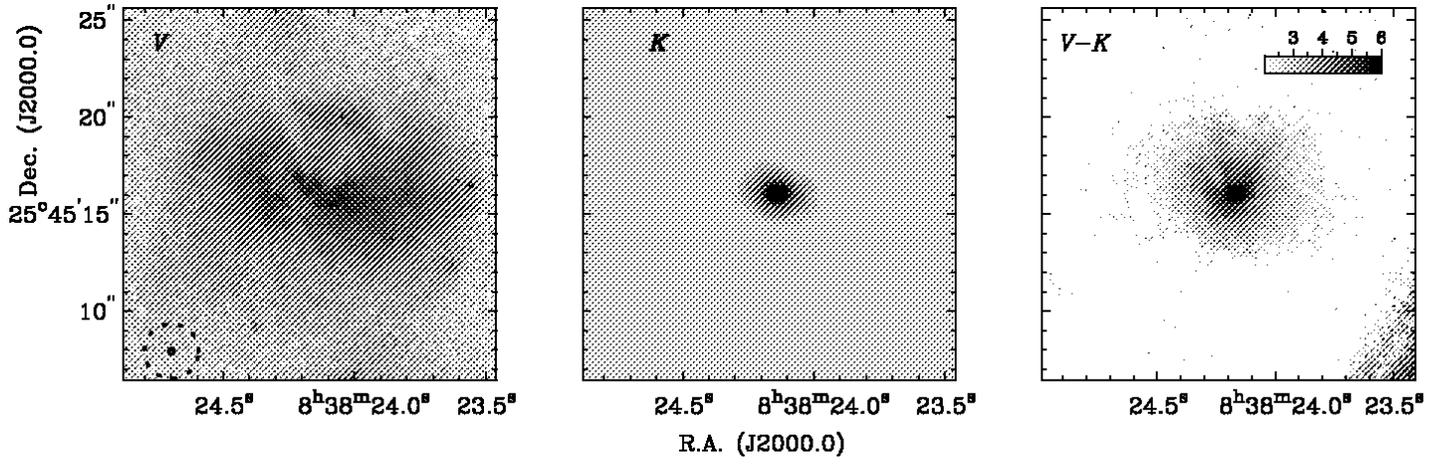}
\caption{As Figure~\ref{f:fortyfortyeightvk}, but for NGC\,2623 (fov:
6.8\,kpc $\times$ 6.8\,kpc). \label{f:twentysixtwentythreevk}}
\end{figure}

\begin{figure}
\includegraphics[angle=270,scale=0.7,clip=t]{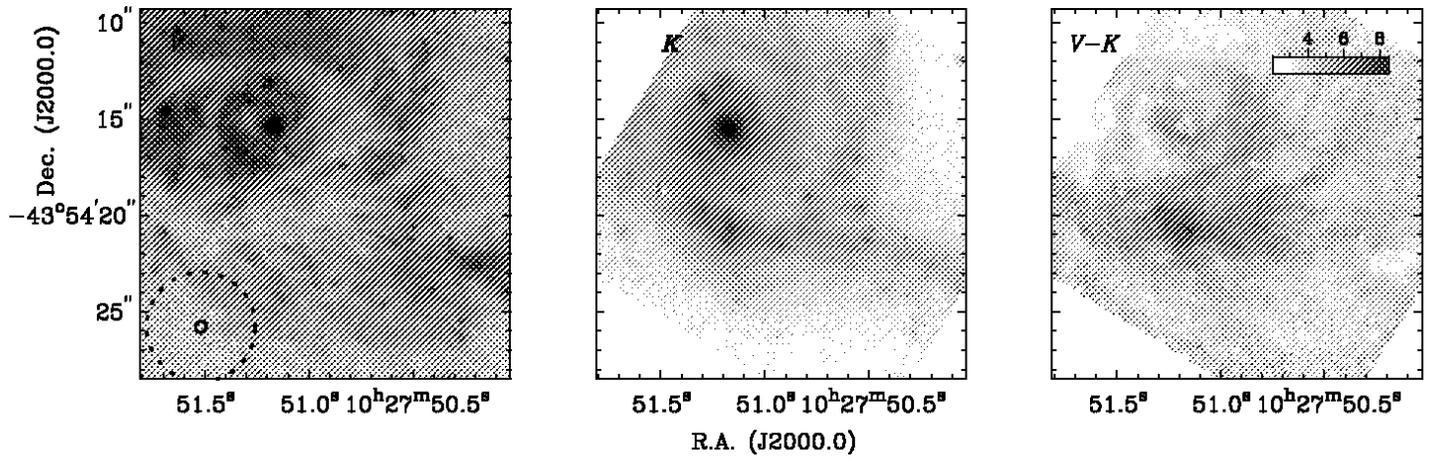}
\caption{As Figure~\ref{f:fortyfortyeightvk}, but for NGC\,3256 (fov:
3.4\,kpc $\times$ 3.4\,kpc). \label{f:thirtytwofiftysixvk}}
\end{figure}

\begin{figure}
\includegraphics[angle=270,scale=0.7,clip=t]{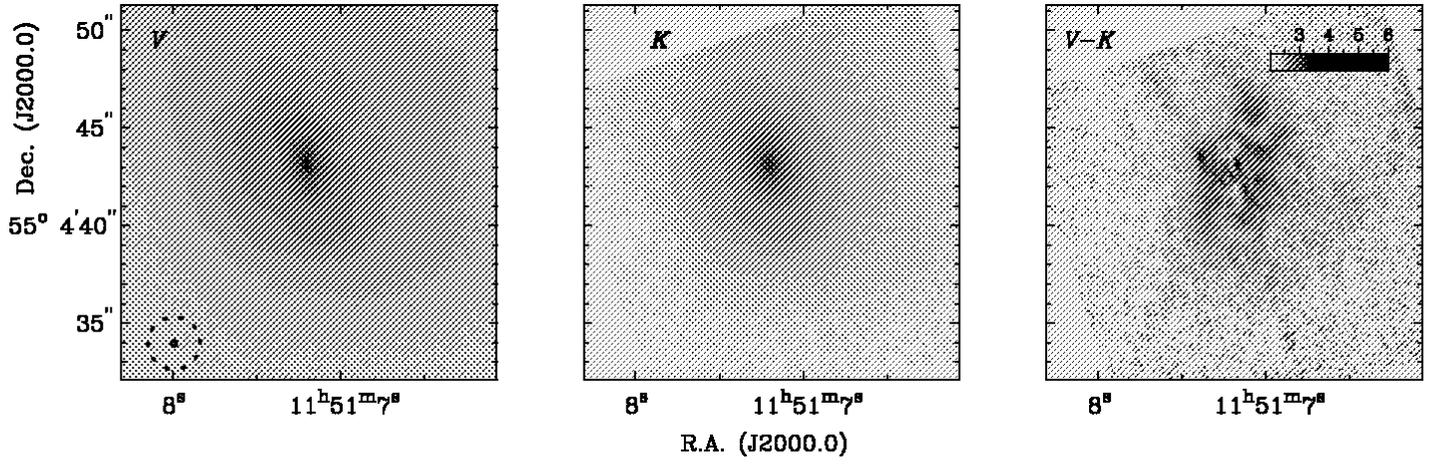}
\caption{As Figure~\ref{f:fortyfortyeightvk}, but for NGC\,3921 (fov:
7.2\,kpc $\times$ 7.2\,kpc). \label{f:thirtyninetwentyonevk}}
\end{figure}

\begin{figure}
\includegraphics[angle=270,scale=0.7,clip=t]{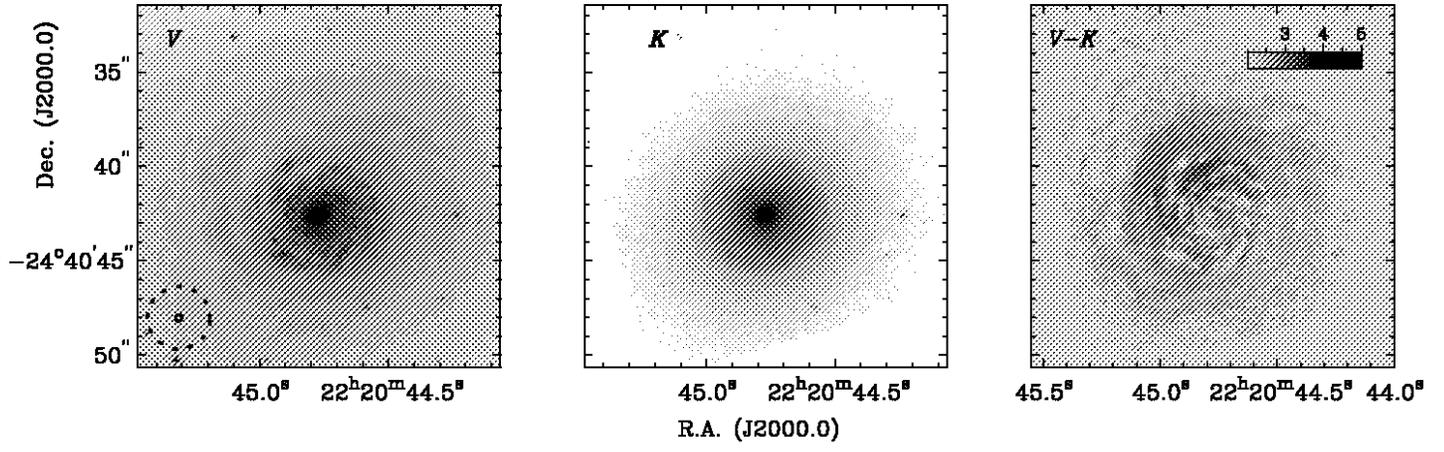}
\caption{As Figure~\ref{f:fortyfortyeightvk}, but for NGC\,7252 (fov:
5.8\,kpc $\times$ 5.8\,kpc). \label{f:seventytwofiftytwovk}}
\end{figure}

\begin{figure}
\includegraphics[angle=270,scale=0.7,clip=t]{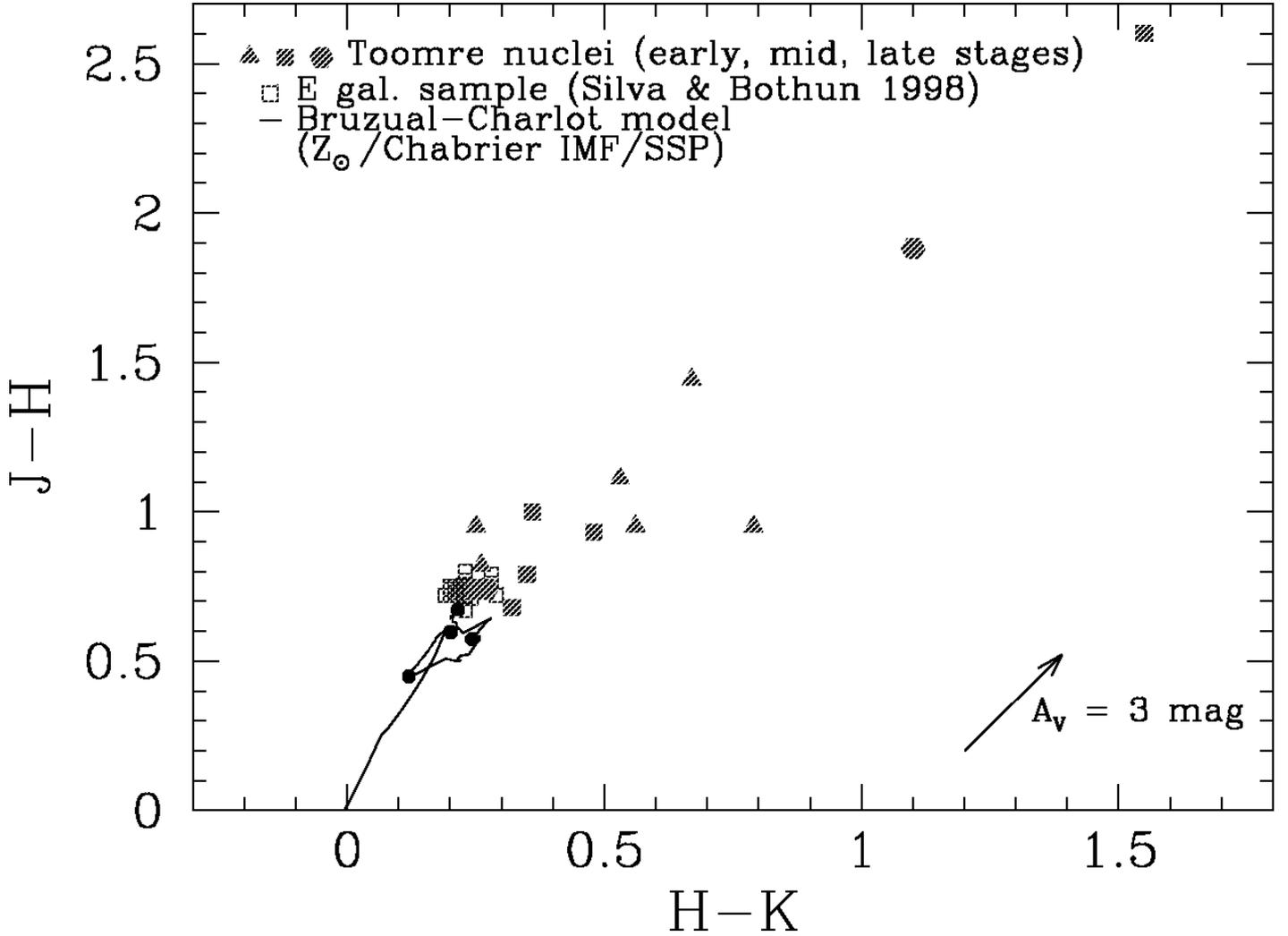}
\caption{$J-H$ vs.~$H-K$ diagram of the Toomre nuclei measured within
a 100\,pc aperture (indicated by red symbols). The sample of
elliptical galaxies studied by \citet{sil98} is also shown (blue open
squares). In addition, the Bruzual-Charlot cluster evolution track for
a Chabrier IMF and solar metallicity is overplotted. The black symbols
indicate cluster ages of $10^7$, $10^8$, $10^9$ and $10^{10}$ yrs. The
indicated reddening vector was calculated from the interstellar
extinction law values given by \citet{rie85}. The random error bars on
our measurements are generally smaller than the symbol sizes. The
actual errors are dominated by systematic calibration uncertainties
($\lta 0.05$\,mag, see \S~\ref{ss:obsreduc}). Note that the symbols
for the two late-stage mergers NGC 3921 and NGC 7252 (at $H-K \approx
0.26$ and $J-H =0.74$) are partially overlapping.
\label{f:jmhhmkonehpeg}}
\end{figure}

\begin{figure}
\includegraphics[angle=270,scale=0.7,clip=t]{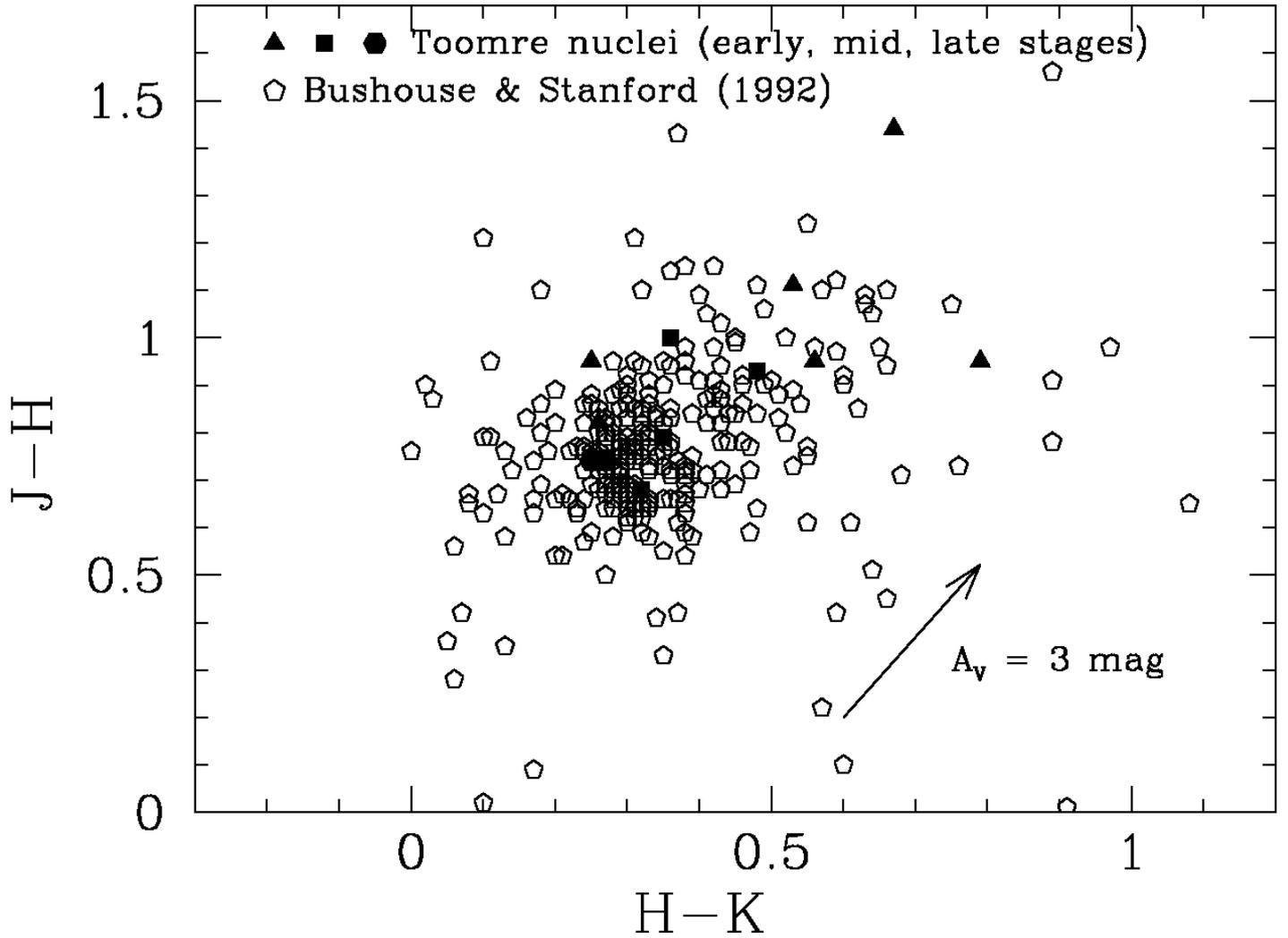}
\caption{$J-H$ vs.~$H-K$ diagram of the Toomre nuclei measured within
a 100\,pc aperture. The merger sample studied by \citet{bus92} is
overplotted (open symbols). Note that we have plotted this diagram
only to $J-H = 1.7$, omitting the two nuclei with the reddest colors
(NGC\,520 S and NGC\,2623; see Figure~\ref{f:jmhhmkonehpeg}). The
indicated reddening vector was calculated from the interstellar
extinction law values given by \citet{rie85}. The random error bars on
our measurements are generally smaller than the symbol sizes. The
actual errors are dominated by systematic calibration uncertainties
($\lta 0.05$\,mag, see \S~\ref{ss:obsreduc}). Note that the symbols
for the two late-stage mergers NGC 3921 and NGC 7252 (at $H-K \approx
0.26$ and $J-H =0.74$) are partially overlapping.
\label{f:jmhhmkonehpms}}
\end{figure}

\begin{figure}
\includegraphics[angle=270,scale=0.7,clip=t]{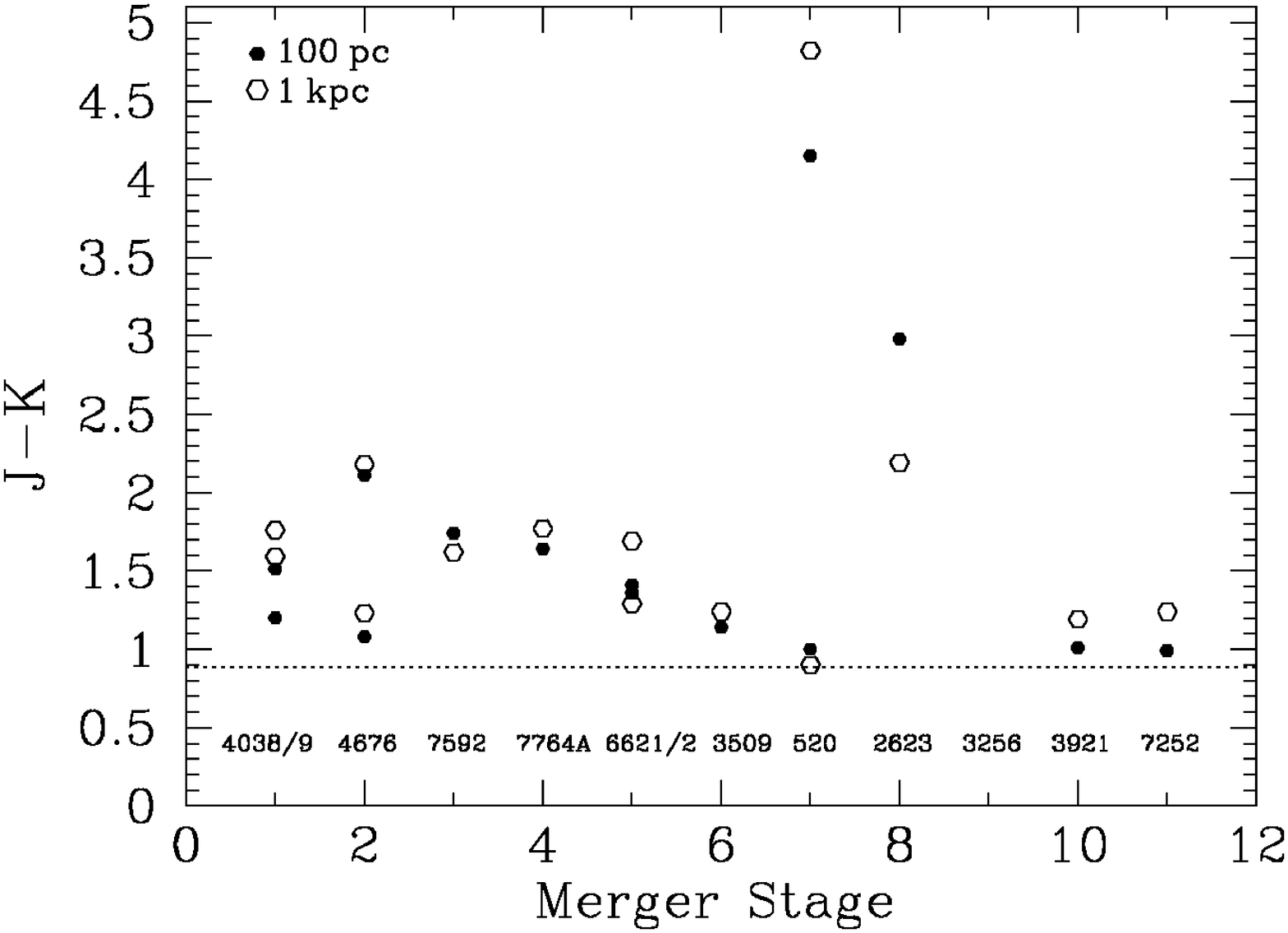}
\caption{$J-K$ color measured within 100\,pc (solid symbols) and
1\,kpc (open symbols) apertures as a function of the Toomre sequence
merger stage. The random uncertainties in the colors are generally
smaller than the symbol sizes. The true uncertainties in the colors
are dominated by systematic uncertainties in the absolute photometric
calibration (see \S~\ref{ss:obsreduc}). The dotted horizontal line
indicates the $J-K$ color for a $10^{10}$\,yr population of solar
metallicity \citep{bru03}, which provides a reasonable fit to the
actual colors measured for E/S0s. Note that for NGC\,3256 there are no
$J$-band measurements available.\label{f:jminuskonehpc}}
\end{figure}

\begin{figure}
\includegraphics[angle=270,scale=0.7,clip=t]{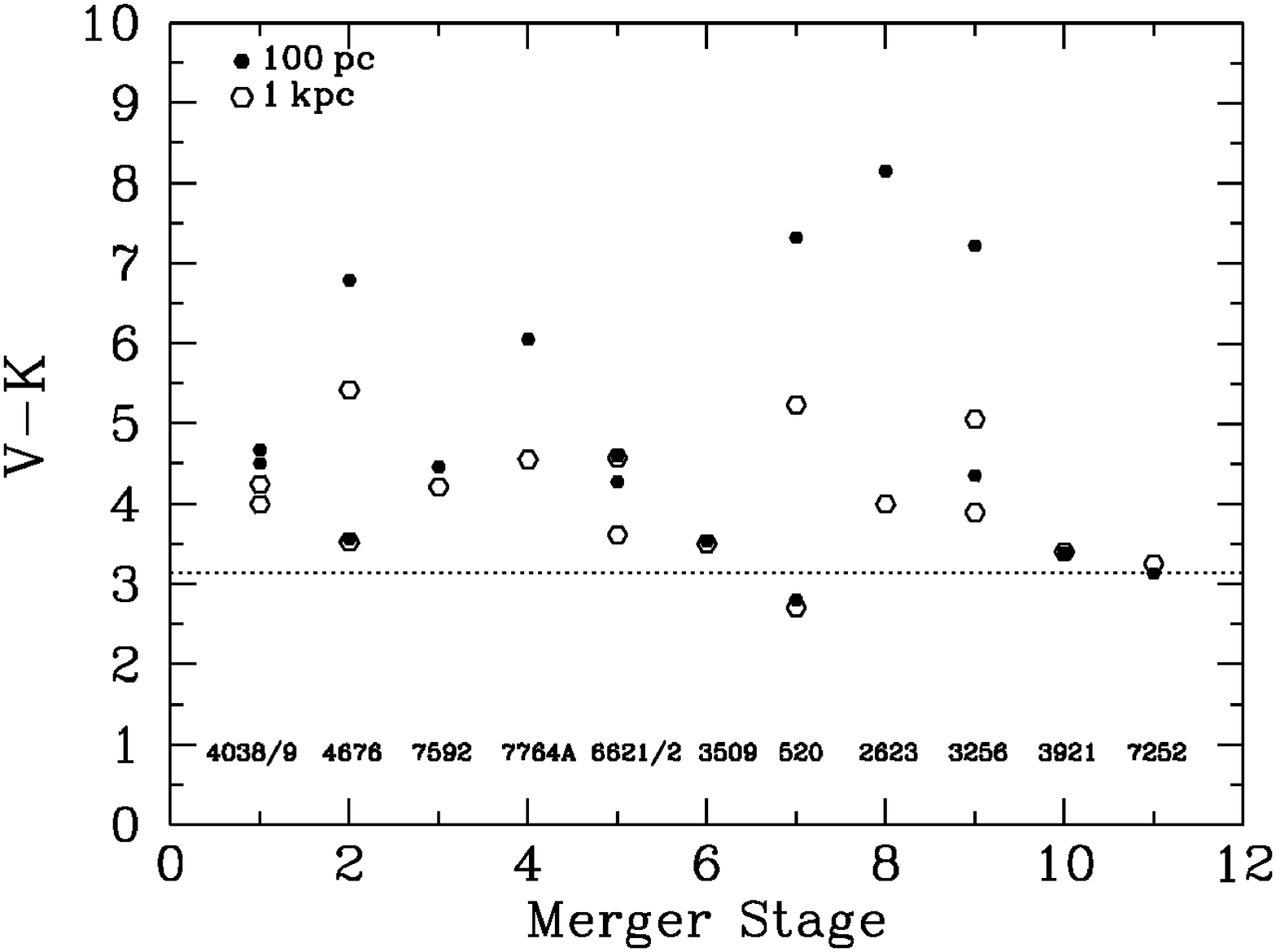}
\caption{$V-K$ color measured within 100\,pc (solid symbols) and
1\,kpc (open symbols) apertures as a function of the Toomre sequence
merger stage. The random uncertainties in the colors are generally
smaller than the symbol sizes. The true uncertainties in the colors
are often dominated by systematic uncertainties in the absolute
photometric calibration (see \S~\ref{ss:obsreduc}). The dotted
horizontal line indicates the $V-K$ color for a $10^{10}$\,yr
population of solar metallicity \citep{bru03}, which provides a
reasonable fit to the actual colors measured for E/S0s.
\label{f:vminuskonehpc}}
\end{figure}

\begin{figure}
\includegraphics[angle=270,scale=0.7,clip=t]{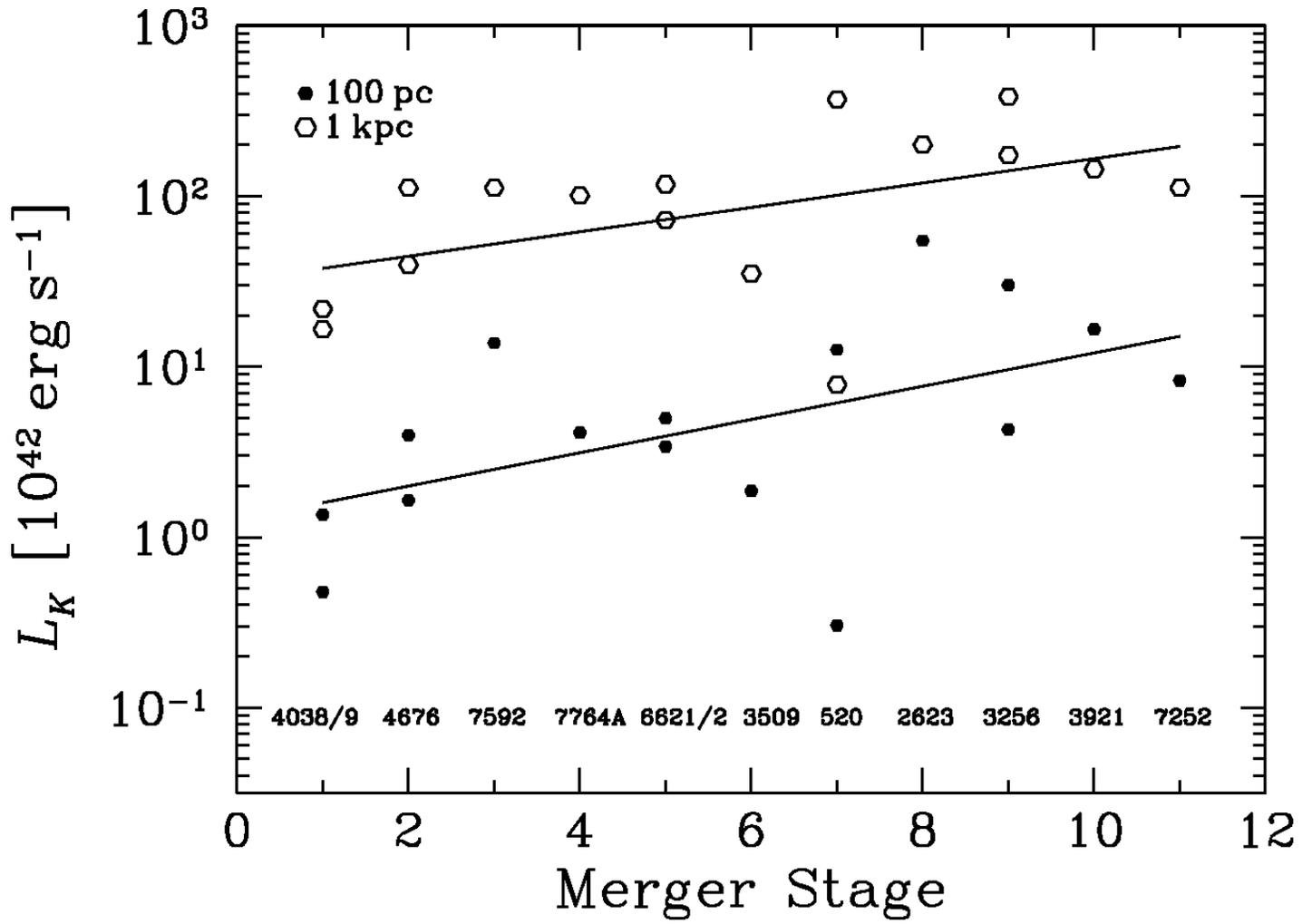}
\caption{$K$-band luminosity (corrected for dust extinction) within a
100\,pc (solid symbols) and 1\,kpc (open symbols) aperture as a
function of merger stage. The luminosities were corrected for dust
extinction based on the observed $J-K$ color, as described in the
text. The random uncertainties in the luminosities are generally
smaller than the symbol sizes. The true uncertainties are dominated by
uncertainties in the galaxy distances. The solid lines are linear
least-squares fits intended to guide the eye. There is a trend of
increasing nuclear $K$-band luminosity as a function of merger stage.
\label{f:toomrelumikonehpc}}
\end{figure}

The two-dimensional spatial variations of the $V-K$ color in the
central regions of Toomre Sequence galaxies are quite complex, as
evident from the gray-scale representations in
Figures~\ref{f:fortyfortyeightvk}--\ref{f:seventytwofiftytwovk}. To
obtain more easily interpretable quantitative insight we have
therefore determined the radial $V-K$ color gradients for all of the
nuclei, measured in concentric circular annuli centered around each
nucleus and shown in Figure~\ref{f:vminuskprofonetwo}. The annuli are
contiguous, and grow in radius by a factor of $1.1$ at each step. The 
color profiles are followed out to a distance of 5$\arcsec$ from each 
nucleus (which corresponds to between 300\,pc and 3.0\,kpc for the 
distances of the sample galaxies). No correction was made for the PSF 
differences between the $V$ and $K$-bands, so $V-K$ features in the 
central $\sim 0\farcs2$ should not be trusted.

The last two merger remnants in the sequence (NGC\,3921 and NGC\,7252)
and NGC\,3509 all have rather shallow color gradients, with a color
change of less than one magnitude within 5$\arcsec$. Most nuclei have
steeper gradients, with a change in color of up to six magnitudes
within 5$\arcsec$. The nuclei with the strongest color gradient are
mostly the mid-stage mergers. The majority of the nuclei have nuclear
colors of $V-K \sim 4-5$, except for the very dusty systems
(NGC520\,S, NGC2623 and NGC\,3256\,S) which have much redder nuclear
colors of $V-K \sim 7-8$ (cf. Fig.~\ref{f:vminuskonehpc}).

Several of the nuclei become considerably redder inward of about
$5\arcsec$ from the nucleus (e.g., NGC\,4676\,A, NGC\,7764A,
NGC\,520\,S, NGC\,2623, NGC\,3256\,S), probably indicative of a strong
concentration of dust in the central regions. This dust complicates
the identification of potential population gradients from the observed
color gradients. NGC\,6621 becomes bluer in its central $\sim
0\farcs6$, but the morphology of the $V-K$ image (see
Figure~\ref{f:sixtysixtwentyonevk}) suggests that this may be due to
an absence of dust rather than the presence of young stars. A few
nuclei are bluer in their central $\sim 0\farcs2$ (e.g., NGC\,520N,
NGC\,7252), but this may be due to PSF differences between $V$ and $K$
rather than differences in stellar populations. On the whole,
Figure~\ref{f:vminuskprofonetwo} certainly shows no evidence for the
bluer nuclear colors that might be expected if the galaxies all
had recent nuclear star formation. On the other hand, this scenario is
certainly not ruled out by the data, given that large quantities of
dust affect most of the color gradients.

\section{SURFACE BRIGHTNESS PROFILES}
\label{s:sfbprof}

\subsection{Analysis}
\label{ss:sbanalysis}

Surface brightness profiles for the Toomre nuclei were derived
primarily from the $K$ band imaging, as they are least affected by
dust and star formation. The images were run through 20 iterations of
Lucy-Richardson deconvolution \citep{ric72,luc74} using the VISTA
software package \citep{lau86}, where the use of a higher number of 
iterations did not result in a significantly different deconvolved 
image. We used TinyTim \citep{kri01} to generate a PSF for each nucleus, 
modeled at the specific position of each nucleus on the detector array 
and using a K-type star for the PSF spectral distribution. For 
NGC\,2623 we used the drizzled images \citep{sco00} with half the 
pixel size of the original data. In each case, the PSF was generated 
out to a radius of 3 arcseconds, which incorporates the low-level 
emission in the extended wings, and was tapered at the edges with a 
Gaussian taper. A few of the galaxies have nuclei that are compact 
enough to produce a strong Airy ring and diffraction spike pattern. We 
found that these artifacts were in general successfully mitigated by 
our deconvolution procedure, although some residuals do remain 
(especially for the most compact sources, such as NGC\,2623).

The nuclear positions were determined by finding the centroid in a
small box, typically with a size of about $7\times7$ pixels. We then
fitted ellipses to the isophotes of the two-dimensional convolved and
deconvolved images. The surface brightness profile was derived only
for the nuclear region (usually out to a radius of a few arcseconds).
We did not attempt to mask out dust or stellar clusters within the
area where the surface brightness profile was derived, since the
effects of dust are subtle in most cases and there were usually no
strong point sources within the few arcsecond radius that could have
interfered with the profile. In cases where such sources exist
(e.g.,\,NGC\,6621), the profile was only extracted within the
innermost nuclear area. The effects of subtle dust features and
circumnuclear spiral arms are still visible as small-scale wiggles in
some of the luminosity profiles (e.g.,\,NGC\,3256 and NGC\,7252).

Following closely the procedure used by \citet{lai03a}, for radii
smaller than 0\farcs5 we used the VISTA {\tt profile} task
\citep{lau85} on the deconvolved image to find the surface brightness
profile. This task keeps the center fixed and fits ellipses by
sampling the light profile in a circle with a radius of 1, 2, 3, etc.
pixels. The {\tt snuc} task \citep{lau86} in VISTA was used on the
deconvolved image between the radii of 0\farcs5 and $1\arcsec$ to
obtain the surface brightness profile. Outside of $1\arcsec$ the
effects of the PSF are negligible (although diffraction spikes from a
very compact nucleus may still be visible there). At these radii we
therefore used the {\tt snuc} task on the surface brightness profile
inferred from the original (not PSF-deconvolved) images, motivated by
S/N considerations. The transition at 0\farcs5 was based on experience
which indicated that the {\tt profile} task is more robust in the
interpolation of pixels at very small radii (Tod Lauer, priv. comm.).
The results of the analysis are shown in Figure~\ref{f:sbpone} for
twelve Toomre sequence nuclei. The remaining four nuclei (NGC\,4038,
NGC\,7592 E, NGC\,7764A, NGC\,520 S) were too complex or otherwise
unsuitable for ellipse fitting and profile extraction. The first panel
shows for comparison the profile of the NICMOS PSF itself, indicated
by star symbols. Figure~\ref{f:sbpone} also shows for each galaxy the
ellipticity and position angle of the fitted ellipses, which can
potentially be used to trace structures such as nuclear bars and
spiral arms. The random uncertainties for all profiles are usually smaller
than the size of the plotted symbols and hence are not shown.

A common analytic function to fit to the luminosity surface density
$\Sigma$ of elliptical galaxies is the so-called ``Nuker law''
\citep[e.g.,][]{lau95,byu96}:
\begin{equation}
\label{nukerlaw}
  \Sigma(r) = \Sigma_b\,2^{\frac{\beta - \gamma}{\alpha}}
              \> (r/r_{\rm b})^{-\gamma} \>
              (1+ [r/r_{\rm b}]^{\alpha})^{\frac{\gamma - \beta}
              {\alpha}} .
\end{equation}
The surface brightness $\mu$ follows from
\begin{equation}
\label{surfdens}
  \mu {\rm [mag/\square^{\arcsec}]} =
      -2.5 \> \log\,\Sigma({\rm{L_{\sun}/ pc^2}}) + 21.572 + M_{\sun}
      ,
\end{equation}
where the solar absolute magnitude in the $K$-band is $M_{K,\odot} =
3.28$ \citep{bin98}. The Nuker law represents a power-law with a break
at the radius $r_{\rm b}$, with the parameter $\alpha$ measuring the
sharpness of the break. The asymptotic power-law slope is $\gamma$ at
small radii and $\beta$ at large radii, while $\Sigma_b$ is the
luminosity surface density at $r_{\rm b}$. The best-fitting Nuker laws
are plotted in Figure~\ref{f:sbpone} as solid curves, and the fit
parameters are listed in Table~\ref{t:nukerfits}, where $\mu_b$ is
derived from the fitted parameter $\Sigma_b$.

As a diagnostic in the discussion below we generally use the quantity
$\Gamma_{0.15}$, defined as the power-law slope at a fixed, resolved
radius of 0\farcs15. This is a slightly more robust measure of the
central brightness profile slope than the fit parameter $\gamma$,
although the two are generally in good agreement \citep{lai03a}. Both
quantities are listed in Table~\ref{t:nukerfits}. Galaxies with slope
$\gta 0.5$ are traditionally called power-law galaxies and those with
slope $\lta 0.3$ are traditionally called core galaxies
\citep{faber97}. Whether these values bracket truly distinct classes
of galaxies or whether there is simply a continuum in cusp slopes
among galaxies continues to be debated
\citep{rav01,res01,lau05,fer06}. Here, we merely adopt the accepted
terminology, without necessarily implying that galaxies come in two
physically distinct classes.

As a consistency check we also derive surface brightness profiles for
a few galaxies from the $J$- and $H$-band images, yielding very
similar results to the $K$-band (apart from an overall color
difference). As the $J$- and $H$-band data have a somewhat narrower
PSF than the $K$-band data, the good agreement therefore indicates
that uncertainties in the $K$-band surface brightness profiles due to
residual problems associated with PSF deconvolution are small. In the
remainder of the paper we discuss only the surface brightness profiles
derived from the $K$-band images.

\subsection{Individual Galaxy Profiles}
\label{ss:sbdescription}

The profiles for the nuclei NGC\,3256\,N, NGC\,3509, NGC\,4039,
NGC\,4676\,B, NGC\,7252 and NGC\,7592\,W are essentially pure power-laws 
with an almost constant slope through the fitted region. The 
$\Gamma_{0.15}$ values vary between 0.94 and 1.58, therefore belonging 
comfortably to the regime of power-law galaxies. Some deviations from 
the best fit are seen in the profiles of NGC\,3256, NGC\,7252, and NGC\,7592, 
most likely due to intrinsic structure in these nuclei. The first two 
of these galaxies have circumnuclear spiral structure, whereas 
NGC\,7592 has an end-brightened bar-like structure. In all cases, 
residual structure from the diffraction pattern may play some role as 
well. Either way, in all three cases the overall slope remains 
relatively unchanged from large to small radii within the fitting area.

The $\Gamma_{0.15}$ values of NGC\,2623, NGC\,3921, NGC\,4676 A, and
NGC\,6621 are large enough to place them in the realm of power-law
galaxies. However, the surface brightness profiles of these galaxies
do show more structure than a pure power law. NGC\,2623, for example,
shows several wiggles in its profile. Nonetheless, the brightness
profile is steep throughout the entire nuclear region, with
$\Gamma_{0.15}=1.42$ and no break toward a shallow core. The profile
of NGC\,3921 appears to have a shallow break around 0\farcs25, but
with an inner slope ($\Gamma_{0.15} = 1.13$) that is still in the
power-law regime. NGC\,4676\,A has a clearer break in the power-law
slope, around 0\farcs43, but its $\Gamma_{0.15}$ value of 0.65 also
still places it comfortably among power-law galaxies. Some curvature
around 0\farcs4 is seen in the profile of NGC\,6621, but the
$\Gamma_{0.15}$ value of 0.68 indicates that this nucleus too belongs
to the power-law group.

NGC\,6622 is the only galaxy for which the profile turns over to a
power-law slope $\gamma= 0.24$ that is characteristic of core
galaxies. However, the turn-over happens inside of 0\farcs1, the
characteristic resolution of our data. So while this galaxy may have a
core-type profile, it is certainly not an unambiguous case. In the
following we do not discuss the parameter $r_b$ of the NUKER profile
fits. This parameter generally has physical meaning only for galaxies
with well resolved cores, and none of the Toomre sequence nuclei fall
in this category.

The only nucleus that does not fit within the class of either the core
or the power-law galaxies is one nucleus in the mid-stage merger
NGC\,520\,N. This nucleus has a very shallow profile out to several
arcsec from the center, but with a strong upturn in the central $\sim$
0\farcs15. The bright central source responsible for this upturn is
clearly visible in the $K$-band image (Figure~\ref{f:fivetwentybvk}).
This profile and nuclear morphology are very similar to what is seen
in the centers of many late-type spiral galaxies \citep[e.g.,][]
{boe02}, where it has been established  that the upturn is due to the
presence of a nuclear star cluster \citep[e.g.,][]{boe04,ros06}. Since
more than half of the spiral galaxies have such nuclear star clusters,
it would not be surprising to find one among the Toomre sequence
galaxies as well.

Because the galaxies at the earliest stage in the Toomre sequence have
not yet experienced the full merger evolution, their nuclei would be
expected to still have brightness profiles similar to those of spiral
galaxies. At {\it HST} resolution, the ``classical'' $R^{1/4}$ bulges
in early-type spirals tend to have power-law brightness profiles with
slopes $0.5 \leq \gamma \leq 1$ \citep[e.g.,][]{sei02}. This is indeed
similar to what we find in the earliest stage mergers of the Toomre
sequence (NGC\,4039, NGC\,4676 A, NGC\,4676 B).

\subsection{Comparison to Elliptical Galaxies}
\label{ss:powerlaws}

As merger remnants may evolve into elliptical galaxies, it is of
interest to compare the nuclear properties of these two classes of
galaxies. We used the results of \citet[][hereafter R01]{rav01} as a
suitable comparison sample of elliptical and S0 galaxies. While many
other {\it HST} studies have addressed the surface brightness profiles
of early-type galaxies in the optical regime, the R01 study is one of
the few that use observations with NICMOS in the near-infrared. We
transformed their $H$-band results to the $K$-band assuming a fixed
color $H-K = 0.215$, as appropriate for normal early-type galaxies
(see Figure~\ref{f:jmhhmkonehpeg}).

Previous studies have found that the properties of the nuclear
brightness profiles of elliptical galaxies correlate strongly with the
total galaxy luminosity \citep[e.g.,][]{faber97}. To compare the
properties of merger remnants and elliptical galaxies we therefore
calculated the total absolute $K$-band magnitudes ($M_K$) for both our
own sample and that of R01, using the apparent $K$-band magnitudes
from the Two Micron All Sky Survey (2MASS) and distances based on the
observed systemic velocity. The results for our own sample are listed
in Table~\ref{t:nukerfits}.

Figure~\ref{f:gammavskmag} shows $\Gamma_{0.15}$, the power-law slope
at 0\farcs15, versus $M_K$. Optical studies have found that galaxies
with $M_V \gta -20.5$ ($M_K \gta -23.6$)\footnote{The absolute $K$-band 
magnitudes listed in parentheses assume $V-K = 3.1$, as typical 
for an old stellar population \citep{bru03}.} are generally power-law 
galaxies, that galaxies with $M_V \lta -22.0$ ($M_K \lta -25.1$) are 
generally core galaxies, and that galaxies of intermediate 
luminosities can be of either type \citep[e.g.,][]{faber97}. The data 
from R01, shown in Figure~\ref{f:gammavskmag}, support these trends 
(e.g., beyond $M_{K} \sim -24.7$ there are only core galaxies in their 
sample, but no power-laws). While these trends appear more pronounced 
in some studies of other samples \citep[e.g.,][]{faber97}, it is clear 
that the correlation has significant scatter.

Figure~\ref{f:gammavskmag} also shows the Toomre sequence nuclei in
comparison to the E/S0 sample. As discussed already in
\S~\ref{ss:sbdescription}, the Toomre nuclei correspond almost
exclusively to power-law profiles; only NGC\,6622 may possibly be a
core-type galaxy. The Toomre sequence galaxies occupy the range of
absolute magnitudes $M_K$ in which both power-law E/S0 and core-type
E/S0 galaxies can be found; it is therefore no surprise to find power-law 
profiles among the nuclei of the Toomre sequence galaxies. However, it 
is of interest that so few of the nuclei have a core-type profile and that 
many have $\Gamma_{0.15}$ values that are high relative to E/S0s. We 
comment on this below in the context of popular theories for the formation 
of cores in E/S0 galaxies.

Additionally, we have studied the luminosity surface density $\Sigma$
at 100\,pc as a function of the total absolute $K$-band magnitude,
shown in Figure~\ref{f:ldensvskmag}. The majority of the Toomre nuclei
have a higher luminosity surface density than the E/S0's, although
there are a few in the overlapping region. The most luminous Toomre
galaxies also tend to have the highest surface luminosity density.
Note that neither sample was corrected for dust extinction in the $K$-band 
(see \S~\ref{ss:mergerstage} for a discussion of the typical 
amount of extinction for the Toomre nuclei). However, 
Figure~\ref{f:jmhhmkonehpeg} shows that the Toomre nuclei have more 
dust extinction than typical E/S0 galaxies. Therefore, any correction 
for dust extinction would in fact increase the systematic difference 
in luminosity surface density between the two samples.

Figure~\ref{f:gammavskmag} shows that, on the whole, the Toomre
Sequence nuclei tend to have slopes that are significantly steeper
than what is seen in E/S0 galaxies. The four latest-stage merger
remnants all have 1.1 $\lta \Gamma_{0.15} \lta$ 1.4. In principle this
could be a resolution effect, as the median distance of the Toomre
galaxies is 77.8\,Mpc, versus 17.0\,Mpc for the galaxies in the sample
of R01. Early-type galaxies in general tend to have steeper
logarithmic surface brightness profiles at larger radii, so that at a
fixed angular scale, one would naturally expect to find somewhat
steeper slopes in a more distant sample. However,
Figure~\ref{f:ldensvskmag} shows that this is unlikely to be the whole
explanation in the present case. The Toomre nuclei tend to have a
higher luminosity surface density than the E/S0's at fixed physical
radius (i.e. at 100\,pc), and thus must be intrinsically brighter at
small radii. It then seems plausible that they have intrinsically
steeper surface brightness slopes as well. A similar result was
recently reported in an {\it HST}/WFPC2 optical study by
\citet{yan04}, who found that E+A galaxies (shown to be evolved merger
remnants) had power-law profiles and higher surface brightnesses than
E/S0s. \citet{yan04} attributed the higher surface brightnesses to the
recent star formation in E+As and argued that, once the A stellar
population faded, the E+A galaxies would have surface brightness
profiles typical of E/S0s. Our results also connect naturally with the
finding of \citet{rot04} that many merger remnants have higher
luminosity densities near their centers than would be expected on the
basis of inward extrapolation of the surface brightness profile at
large radii.

The steep $K$-band profiles and high luminosity surface densities seen
in the Toomre galaxies cannot be attributed to the presence of bright
non-thermal point sources. We measure the profile slope at 0\farcs15
and the luminosity surface density at 100\,pc; the latter corresponds
to 0\farcs26 at the median distance of the sample galaxies. Therefore,
both of these quantities are defined outside of the direct influence
of a central non-thermal point source, if one existed. This is true in
particular because we measure these quantities on the deconvolved
profiles, not the observed profiles. Moreover, there is little
independent reason to suspect the presence of non-thermal point
sources. Most of the surface brightness profiles do not show an upturn
near 0\farcs1. The only exception is NGC\,520\,N, but for this galaxy
a nuclear star cluster is the likely cause. While some of the galaxy
images (e.g., NGC\,2623 and NGC\,3921) do show the NICMOS PSF
diffraction pattern around the galaxy center, this merely implies that
the galaxy light must be very centrally concentrated (as indeed our
analysis confirms), not that a central point source need be present.
Finally, non-thermal point sources at optical wavelengths are
generally associated with powerful Active Galactic Nuclei (AGN), and
radio galaxies in particular \citep{ver02}. By contrast, most of the
Toomre galaxies do not host strong AGN (see Table~\ref{t:ts}).

The results of our surface brightness profile analysis have a natural
explanation in the context of our understanding of the merger
process. As two galaxies interact, gravitational torques remove
angular momentum from the gas and drive nuclear inflows
\citep{nog88,hos96,bar96}. This transports gas to the centers of the
interacting galaxies, and subsequently to the center of the merger
remnant, triggering nuclear starbursts. The newly formed stars
increase the stellar density in the nucleus and steepen the surface
mass density profile \citep{hos94}. As the young stellar population
will have a lower mass-to-light ratio than the older stars in the
galaxy, the steepening of the projected luminosity profile will be
even stronger. All this is qualitatively consistent with the results
shown in Figures~\ref{f:gammavskmag} and~\ref{f:ldensvskmag}. The
formation of new stars in a progressing merger may also be responsible
for the observed trend of increasing nuclear luminosity with advancing
merger stage shown in Figure~\ref{f:toomrelumikonehpc}. NGC\,3921 and
NGC\,7252 are known to have ``post-starburst'' A-star features in
spectra of their nuclei, which provides direct evidence that stars
recently formed in these mergers \citep{sch82,liu95a,liu95b}.

To quantitatively compare the mass densities of ongoing mergers and
elliptical galaxies on the basis of their luminosity densities, one
needs to take into account potential differences in the ages of their
stellar populations as well as any radial gradients in the average age
within the galaxies. This is not possible in detail here, as we have
no estimates available of the stellar population ages in the Toomre
sequence galaxies. However, a simple estimate can be made on the basis
of the fact that the dynamical merging age of several of the
late-stage Toomre sequence galaxies has been estimated to be of the
order of 1\,Gyr \citep[e.g.,][]{hib96}. A stellar population fades in
the $K$-band by $\sim 1.5$ mag between ages of $10^9$ and $10^{10}$
yrs \citep{bru03}, corresponding to $0.6$ dex in luminosity. As the
nuclei likely contain a mix of young and old stellar populations, this
represents an upper limit to the amount of fading the nuclei may
experience. Such significant fading would be sufficient to explain
much of the excess luminosity density compared to E/S0 galaxies for
most of the Toomre sequence nuclei (see
Figure~\ref{f:ldensvskmag}). Also, simulations show that the young
stars formed in mergers are likely to be more concentrated toward the
center than the underlying older stars. A fading of the young
population with time will therefore decrease not only the amplitude
but also the power-law slope of the total projected luminosity density
profile, moving the merger remnants toward the locus of E/S0 galaxies
in Figure~\ref{f:gammavskmag}. The effects combined suggest that, if
left to evolve for several more Gyrs, it is quite possible that the
properties of the Toomre sequence nuclei would look more similar to
the nuclei of normal E/S0 galaxies. The Toomre sequence remnants would
then populate the steep power-law, high luminosity density ends of the E/S0 
distributions shown in Figures~\ref{f:gammavskmag} and \ref{f:ldensvskmag}, 
respectively.

One possible surprise is that the observed surface brightness profiles
are smooth and do not show any obvious kink or upturn. This differs
from the profiles obtained in the merger simulations of \citet{hos94}
and later by \citet{spr00}, which had a pronounced ``cusp'' inward of
a few hundred parsecs due to the presence of the young starburst
population (although see \citet{spr05} for a discussion of how black
hole feedback may moderate these effects). Most of the surface
brightness profiles presented here cover radii in the 10\,pc to 1\,kpc
range, so we would have detected any such cusps if they existed.
However, it should be noted that pronounced kinks are not necessarily
a generic feature of the merger simulations. They can be artifacts of
the limited resolution of the simulations, and their possible presence
depends sensitively on the adopted (very uncertain) prescriptions for
star formation and feedback. In any case, our observations indicate
that if there is indeed a centrally concentrated population of young
stars, then either its brightness profile connects seamlessly with
that of the underlying older stars, or it is more extended (beyond the
kpc range) than was predicted in the \citet{hos94} models.

A predominant theory for the formation of cores in elliptical galaxies
is through the evolution of supermassive black holes involved in a
merger. If the two merging galaxies both host a central black hole,
then these black holes will sink to the center of the merger remnant
through dynamical friction and form a binary system that interacts
with the surrounding stars. These interactions lead to a hardening of
the binary and an ejection of stars from the galaxy center, thus
excavating a core \citep[e.g.,][]{mil01}. The formation of the core
happens quickly, in $10^6 - 10^7$ years, so that in this scenario one
might expect to find cores in late-stage merger remnants. In fact,
there is little evidence for cores in the latest-stage mergers of the
Toomre sequence. However, our findings do not rule out this scenario,
for several reasons: (a) the original merging partners may not both
have had black holes; (b) a core may be present, but may not be of 
sufficient size to resolve with our observations; (c) the time scale 
for black hole binary coalescence may have been shortened substantially 
by the presence of copious amounts of gas \citep{esc05} so that there 
was little time for the scattering of stars and the formation of a core; 
(d) a core may have formed, but was subsequently filled in by stars 
formed from gas driven to the galaxy center during the merger; 
or (e) perhaps some minimum galaxy mass is essential for the formation 
of a core and the Toomre systems are not massive enough. Note with respect 
to the last point that galaxy luminosity may not be a good proxy for galaxy 
mass. The Toomre sequence galaxies do occupy the same range of absolute 
magnitudes $M_K$ in which E/S0 galaxies are sometimes found to have core 
profiles. Nonetheless, if the Toomre sequence galaxies have younger 
populations on a galaxy-wide scale (and not just on the nuclear scale 
discussed above), the fading of these young populations would make the galaxy 
fainter with time. One should then compare to E/S0 galaxies that are fainter 
in $M_K$, which have core-type profile much more rarely.

\section{SUMMARY}
\label{s:summary}

We have imaged the central regions of the 11 interacting and merging
galaxies of the Toomre sequence in $J$, $H$, and $K$, using the NIC2
camera on {\it HST}/NICMOS, and augmented these data with {\it HST}
archival data in similar bands for those galaxies with existing NIR 
data. Compared to optical data, NIR data have the advantage that the 
images are much less affected by dust. The observations therefore offer 
a relatively unimpeded view into the nuclei, especially when compared to 
the results of our previous {\it HST}/WFPC2 study presented in Paper~I. 
With our previous optical data it was difficult or impossible in several 
cases to identify the true galaxy nucleus/nuclei; by contrast, the near-IR 
data presented here generally reveal the nuclei clearly. In most cases we 
confirm the galaxy nucleus/nuclei reported from data in other wavebands, 
such as radio, X-ray or millimeter (CO) wavelengths. The nuclear positions
identified from the different observations are generally consistent
with each other. In NGC\,7764A we detect a double nucleus for the
first time, with a separation of $0\farcs43$ (260\,pc), similar to the
separation of the two nuclei in the merging galaxy Arp\,220. The
presence of a double nucleus in NGC\,7764A is consistent with Toomre's
initial assumption of the system being in the final merging stage
prior to coalescence. In NGC\,3256\,S we find a highly extincted near-IR 
source coincident with the secondary nucleus previously identified 
on the basis of radio and X-ray data. The extinction-corrected $K$-band 
luminosity of this source is similar to that of the nuclei in other 
Toomre sequence galaxies.

Aperture photometry of the Toomre sequence nuclei reveals that their
colors are consistent with the observed colors of elliptical galaxies
and the predicted colors of intermediate age or old stellar
populations, but often with considerable extinction.  There is a
marginal trend for the nuclei to become bluer as a function of the
merger stage in the Toomre sequence. Because stellar population
differences alone cannot account for the range of colors we observe in
the nuclei, we attribute this trend to indicate a dispersal of dust
during the late stages of merging. This is qualitatively consistent
with the predictions of models of merging galaxies with black holes in
which AGN feedback is invoked to disperse obscuring dust as the merger
progresses \citep[e.g.,][]{hop05}. It will be valuable to construct
numerical models that make quantitative predictions for the dust and
gas content of interacting and merging galaxies as a function of
merger (st)age, for comparison to our data.

There is also a trend for the Toomre sequence nuclei to become more
luminous (within an aperture of fixed physical size, and after
correction for extinction) as a function of the merger stage along the
sequence. As long as there are no systematic errors in the analysis
which depend on the merger stage, this trend is statistically
significant at the 2$\sigma$ level.  This result suggests that, as the
merger proceeds, either the stellar density must increase, or the
average stellar age must decrease (or both). As a cautionary note we
remark that this result is based on the assumed merger stage, which
may not necessarily reflect the true merger (st)age, given how the
Toomre sequence was initially defined in the first place. In principle
it is possible to test these scenarios in more detail using population
synthesis models to fit the broad-band colors. However, the results of
such an analysis are likely to be quite uncertain given the large
amounts of extinction and the uncertainties in the dust geometry. We
therefore postpone a discussion of the stellar populations to a
subsequent paper that will present {\it HST}/STIS spectra, obtained in
the context of our investigation.

We have calculated radial $V-K$ color gradients around all the nuclei
in our sample. The latest two merger remnants have rather shallow
color gradients, whereas most other nuclei have steeper gradients. The
nuclei with the strongest color gradients are mostly found in the
mid-stage mergers. Several of the nuclei become considerably redder
inward of about 5$\arcsec$ from the nucleus, indicating strong
concentrations of dust in the nuclei.

Finally, we have derived the $K$-band surface brightness profiles for
those Toomre sequence nuclei for which the morphology allows a
meaningful isophotal analysis (12 out of the total 18). The profiles
were fit with a so-called ``Nuker law'' to facilitate a comparison
with other samples of galaxies observed with {\it HST}. The majority
of the nuclei have steep profiles that can be characterized as
power-law profiles. In general, the Toomre sequence galaxies tend to
have steeper profiles and higher central luminosity surface densities
than E/S0's. These findings can both be qualitatively explained if the
Toomre sequence galaxies have young stellar populations concentrated
toward their centers. Such populations are expected on the basis of
$N$-body simulations of spiral galaxy mergers, in which gas flows
toward the center of the merger remnant trigger nuclear starbursts.
This process may also be responsible for the observed trend in the
Toomre sequence of increasing nuclear luminosity with advancing merger
stage. If left to evolve and fade for several more Gyrs, it is quite
possible that the properties of the Toomre sequence nuclei would look
similar to the nuclei of normal E/S0 galaxies. Our results
therefore support the view that mergers of spiral galaxies lead to the
formation of early-type galaxies.


\acknowledgments Support for proposal \#9402 was provided by NASA
through a grant from the Space Telescope Science Institute, which is
operated by the Association of Universities for Research in Astronomy,
Inc., under NASA contract NAS 5-26555. We thank Nick Scoville for kindly 
providing us with the drizzled NGC\,2623 images, and Daisuke Iono for 
providing us with the unpublished coordinates of the CO measurements of the 
nuclei of NGC\,6621 and NGC\,7592. We would also like to thank the anonymous
referee for a very constructive and detailed report, which helped improving 
the clarity and contents of the paper. RvdM carried out part of this research 
at the Kavli Institute for Theoretical Physics in Santa Barbara, supported 
in part by the National Science Foundation under Grant No.~PHY99-07949. AIZ 
is grateful for the hospitality and support of the Aspen Center for Physics, 
the Kavli Institute for Theoretical Physics, and the NYU Physics Department 
and Center for Cosmology and Particle Physics during her sabbatical year.
JCM thanks Svedka Vodka for support during the final editing of this paper, 
and JR thanks Coca-Cola for support during the entire time of the Toomre 
sequence project. This research has made use of the NASA/IPAC Extragalactic 
Database (NED) which is operated by the Jet Propulsion Laboratory, California
Institute of Technology, under contract with the National Aeronautics
and Space Administration. This publication makes use of data products
from the Two Micron All Sky Survey, which is a joint project of the
University of Massachusetts and the Infrared Processing and Analysis
Center/California Institute of Technology, funded by the National
Aeronautics and Space Administration and the National Science
Foundation.

\clearpage



\clearpage



\clearpage

\clearpage

\clearpage

\clearpage

\clearpage

\clearpage

\clearpage

\clearpage

\clearpage

\clearpage

\clearpage

\clearpage

\clearpage

\clearpage

\clearpage

\clearpage

\clearpage

\clearpage

\clearpage

\clearpage

\clearpage

\begin{figure} \includegraphics[angle=270,scale=0.60,clip=t]{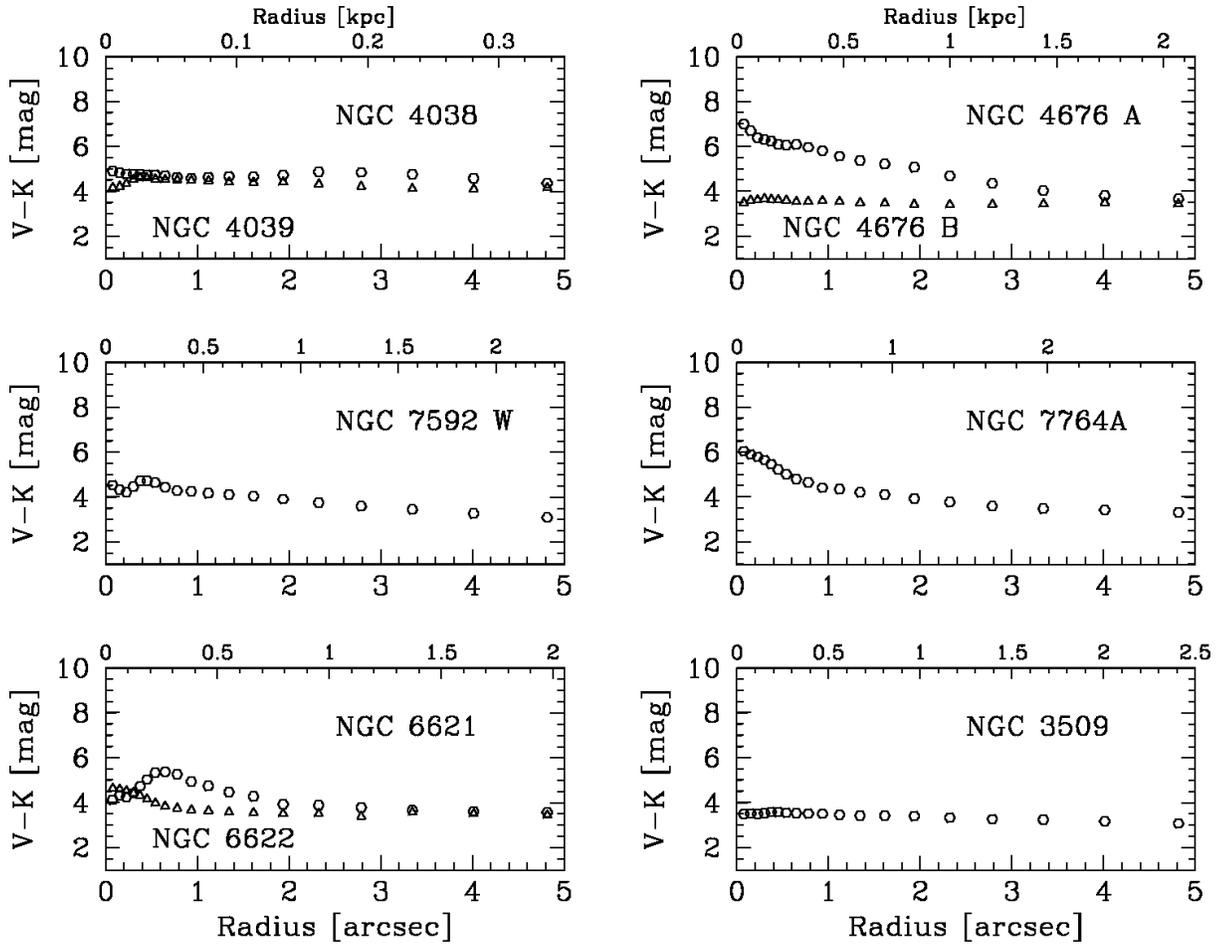}
\caption{Radial $V-K$ color gradients for the regions around each of
the Toomre sequence nuclei, measured using circular annuli. The
results for different nuclei in the same interacting system are shown
using different symbols in one panel, as labeled. Note that no
corrections were made for differences in the PSF FWHM between the $V$
and $K$ bands, leading to spurious artifacts at radii smaller than
0\farcs2. The x-axis measures radius both in arcseconds at the bottom
and kiloparsecs at the top.
\label{f:vminuskprofonetwo}}
\end{figure}


\clearpage

\setcounter{figure}{21}
\begin{figure}
\includegraphics[angle=270,scale=0.60,clip=t]{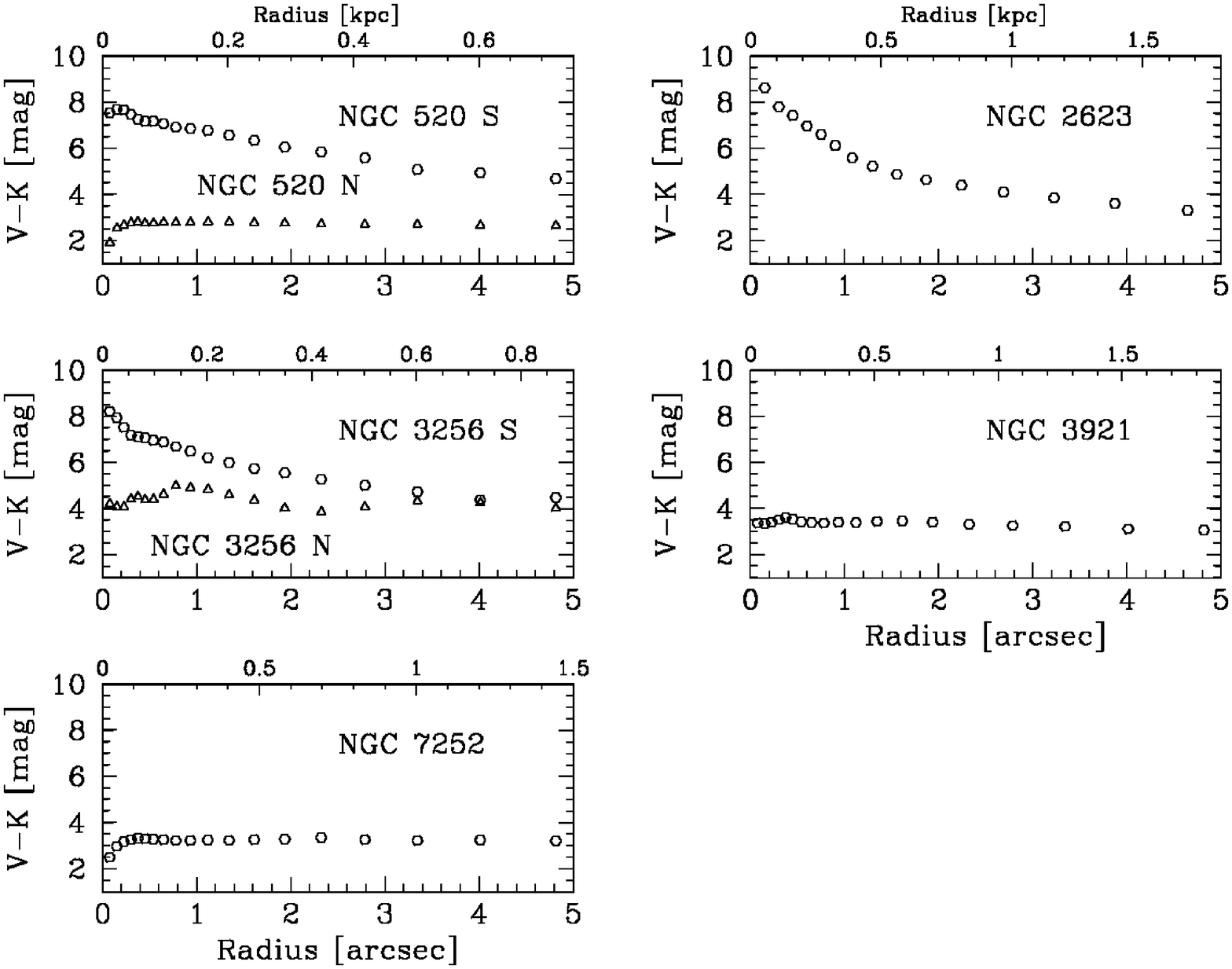}
\caption{(continued)}
\end{figure}


\clearpage

\begin{figure}
\includegraphics[angle=0,scale=0.8,clip=t]{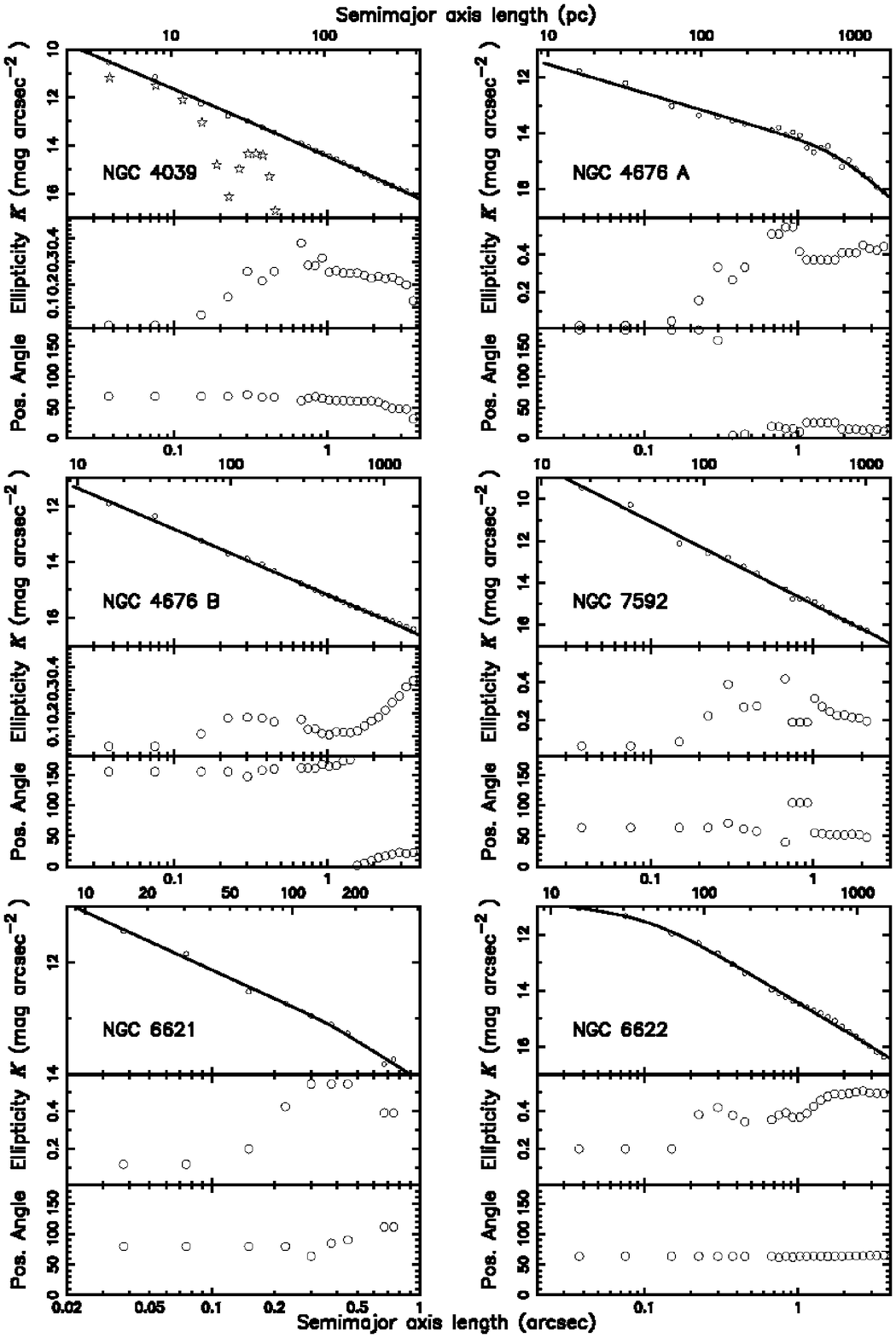}
\caption{Major axis surface brightness profiles of the nuclear regions
of selected Toomre sequence galaxies, where a fit was possible. The
sub-panels show the surface brightness profiles overplotted with a
Nuker model fit (solid line). The host galaxies of the individual
nuclei are labeled.  For comparison, we plot the profile of the NICMOS
PSF in the first panel (star symbols, arbitrarily normalized).
\label{f:sbpone}}
\end{figure}


\clearpage

\setcounter{figure}{22}
\begin{figure}
\includegraphics[angle=0,scale=0.8,clip=t]{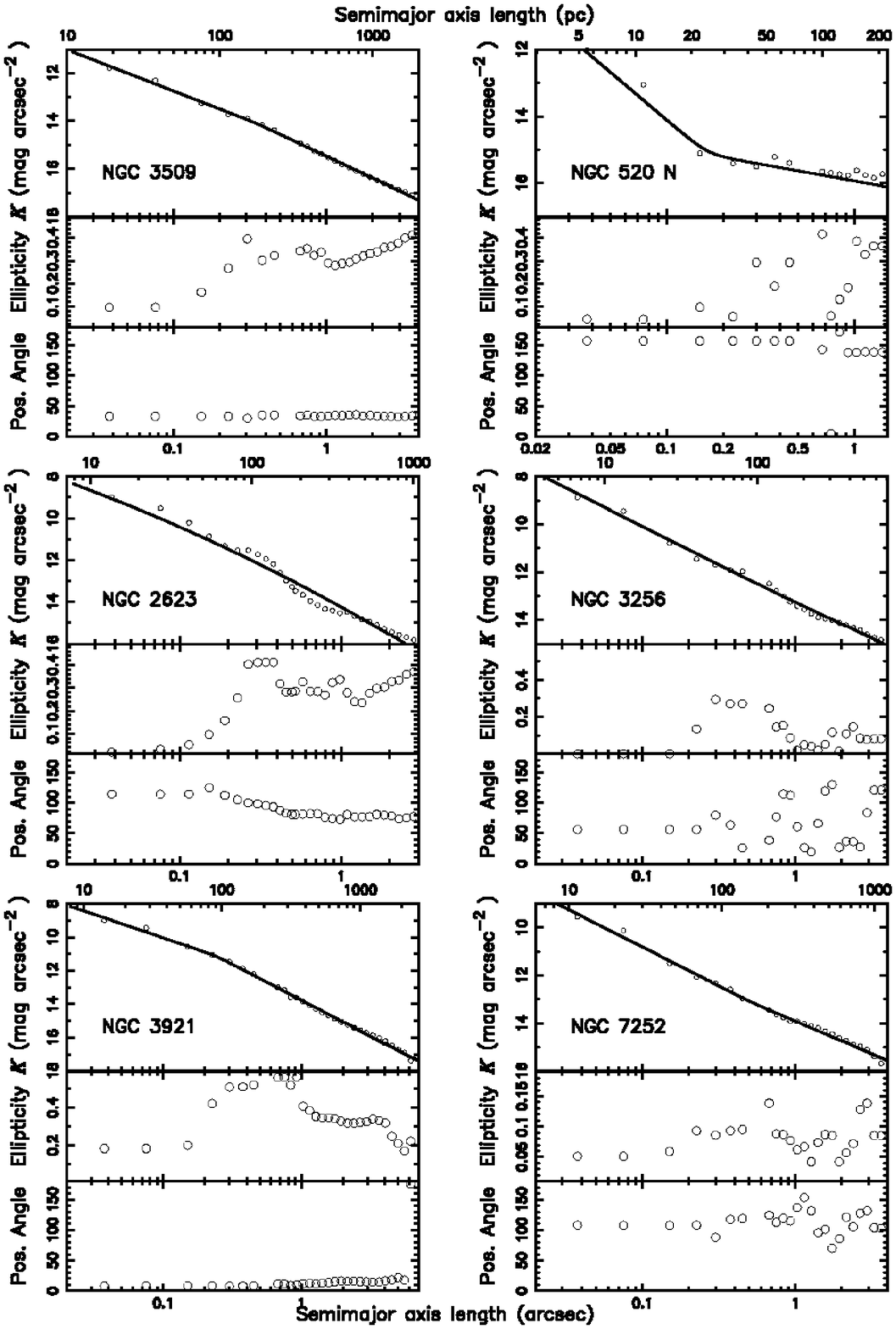}
\caption{(continued)}
\end{figure}


\clearpage

\begin{figure}
\includegraphics[angle=270,scale=0.7,clip=t]{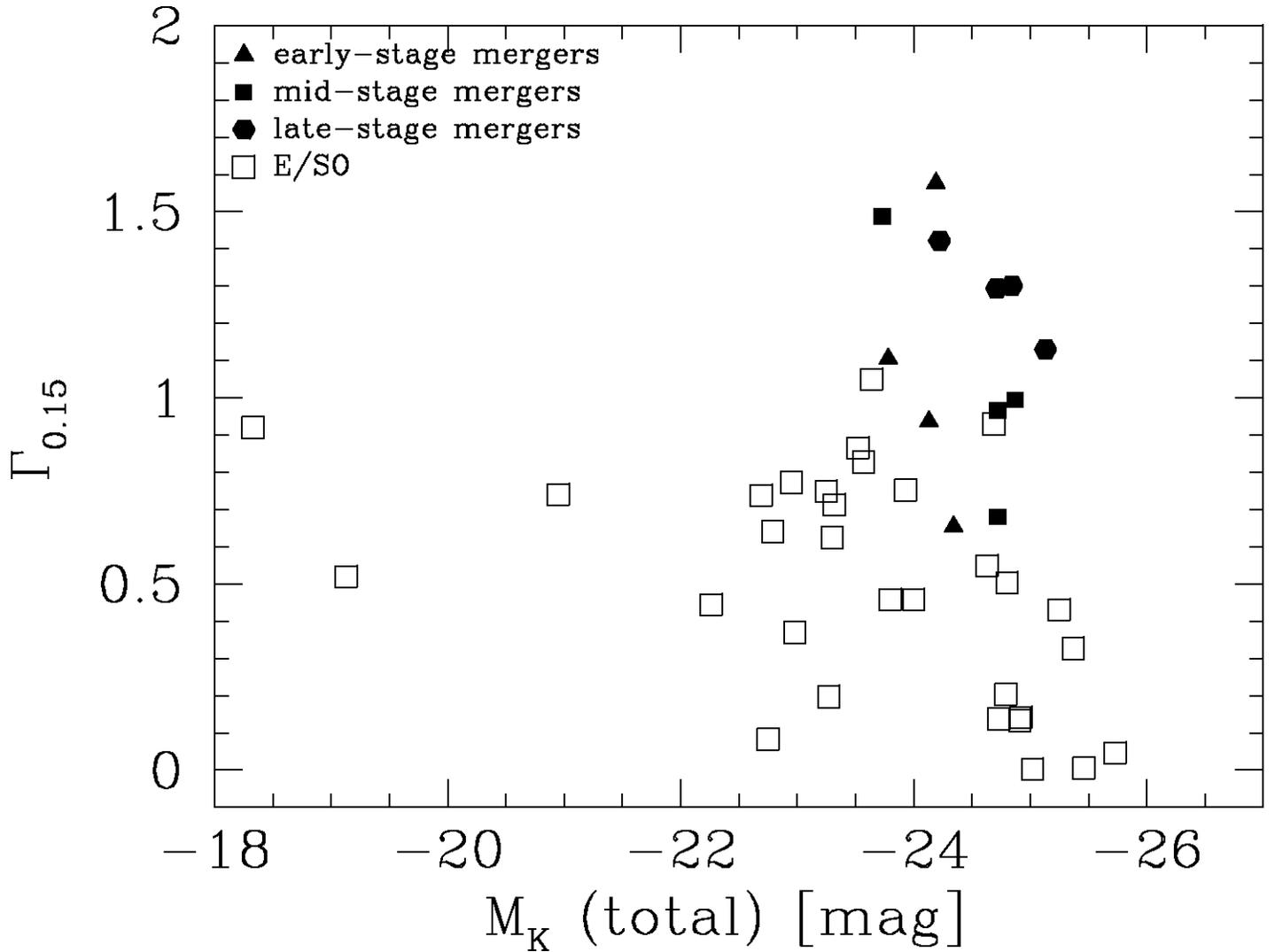}
\caption{Relation between the best-fit power-law slope at 0\farcs15
($\Gamma_{0.15}$) and the total absolute $K$-band magnitude. For
comparison, the early-type E/S0 galaxies from the \citet{rav01} sample
are plotted as open squares. The random uncertainties in
$\Gamma_{0.15}$ are small. However, the true uncertainties in
$\Gamma_{0.15}$ are dominated by systematic uncertainties that are
difficult to quantify, having to do primarily with the accuracy of PSF
deconvolution. Nonetheless, these uncertainties can be kept to a
minimum by measuring $\Gamma$ at a radius larger than the
FWHM of HST's PSF. For the adopted radius of 0\farcs15
(i.e. $\Gamma_{0.15}$), the uncertainties are small enough so as to
not affect the overall distribution of points in the plot. The random
uncertainties in the absolute magnitudes $M_K$ are generally smaller
than the symbol sizes. The true uncertainties in $M_K$ are dominated
by uncertainties in the galaxy distances.
\label{f:gammavskmag}}
\end{figure}


\clearpage

\begin{figure}
\includegraphics[angle=270,scale=0.7,clip=t]{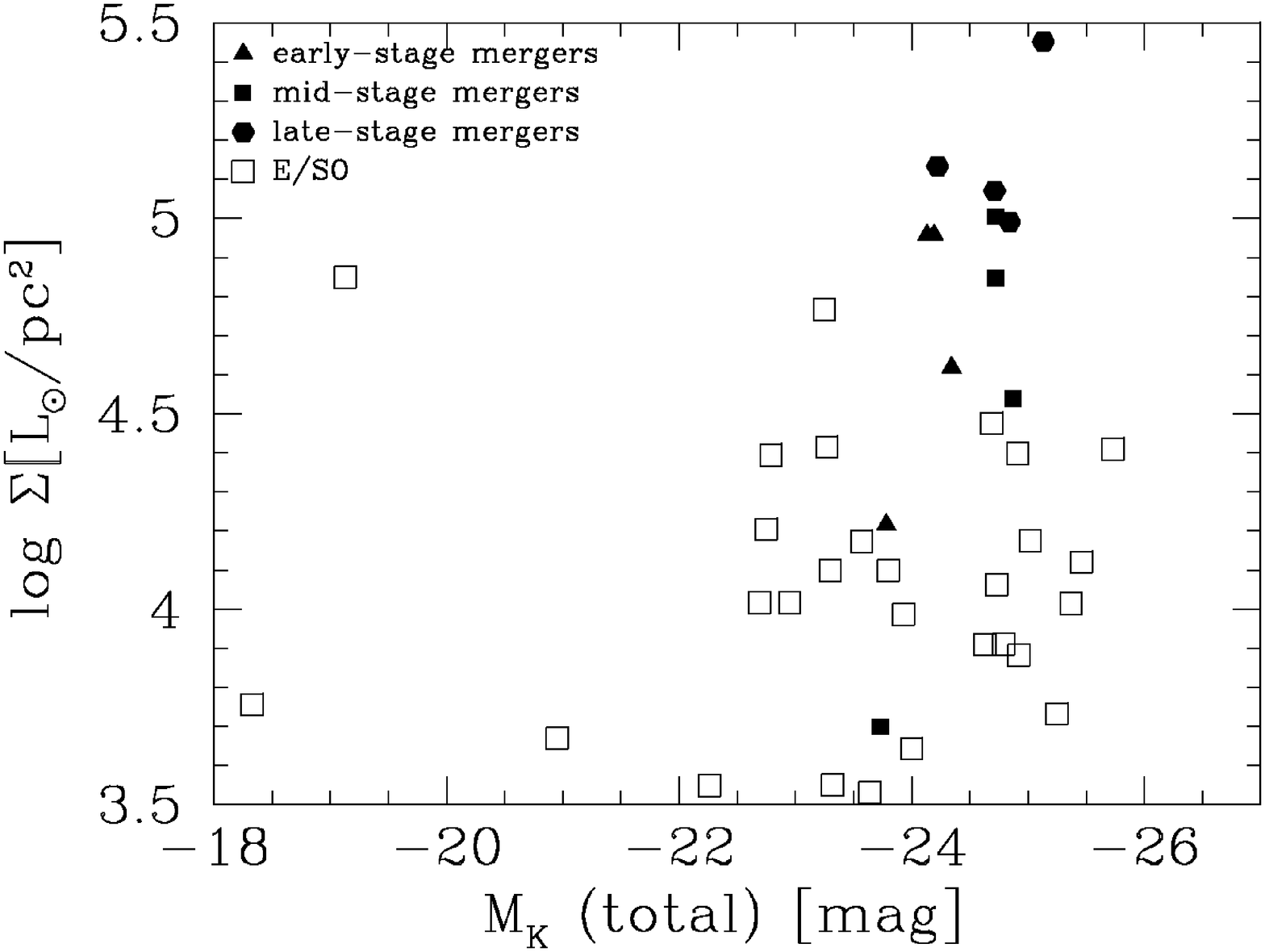}
\caption{Relation between the luminosity surface density $\Sigma$ and
the total absolute $K$-band magnitude. The random uncertainties in $\Sigma$ 
are small. However, the true uncertainties in $\Sigma$ are dominated by 
systematic uncertainties that are difficult to quantify, having to do 
primarily with the accuracy of PSF deconvolution. Nonetheless, these 
uncertainties can be kept to a minimum by measuring $\Sigma$ at a radius 
larger than the FWHM of HST's PSF. For the adopted radius of 100\,pc, the 
uncertainties are small enough so as to not affect the overall distribution 
of points in the plot. The symbols are as in Figure~\ref{f:gammavskmag}, 
which also discusses the uncertainties in $M_K$. \label{f:ldensvskmag}}
\end{figure}


\clearpage




\begin{deluxetable}{cllcccccccc}
\tabletypesize{\scriptsize}
\tablecaption{The Nuclei of the Toomre Sequence \label{t:ts}}
\tablewidth{0pt}
\tablehead{
Seq. No. & Galaxy & Alt. name & R.A.~(J2000.0) & Dec.~(J2000.0) &
Comments
on & $cz$ & Distance & 0\farcs1 & Class\\
& & & [hh mm ss.ss] & [dd\degr mm\arcmin ss\farcs ss] &
nuc. morph. & [km~s$^{-1}$] & [Mpc] & [pc] & \\
\colhead{(1)} & \colhead{(2)} & \colhead{(3)} & \colhead{(4)} &
\colhead{(5)} & \colhead{(6)} & \colhead{(7)} & \colhead{(8)} &
\colhead{(9)} & \colhead{(10)}
\\
}
\startdata
Early-stage & & & & & & & & & \\
\hline
1 & NGC\,4038 & ARP\,244 & 12 01 53.06 & $-$18 52 03.6 & c,d,f & 1616
&
13.8 & 7 & ... \\
1 & NGC\,4039 & ARP\,244 & 12 01 53.61 & $-$18 53 10.7 & p & 1624 &
13.8 &
7 & ... \\
2 & NGC\,4676 A & ARP\,242 & 12 46 10.09 & +30 43 54.6 & p+s, b? &
6613
& 88.2 & 43 & LINER \\
2& NGC\,4676 B & ARP\,242 & 12 46 11.21 & +30 43 22.2 & p & 6613 &
88.2 &
43 & LINER \\
3 & NGC\,7592 W & VV\,731 & 23 18 21.78 & $-$04 24 58.0 & p+r & 7280 &
97.1 & 47 & Sy2 \\
3 & NGC\,7592 E & VV\,731 & ... & ... & c,e,f & 7280 &
97.1 & 47 & ... \\
4 & NGC\,7764A E & AM\,2350-410 & 23 53 23.75 & $-$40 48 25.2 & p+f &
9162 & 122.2 & 59 & ... \\
4 & NGC\,7764A W & AM\,2350-410 & 23 53 23.72 & $-$40 48 25.4 & p &
9162
& 122.2 & 59 & ... \\
\hline
Mid-stage & & & & & & & & & \\
\hline
5 & NGC\,6621 & ARP\,81 & 18 12 55.34 & +68 21 48.8 & p + c,e & 6191 &
84.4
& 41 & ... \\
5 & NGC\,6622 & ARP\,81 & 18 12 59.40 & +68 21 15.0 & p & 6466 & 84.4
& 41 &
... \\
6 & NGC\,3509 & ARP\,335 & 11 04 23.63 & +04 49 43.9 & p & 7704 &
102.7 &
50 & ... \\
7 & NGC\,520 S & ARP\,157 & 01 24 34.91 & +03 47 29.1 & p & 2281 &
30.4 &
15 & ... \\
7 & NGC\,520 N & ARP\,157 & 01 24 33.29 & +03 48 02.6 & p?,c & 2281 &
30.4 & 15 & ... \\
\hline
Late-stage & & & & & & & & & \\
\hline
8 & NGC\,2623 & ARP\,243 & 08 38 24.14 & +25 45 16.1 & p & 5535 & 73.8
& 36 &
LINER \\
9 & NGC\,3256 N & VV\,65 & 10 27 51.17 & $-$43 54 15.6 & p & 2738 &
36.5
& 18 & ... \\
9 & NGC\,3256 S & VV\,65 & 10 27 51.16 & $-$43 54 20.8 & e & 2738 &
36.5
& 18 & ... \\
10 & NGC\,3921 & ARP\,224 & 11 51 07.19 & +55 04 43.1 & p & 5838 &
77.8 & 38
& LINER \\
11 & NGC\,7252 & ARP\,226 & 22 20 44.77 & $-$24 40 42.6 & p,s & 4688 &
62.5
& 30 & ... \\
\enddata
\tablecomments{Column~(1) lists the Toomre sequence number. Cols.~(2)
and (3) list the galaxy name and alternate name. Cols.(4) and (5) give
the coordinates of the nuclei as measured on our NICMOS F205W and
F222M images. The $1\sigma$ absolute accuracy of these coordinates is
$\sim$1\farcs6. Col.~(6) describes the nuclear morphology (as seen in
the field of view of the NIC2 images, which typically extend to $\sim
10\arcsec$ from the galaxy nucleus), with the following nomenclature:
b = bar; c = complex; d = dusty; e = extended; f = filamentary; p =
point-like (compact); r = ring; s = nuclear spiral. Col.~(7) lists the
heliocentric velocity, taken from the NASA Extragalactic Database
(NED), except for NGC\,4676, whose value was taken from the Lyon
Extragalactic Database (LEDA). Col.~(8) lists the distance, calculated
using $H_{0} = \rm{75\,km\,s^{-1} \,Mpc^{-1}}$. For the NGC\,6621/22
system we have calculated the mean distance by averaging both
distances calculated from the velocities. For NGC 4038/4039 we adopted
the distance derived from a measurement of the tip of the red giant
branch \citep{sav04}. Col.~(9) gives the physical scale in parsec
corresponding to 0\farcs1. Col.~(10) lists the spectral classification
for all galaxies with possible AGN contributions. These
classifications were compiled from various literature sources, as
described in Paper~I. We note that this classification is by no means
complete, since not all galaxies in the Toomre sequence have been
studied in detail (or at all) spectroscopically. The position of the
true nucleus of NGC\,7592 E is difficult to assess from our NIR
images, as it has an extended morphology with two maxima, and other
parts of the circumnuclear region are not covered by our aperture. The
horizontal lines separate the groups that we denote as early-stage,
mid-stage, and late-stage mergers.}
\end{deluxetable}


\begin{deluxetable}{lccccc}
\tabletypesize{\scriptsize}
\tablecaption{NICMOS observations \label{t:obs}}
\tablewidth{0pt}
\tablehead{
Galaxy & dataset identifier & Date & Filter & $\lambda_{\rm cen}$ &
$t_{\rm exp}$ \\
& ({\it HST} archive) & [mm/dd/yyyy] & & [$\mu$m] & [s] \\
\colhead{(1)} & \colhead{(2)} & \colhead{(3)} & \colhead{(4)} &
\colhead{(5)} & \colhead{(6)} \\
}
\startdata
NGC\,4038 & N8FR01010 & 06/12/2003 & F110W & 1.1264  & 160 \\
NGC\,4038 & N8FR01020 & 06/12/2003 & F160W & 1.6037  & 160 \\
NGC\,4038 & N8FR01030 & 06/12/2003 & F205W & 2.0658  & 448 \\
NGC\,4039 & N8FR02010 & 05/26/2003 & F110W & 1.1264  & 160 \\
NGC\,4039 & N8FR02020 & 05/26/2003 & F160W & 1.6037  & 160 \\
NGC\,4039 & N8FR02030 & 05/26/2003 & F205W & 2.0658  & 448 \\
NGC\,4676 A & N8FR03010 & 06/12/2003 & F110W & 1.1264 & 192 \\
NGC\,4676 A & N8FR03020 & 06/12/2003 & F160W & 1.6037 & 160 \\
NGC\,4676 A & N8FR03030 & 06/12/2003 & F205W & 2.0658 & 448 \\
NGC\,4676 B & N8FR04010 & 06/21/2003 & F110W & 1.1264 & 192 \\
NGC\,4676 B & N8FR04020 & 06/21/2003 & F160W & 1.6037 & 160 \\
NGC\,4676 B & N8FR04030 & 06/21/2003 & F205W & 2.0658 & 448 \\
NGC\,7592 E+W & N8FR05010 & 10/04/2002 & F110W & 1.1264 & 160 \\
NGC\,7592 E+W & N8FR05020 & 10/04/2002 & F160W & 1.6037 & 160 \\
NGC\,7592 E+W & N8FR05030 & 10/04/2002 & F205W & 2.0658 & 448 \\
NGC\,7764A E+W & N8FR06010 & 04/24/2003 & F110W & 1.1264 & 192 \\
NGC\,7764A E+W & N8FR06020 & 04/24/2003 & F160W & 1.6037 & 192 \\
NGC\,7764A E+W & N8FR06030 & 04/24/2003 & F205W & 2.0658 & 448 \\
NGC\,6621 & N8FR07010 & 08/30/2002 & F110W & 1.1264 & 224 \\
NGC\,6621 & N8FR07020 & 08/30/2002 & F160W & 1.6037 & 224 \\
NGC\,6621 & N8FR07030 & 08/30/2002 & F205W & 2.0658 & 448 \\
NGC\,6622 & N8FR08010 & 08/28/2002 & F110W & 1.1264 & 224 \\
NGC\,6622 & N8FR08020 & 08/28/2002 & F160W & 1.6037 & 224 \\
NGC\,6622 & N8FR08030 & 08/28/2002 & F205W & 2.0658 & 448 \\
NGC\,3509 & N8FR09010 & 10/30/2002 & F110W & 1.1264 & 160 \\
NGC\,3509 & N8FR09020 & 10/30/2002 & F160W & 1.6037 & 160 \\
NGC\,3509 & N8FR09030 & 10/30/2002 & F205W & 2.0658 & 448 \\
NGC\,520 S & N8FR10010 & 01/18/2003 & F110W & 1.1264 & 160 \\
NGC\,520 S & N8FR10020 & 01/18/2003 & F160W & 1.6037 & 160 \\
NGC\,520 S & N8FR10030 & 01/18/2003 & F205W & 2.0658 & 448 \\
NGC\,520 N & N8FR11010 & 01/18/2003 & F110W & 1.1264 & 160 \\
NGC\,520 N & N8FR11020 & 01/18/2003 & F160W & 1.6037 & 160 \\
NGC\,520 N & N8FR11030 & 01/18/2003 & F205W & 2.0658 & 448 \\
NGC\,2623 & N48H23010 & 11/19/1997 & F110W & 1.1264 & 88 \\
NGC\,2623 & N48H23020 & 11/19/1997 & F160W & 1.6037 & 88 \\
NGC\,2623 & N48H23030 & 11/19/1997 & F222M & 2.2177 & 120 \\
NGC\,3256 & N4G601080 & 11/28/1997 & F160W & 1.6037 & 48 \\
NGC\,3256 & N4G601090/91& 11/28/1997 & F222M & 2.2177 & 80 \\
NGC\,3921 & N49J09040 & 01/10/1998 & F110W & 1.1264 & 160 \\
NGC\,3921 & N49J09020/60 & 01/10/1998 & F160W & 1.6037 & 192 \\
NGC\,3921 & N49J09080/81/82 & 01/10/1998 & F205W & 2.0658 & 256 \\
NGC\,7252 & N49J12040 & 11/11/1997 & F110W & 1.1264 & 128 \\
NGC\,7252 & N49J12020/60 & 11/11/1997 & F160W & 1.6037 & 128 \\
NGC\,7252 & N49J12080/81/82 & 11/11/1997 & F205W & 2.0658 & 256 \\
\enddata
\tablecomments{Col.~(1) lists the galaxy identifier. Col.~(2) lists
the
dataset identifier from the {\it HST} data archive. Col.~(3) lists the
date
of observations. Col.~(4) lists the used {\it HST}/NICMOS filter, and
Col.~(5) lists the central wavelength. Col.~(6) lists the exposure
time. Note
that for NGC\,3256 no F110W observations are available in the archive.
The
$K$-band observations for NGC\,2623 and NGC\,3256 were performed with
the
F222M filter which covers the CO continuum
($\lambda\lambda$2.15--2.29$\mu$m).}
\end{deluxetable}

\begin{deluxetable}{lccccccccccc}
\tablecolumns{12} 
\tabletypesize{\scriptsize}
\tablecaption{Positional offsets of the nuclei compared to other
wavebands \label{t:pos}}
\tablewidth{0pt}
\tablehead{
\multicolumn{1}{l}{} & \multicolumn{3}{c}{NIR--X-ray} & \multicolumn{4}{c}{
NIR--radio-cont.} & \multicolumn{4}{c}{NIR--CO} \\ 
\cline{2-4} \cline{6-8} \cline{10-12}\\ 
Galaxy & $\Delta$R.A. & $\Delta$Dec & $\Delta\,(\alpha, \delta)$ & &
$\Delta$R.A. & $\Delta$Dec. & $\Delta\,(\alpha, \delta)$ & &
$\Delta$R.A. & $\Delta$Dec. & $\Delta\,(\alpha, \delta)$ \\
& [\arcsec] & [\arcsec] & [\arcsec] & & [\arcsec] & [\arcsec] &
[\arcsec] & & [\arcsec] & [\arcsec] & [\arcsec]\\
\colhead{(1)} & \colhead{(2)} & \colhead{(3)} & \colhead{(4)} & \colhead{} &
\colhead{(5)} & \colhead{(6)} & \colhead{(7)} & \colhead{} & \colhead{(8)} &
\colhead{(9)} & \colhead{(10)}\\
}
\startdata
NGC\,4038 & $+1.05^a$ & $+0.42^a$ & 1.8 & & $+0.60^g$ & $+1.30^g$ & 1.6 & 
& ... & ... & ... \\
NGC\,4039 & $+0.11^a$ & $-0.30^a$ & 1.8 & & $+1.35^g$ & $+0.30^g$ & 1.6 & 
& ... & ... & ... \\
NGC\,4676 A & \,~~$0.00^b$ & $-1.20^b$ & 1.8 & & $+0.24^h$ & $-0.67^h$ &
1.6 & & $+0.75^o$ & $-1.36^o$ & 1.7 \\
NGC\,4676 B & $-0.30^b$ & $+0.20^b$ & 1.8 & & $-1.50^h$ & $-0.86^h$ &
1.6 & & $-0.45^o$ & $-1.78^o$ & 1.7 \\
NGC\,7592 W & ... & ... & ... & & $+0.90^i$ & $+0.73^i$ & 1.6 & &
$-0.75^p$ & $+1.20^p$ & 1.6 \\
NGC\,7592 E & ... & ... & ... & & ... & ... & ... & & ... & ... & ... \\
NGC\,7764A E & ... & ... & ... & & ... & ... & ... & & ... & ... & ...\\
NGC\,7764A W & ... & ... & ... & & ... & ... &  ... & & ... & ... & ...\\
NGC\,6621 & ... & ... & ... & & $+0.60^j$ & $+0.80^j$ & $\leq 2.4$ & &
$-2.25^p$ & $-0.70^p$ & 1.6 \\
NGC\,6622 & ... & ... & ... & & ... & ... & ... & & ... & ... & ... \\
NGC\,3509 & ... & ... & ...& & $+0.15^k$ & $+0.80^k$ & $\leq 1.9$ & & ...
& ... & ... \\
NGC\,520 S & $+4.95^c$ & $+0.70^c$ & $\geq 35.0$ & & $+0.15^l$ &
$-0.77^l$ & 1.6 & & $+1.05^o$ & $-0.70^o$ & $\leq 1.7$ \\
NGC\,520 N & ... & ... & ... & & ... & ... & ... & & ... & ... & ... \\
NGC\,2623 & $+0.15^d$ & $+2.77^d$ & 1.8 & & $+0.83^h$ & $-0.59^h$ & 1.6
& & $+1.05^q$ & $-0.37^q$ & $\leq 1.7$ \\
NGC\,3256 N & $-1.20^e$ & $+1.40^e$ & 1.8 & & $-0.90^m$ & $+1.60^m$ &
1.6 & & ... & ... & ... \\
NGC\,3256 S & $-0.90^e$ & $+1.40^e$ & 1.8 & & $-0.90^m$ & $+1.60^m$ &
1.6 & & ... & ... & ... \\
NGC\,3921 & $-0.30^f$ & $+1.31^f$ & 2.6 & & $+1.11^h$ & $+0.09^h$ & 1.6
& & $+0.45^o$ & $+1.57^o$ & $\leq 1.7$ \\
NGC\,7252 & $+0.15^f$ & $+1.60^f$ & 2.6 & & $+1.20^n$ & $-2.10^n$ &
$\leq 2.4$ & & $+0.75^r$ & $-0.44^r$ & $\leq 1.9$ \\
\enddata
\tablecomments{Col.~(1) lists the galaxy name. Cols.~(2) and (3) list
the positional offsets between the HST and X-ray detected nucleus.
Col.~(4) lists the uncertainty associated with this positional offset.
Cols.~(5) and (6) list the positional offsets between HST and radio
continuum measurements.  Col.~(7) lists the uncertainty associated
with this positional offset.  Cols.~(8) and (9) list the positional
offsets between the HST and CO measurements.  Col.~(10) lists the
uncertainty associated with this positional offset. The HST positions
adopted for the nuclei are the NICMOS positions listed in
Table~\ref{t:ts}, except for NGC\,3921 for which we used the WFPC2
position from Paper~I (due to evidence for an $\sim 4''$ absolute
astrometric error in the corresponding NICMOS dataset). The sources
for the X-ray measurements are Einstein/Chandra/XMM-Newton, for the
radio continuum measurements the VLA, and for the CO measurements
various telescopes. The listed uncertainties associated with
positional offsets are the quadrature sums of the HST astrometric
accuracy and the astrometric accuracies of the observations in the
other wavebands. Astrometric accuracies were obtained from the
following sources: HST, \citet{pta06}; Chandra,
\citet{ald02,sch02}; XMM-Newton/Einstein, S.~Snowden, priv. comm.;
radio continuum and CO, the original literature sources of the data.
See the text for a discussion of the X-ray and radio nucleus positions
adopted for NGC\,4038. References: ($a$)
\citet{zez02b}, ($b$) \citet{rea03}, ($c$) \citet{fab92}, ($d$) own
measurement from archival data, ($e$) \citet{lir02}, ($f$)
\citet{nol04}, ($g$) \citet{nef00}, ($h$)
\citet{mcm02}, ($i$) \citet{con90}, ($j$) \citet{con96}, ($k$)
Hibbard,
unpublished, ($l$) \citet{car90}, ($m$) \citet{nef03}, ($n$)
\citet{hib94},
($o$) \citet{yun01}, ($p$) Iono, priv. comm., ($q$) \citet{bry99},
($r$)
\citet{wan92}.}
\end{deluxetable}

\begin{deluxetable}{lccccccc}
\tabletypesize{\scriptsize}
\tablecaption{NIR photometry of the Toomre sequence
nuclei in 100\,pc apertures\label{t:colors}}
\tablewidth{0pt}
\tablehead{
Galaxy & $J$ & $H$ & $K$ & $J-H$ & $J-K$ & $H-K$ & $V-K$ \\
& [mag] & [mag] & [mag] & [mag] & [mag] & [mag] & [mag] \\
\colhead{(1)} & \colhead{(2)} & \colhead{(3)} & \colhead{(4)} &
\colhead{(5)} & \colhead{(6)} & \colhead{(7)} & \colhead{(8)} \\
}
\startdata
NGC\,4038 & 16.26 & 15.31 & 14.75 & 0.95 & 1.51 & 0.56 & 4.67 \\
NGC\,4039 & 14.62 & 13.67 & 13.42 & 0.95 & 1.20 & 0.25 & 4.50 \\
NGC\,4676 A & 18.36 & 16.92 & 16.25 & 1.44 & 2.11 & 0.67 & 6.79 \\
NGC\,4676 B & 17.60 & 16.78 & 16.52 & 0.82 & 1.08 & 0.26 & 3.56 \\
NGC\,7592 W & 16.60 & 15.65 & 14.86 & 0.95 & 1.74 & 0.79 & 4.46 \\
NGC\,7764A & 18.25 & 17.14 & 16.61 & 1.11 & 1.64 & 0.53 & 6.05 \\
NGC\,6621 & 17.27 & 16.34 & 15.86 & 0.93 & 1.41 & 0.48 & 4.27 \\
NGC\,6622 & 16.77 & 15.77 & 15.41 & 1.00 & 1.36 & 0.36 & 4.59 \\
NGC\,3509 & 17.90 & 17.11 & 16.76 & 0.79 & 1.14 & 0.35 & 3.53 \\
NGC\,520 S & 18.18 & 15.58 & 14.03 & 2.60 & 4.15 & 1.55 & 7.32 \\
NGC\,520 N & 16.99 & 16.31 & 15.99 & 0.68 & 1.00 & 0.32 & 2.80 \\
NGC\,2623 & 16.56 & 14.68 & 13.58 & 1.88 & 2.98 & 1.10 & 8.15 \\
NGC\,3256 N & ... & 13.08 & 12.53 & ... & ... & 0.55 & 4.35 \\
NGC\,3256 S & ... & 16.27 & 14.65 & ... & ... & 1.62 & 7.22 \\
NGC\,3921 & 14.71 & 13.97 & 13.70 & 0.74 & 1.01 & 0.27 & 3.39 \\
NGC\,7252 & 14.95 & 14.21 & 13.96 & 0.74 & 0.99 & 0.25 & 3.13 \\
\enddata
\tablecomments{Col.~(1) lists the galaxy identifier. Cols.~(2)-(4)
list
the $J$, $H$ and $K$-band magnitudes in a 100\,pc aperture.
Cols.~(5)-(8)
lists the $J-H$, $J-K$, $H-K$ and $V-K$ colors. Note that for
NGC\,3256 no
$J$-band observations were performed. For NGC\,7592 and NGC\,7764A
only the
primary nucleus was measured. The photometric random errors on the
magnitudes
and colors are very small for the near-IR bands (with a median of
0.01\,mag)
and are not listed in the table. The actual errors are dominated by
systematic calibration uncertainties ($\lta 0.05$\,mag in each band,
see
\S~\ref{ss:obsreduc}).}
\end{deluxetable}

\begin{deluxetable}{lccccccc}
\tabletypesize{\scriptsize}
\tablecaption{NIR photometry of the Toomre sequence
nuclei in 1\,kpc apertures\label{t:colorsonek}}
\tablewidth{0pt}
\tablehead{
Galaxy & $J$ & $H$ & $K$ & $J-H$ & $J-K$ & $H-K$ & $V-K$ \\
& [mag] & [mag] & [mag] & [mag] & [mag] & [mag] & [mag] \\
\colhead{(1)} & \colhead{(2)} & \colhead{(3)} & \colhead{(4)} &
\colhead{(5)} & \colhead{(6)} & \colhead{(7)} & \colhead{(8)} \\
}
\startdata
NGC\,4038 & 12.53 & 11.61 & 10.77 & 0.92 & 1.76 & 0.84 & 3.99 \\
NGC\,4039 & 12.55 & 11.77 & 10.96 & 0.78 & 1.59 & 0.81 & 4.24 \\
NGC\,4676 A & 14.85 & 13.42 & 12.67 & 1.43 & 2.18 & 0.75 & 5.42 \\
NGC\,4676 B & 14.40 & 13.55 & 13.17 & 0.85 & 1.23 & 0.38 & 3.52 \\
NGC\,7592 W & 14.13 & 13.25 & 12.51 & 0.88 & 1.62 & 0.74 & 4.21 \\
NGC\,7764A & 14.99 & 13.88 & 13.22 & 1.11 & 1.77 & 0.66 & 4.55 \\
NGC\,6621 & 13.89 & 12.82 & 12.20 & 1.07 & 1.69 & 0.62 & 4.57 \\
NGC\,6622 & 13.75 & 12.84 & 12.46 & 0.91 & 1.29 & 0.38 & 3.61 \\
NGC\,3509 & 14.88 & 14.02 & 13.64 & 0.86 & 1.24 & 0.38 & 3.50 \\
NGC\,520 S & 15.62 & 12.56 & 10.80 & 3.06 & 4.82 & 1.76 & 5.23 \\
NGC\,520 N & 13.30 & 13.18 & 12.40 & 0.12 & 0.90 & 0.78 & 2.70 \\
NGC\,2623 & 13.85 & 12.50 & 11.66 & 1.35 & 2.19 & 0.84 & 3.99 \\
NGC\,3256 N & ... & 10.93 & 10.16 & ... & ... & 0.77 & 3.89 \\
NGC\,3256 S & ... & 12.18 & 11.02 & ... & ... & 1.16 & 5.05 \\
NGC\,3921 & 12.66 & 11.83 & 11.47 & 0.83 & 1.19 & 0.36 & 3.40 \\
NGC\,7252 & 12.54 & 11.69 & 11.30 & 0.85 & 1.24 & 0.39 & 3.25 \\
\enddata
\tablecomments{Col.~(1) lists the galaxy identifier. Cols.~(2)-(4)
list
the $J$, $H$ and $K$-band magnitudes in a 1\,kpc aperture.
Cols.~(5)-(8)
lists the $J-H$, $J-K$, $H-K$ and $V-K$ colors. Note that for
NGC\,3256 no
$J$-band observations were performed. For NGC\,7592 and NGC\,7764A
only the
primary nucleus was measured. The photometric random errors on the
magnitudes
and colors are very small (with a median of 0.01 mag) and are not
listed in
the table. The actual errors are dominated by systematic calibration
uncertainties ($\lta 0.05$\,mag in each band, see
\S~\ref{ss:obsreduc}).}
\end{deluxetable}

\begin{deluxetable}{lcccccccc}
\tabletypesize{\scriptsize}
\tablecaption{Surface Brightness Profile Fitting
Parameters\label{t:nukerfits}}
\tablewidth{0pt}
\tablehead{
Galaxy & $\mu_b$ & $r_{b}$ & $r_{b}$ & $\alpha$ &
$\beta$ & $\gamma$ & $\Gamma_{0.15}$ & $M_{\rm K}$ \\
& [K mag\,/$\square^{\arcsec}$] & [\arcsec] & [pc] & & & & & [mag] \\
\colhead{(1)} & \colhead{(2)} & \colhead{(3)} & \colhead{(4)} &
\colhead{(5)} & \colhead{(6)} & \colhead{(7)} & \colhead{(8)}
& \colhead{(9)}\\
}
\startdata
NGC\,4039 & 13.71 & 0.547 & \,~36.6 & \,~0.13 & 1.46 & 0.80 & 1.10 &
$-23.78$ \\
NGC\,4676 A & 14.31 & 0.429 & 183.4 & \,~0.10 & 1.01 & 0.87 & 0.65 &
$-24.34$ \\
NGC\,4676 B & 14.80 & 1.738 & 743.0 & \,~4.18 & 1.86 & 0.65 & 0.94 &
$-24.13$ \\
NGC\,7592 W & 14.07 & 0.576 & 271.3 & 10.00 & 1.57 & 1.58 & 1.58 &
$-24.19$ \\
NGC\,6621 & 13.11 & 0.366 & 146.6 & 10.00 & 0.92 & 0.68 & 0.68 &
$-24.72$ \\
NGC\,6622 & 11.63 & 0.112 & \,~46.9 & \,~2.58 & 1.31 & 0.24 & 0.97 &
$-24.72$ \\
NGC\,3509 & 14.13 & 0.355 & 176.7 & 10.00 & 1.21 & 0.99 & 0.99 &
$-24.87$ \\
NGC\,520 N & 15.04 & 0.165 & \,~24.4 & 10.00 & 0.40 & 1.90 & 1.49 &
$-23.73$ \\
NGC\,2623 & 13.10 & 0.531 & 190.0 & \,~0.18 & 3.21 & 0.00 & 1.42 &
$-24.22$ \\
NGC\,3256 N & 12.72 & 0.673 & 119.1 & \,~1.79 & 1.18 & 1.30 & 1.29 &
$-24.72$ \\
NGC\,3921 & 11.16 & 0.242 & \,~91.2 & 10.00 & 1.72 & 1.12 & 1.13 &
$-25.13$ \\
NGC\,7252 & 13.12 & 0.517 & 156.5 & 10.00 & 1.08 & 1.30 & 1.30 &
$-24.84$ \\
\enddata
\tablecomments{Col.~(1) lists the galaxy name. Col.~(2) lists the
surface brightness in the $K$-band at the break-radius (i.e., the
surface brightness $\mu_b$ corresponding to the luminosity surface
density $\Sigma_b$ in equation~(\ref{nukerlaw})). Cols.~(3) and (4)
list the break-radius of the surface brightness profile fitting in
arcseconds and parsecs, respectively. Cols.~(5)-(7) list the fit
parameters.  Column~(8) lists the calculated power-law slope at a
radius of 0\farcs15.  Col.~(9) lists the total absolute $K$-band
magnitudes, calculated from the apparent $K$-band magnitudes from
NED. The majority of the apparent $K$-band magnitudes are from
2MASS. The ones for the last four merger remnants were taken from
\citet{rot06}.}
\end{deluxetable}




\begin{thebibliography}{}

\bibitem[Aldcroft(2002)]{ald02}
Aldcroft, T. 2002, Chandra Absolute Astrometric Accuracy (Cambridge:
Chandra X-Ray Science Center) [http://cxc.harvard.edu/cal/ASPECT/celmon/]

\bibitem[Arp(1966)]{arp66}
Arp, H. C. 1966, Atlas of Peculiar Galaxies (Pasadena: California 
Institute of Technology)

\bibitem[Barnes(1988)]{bar88}
Barnes, J. E. 1988, \apj, 331, 699

\bibitem[Barnes \& Hernquist(1996)]{bar96}
Barnes, J. E., \& Hernquist, L. 1996, \apj, 471, 115

\bibitem[Bessell \& Brett(1988)]{bes88}
Bessell, M. S., \& Brett, J. M. 1988, \pasp, 100, 1134

\bibitem[Beswick et al.(2003)]{bes03}
Beswick, R. J., Pedlar, A., Clemens, M. S., \& Alexander, P. 2003,
\mnras, 346, 424

\bibitem[Binney \& Merrifield(1998)]{bin98}
Binney, J. J., \& Merrifield, M. 1998, Galactic Astronomy 
(Princeton: Princeton University Press)

\bibitem[B\"oker et al.(1997)]{boe97}
B\"oker, T., Storey, J. W. V., Krabbe, A., \& Lehmann, T. 1997, \pasp, 
109, 827

\bibitem[B\"oker et al.(2002)]{boe02}
B\"oker, T., Laine, S., van der Marel, R. P., Sarzi, M., Rix, H.-W.,
Ho, L. C., \& Shields, J. C. 2002, AJ, 123, 1389

\bibitem[B\"oker et al.(2004)]{boe04}
B\"oker, T., Sarzi, M., McLaughlin, D. E., van der Marel, R. P.,
Rix, H.-W., Ho, L. C., \& Shields, J. C. 2004, AJ, 127, 105

\bibitem[Borne \& Richstone(1991)]{bor91}
Borne, K. D., \& Richstone, D. G. 1991, \apj, 369, 111

\bibitem[Borne et al.(2000)]{bor00}
Borne, K. D., Colina, L., Bushouse, H., \& Lucas, R. A. 2000, \apj, 527, 554

\bibitem[Braine et al.(2001)]{brai01}
Braine, J., Duc, P.-A., Lisenfeld, U., Charmandaris, V., Vallejo, O., 
Leon, S., \& Brinks, E. 2001, \aap, 378, 51

\bibitem[Bruzual \& Charlot(2003)]{bru03}
Bruzual, G., \& Charlot, S. 2003, \mnras, 344, 1000

\bibitem[Bryant \& Scoville(1999)]{bry99}
Bryant, P. M., \& Scoville, N. Z. 1999, \aj, 117, 2632

\bibitem[Bushouse \& Stanford(1992)]{bus92}
Bushouse, H. A., \& Stanford, S. A. 1992, \apjs, 79, 213

\bibitem[Bushouse \& Werner(1990)]{bus90}
Bushouse, H. A., \& Werner, M. W. 1990, \apj, 359, 72

\bibitem[Byun et al.(1996)]{byu96}
Byun, Y.-I., et al. 1996, \aj, 111, 1889

\bibitem[Carollo et al.(1997)]{car97} Carollo, C. M., Stiavelli, M., 
de Zeeuw, P. T., \& Mack, J. 1997, \aj, 114, 2366

\bibitem[Carral et al.(1990)Carral, Turner, \& Ho]{car90}
Carral, P., Turner, J. L., \& Ho, P. T. P. 1990, \apj, 362, 434

\bibitem[Casoli et al.(1991)]{cas91}
Casoli, F., Dupraz, C., Combes, F., \& Kazes, I. 1991, \aap, 251, 1

\bibitem[Chabrier(2003)]{cha03}
Chabrier, G. 2003, \pasp, 115, 763

\bibitem[Condon et al.(1990)]{con90}
Condon, J. J., Helou, G., Sanders, D. B., \& Soifer, B. T. 1990,
\apjs, 73, 359

\bibitem[Condon et al.(1996)]{con96}
Condon, J. J., Helou, G., Sanders, D. B., \& Soifer, B. T. 1996, 
\apjs, 103, 81


\bibitem[Dickinson et al.(2002)]{dic02}
Dickinson, M. E. et al. 2002, in: {\it HST} NICMOS Data Handbook v.~5.0, 
ed. B. Mobasher, (Baltimore: STScI)

\bibitem[Escala et al.(2005)]{esc05}
Escala, A., Larson, R. B., Coppi, P. S., \& Mardones, D. 2005, \apj, 630, 152

\bibitem[Fabbiano et al.(1992)]{fab92}
Fabbiano, G., Kim, D.-W., \& Trinchieri, G. 1992, \apjs, 80, 531

\bibitem[Fabbiano et al.(1997)[Fabbiano, Schweitzer \& Mackie]{fab97}
Fabbiano, G., Schweizer, F., \& Mackie, G. 1997, \apj, 478, 542

\bibitem[Fabbiano et al.(2001)[Fabbiano, Zezas \& Murray]{fab01}
Fabbiano, G., Zezas, A., \& Murray, S. S. 2001, \apj, 554, 1035

\bibitem[Faber et al.(1997)]{faber97}
Faber, S. M., et al. 1997, \aj, 114, 1771

\bibitem[Ferrarese et al.(2006)]{fer06}
Ferrarese, L., et al. 2006, \apjs, 164, 334

\bibitem[Georgakakis et al.(2000)]{geo00}
Georgakakis, A., Forbes, D. A., \& Norris, R. P. 2000, \mnras, 318, 124

\bibitem[Gerssen et al.(2004)]{ger04}
Gerssen, J., van der Marel, R. P., Axon, D., Mihos, J. C., Hernquist, 
L., \& Barnes, J. E. 2004, \aj, 127, 75

\bibitem[Hibbard \& van Gorkom(1996)]{hib96}
Hibbard, J. E., \& van Gorkom, J. H. 1996, \aj, 111, 655

\bibitem[Hibbard \& Yun(1999)]{hib99}
Hibbard, J. E., \& Yun, M. S. 1999, \apj, 522, L93

\bibitem[Hibbard et al.(1994)]{hib94}
Hibbard, J. E., Guhathakurta, P., van Gorkom, J. H, \& Schweizer, F. 
1994, \aj, 107, 67

\bibitem[Hibbard et al.(2001)]{hib01}
Hibbard, J. E., van der Hulst, J. M., Barnes, J. E., \& Rich, R. M. 
2001, \aj, 122, 2969

\bibitem[Hopkins et al.(2005)]{hop05}
Hopkins, P. F., Hernquist, L., Martini, P., Cox, T. J., Robertson, B.,
Di Matteo, T., \& Springel, V. 2005, \apj, 625, L71

\bibitem[Hopkins et al.(2006)]{hop06}
Hopkins, P. F., Hernquist, L., Cox, T. J., Di Matteo, T., Robertson, B., 
\& Springel, V. 2006, \apjs, 163, 1

\bibitem[Iono et al.(2005)]{ion05}
Iono, D., Yun, M. S., \& Ho, P. T. P. 2005, \apjs, 128, 1

\bibitem[Joseph \& Wright(1985)]{jos85}
Joseph, R. D., \& Wright, G. S. 1985, \mnras, 214, 87

\bibitem[Joseph et al.(1984)]{jos84}
Joseph, R. D., Meikle, W. P. S., Robertson, N. A., \& Wright, G. S.
1984, \mnras, 209, 111

\bibitem[Kassin et al.(2003)]{kas03}
Kassin, S. A., Frogel, J. A., Pogge, R. W., Tiede, G. P., \& Sellgren, 
K. 2003, \aj, 126, 1276

\bibitem[Kennicutt et al.(1987)]{ken87}
Kennicutt, R. C., Jr., Roettiger, K. A., Keel, W. C., van der Hulst, 
J. M., \& Hummel, E. 1987, \aj, 93, 1011

\bibitem[Komossa et al.(2003)]{kom03}
Komossa, S., Burwitz, V., Hasinger, G., Predehl, P., Kaastra, J. S., 
\& Ikebe, Y. 2003, \apj, 582, L15

\bibitem[Kotilainen et al.(1996)]{kot96}
Kotilainen, J. K., Moorwood, A. F. M., Ward, M. J., \& Forbes, D. A. 
1996, \aap, 305, 107

\bibitem[Kotilainen et al.(2001)]{kot01}
Kotilainen, J. K., Reunanen, J., Laine, S., \& Ryder, S. D. 2001, 
\aap, 366, 439

\bibitem[Krist \& Hook(2001)]{kri01}
Krist, J., \& Hook, R. 2001, The TinyTim User's Guide, (Baltimore: STScI)

\bibitem[Laine et al.(2003a)]{lai03a} 
Laine, S., van der Marel, R. P., Lauer, T. R., Postman, M., O'Dea, C. P., 
\& Owen, F. N. 2003a, \aj, 125, 478

\bibitem[Laine et al.(2003b)]{lai03b}
Laine, S., van der Marel, R. P., Rossa, J., Hibbard, J. E., Mihos, J. C., 
B\"oker, T., \& Zabludoff, A. I. 2003b, \aj, 126, 2717 [Paper~I]

\bibitem[Lauer(1985)]{lau85}
Lauer, T. R. 1985, \apjs, 57, 473

\bibitem[Lauer(1986)]{lau86}
Lauer, T. R. 1986, \apj, 311, 34

\bibitem[Lauer et al.(1995)]{lau95}
Lauer, T. R., et al. 1995, \aj, 110, 2622

\bibitem[Lauer et al.(2005)]{lau05}
Lauer, T. R., et al. 2005, \aj, 129, 2138

\bibitem[Lira et al.(2002)]{lir02}
Lira, P., Ward, M., Zezas, A., Alonso-Herrero, A., \& Ueno, S. 2002, 
\mnras, 330, 259

\bibitem[Liu \& Kennicutt(1995a)]{liu95a}
Liu, C. T., \& Kennicutt, R. C., Jr. 1995a, \apj, 450, 547

\bibitem[Liu \& Kennicutt(1995b)]{liu95b}
Liu, C. T., \& Kennicutt, R. C., Jr. 1995b, \apjs, 100, 325

\bibitem[Lonsdale et al.(1984)]{lon84}
Lonsdale, C. J., Persson, S. E., \& Matthews, K. 1984, \apj, 287, 95

\bibitem[Lucy(1974)]{luc74} Lucy, L. B. 1974, \aj, 79, 745

\bibitem[McMahon et al.(2002)]{mcm02}
McMahon, R. G., White, R. L., Helfand, D. J., \& Becker, R. H. 2002, 
\apjs, 143, 1

\bibitem[Mihos \& Hernquist(1994)]{hos94}
Mihos, J. C., \& Hernquist, L. 1994, \apj, 431, L9

\bibitem[Mihos \& Hernquist(1996)]{hos96}
Mihos, J. C., \& Hernquist, L. 1996, \apj, 464, 641

\bibitem[Milosavljevi\'c \& Merritt(2001)]{mil01}
Milosavljevi\'c, M., \& Merritt, D. 2001, ApJ, 563, 34

\bibitem[Neff \& Ulvestad(2000)]{nef00}
Neff, S. G., \& Ulvestad, J. S. 2000, \aj, 120, 670

\bibitem[Neff et al.(2003)]{nef03}
Neff, S. G., Ulvestad, J. S., \& Campion, S. D. 2003, \apj, 599, 1043

\bibitem[Noguchi(1988)]{nog88}
Noguchi, M. 1988, \aap, 203, 259

\bibitem[Nolan et al.(2004)]{nol04}
Nolan, L. A., Ponman, T. J., Read, A. M., \& Schweizer, F. 2004, 
\mnras, 353, 221

\bibitem[Norris \& Forbes(1995)]{nor95}
Norris, R. P., \& Forbes, D. A. 1995, \apj, 446, 594

\bibitem[Pasquali et al.(2004)]{pas04}
Pasquali, A., Gallagher, J. S., \& de Grijs, R. 2004, A\&A, 415, 103

\bibitem[Ptak et al.(2006)]{pta06}
Ptak, A. F., Colbert, E. J. M., van der Marel R. P., Roye, E. W., 
Heckman, T. M., \& Towne, B. 2006, \apjs, 166, 154

\bibitem[Rafanelli \& Marziani(1992)]{raf92}
Rafanelli, P., \& Marziani, P. 1992, \aj, 103, 743

\bibitem[Ravindranath et al.(2001)]{rav01}
Ravindranath, S., Ho, L. C., Peng, C. Y., Filippenko, A. V., \& 
Sargent, W. L. W. 2001, \aj, 122, 653 [R01]

\bibitem[Read(2003)]{rea03}
Read, A. M. 2003, \mnras, 342, 715

\bibitem[Rest et al.(2001)]{res01} 
Rest, A., van den Bosch, F. C., Jaffe, W., Tran, H., Tsvetanov, Z., 
Ford, H. C., Davies, J., \& Schafer, J. 2001, \aj, 121, 2431

\bibitem[Richardson(1972)]{ric72} 
Richardson, W. H. 1972, J. Opt. Soc. A., 62, 55

\bibitem[Rieke \& Lebofsky(1985)]{rie85}
Rieke, G. H., \& Lebofsky, M. J. 1985, \apj, 288, 618

\bibitem[Rossa et al.(2006)]{ros06}
Rossa, J., van der Marel, R. P., B\"oker, T., Gerssen, J., 
Ho, L. C., Rix, H.-W., Shields, J. C., \& Walcher C. J. 2006, \aj, 
132, 1074

\bibitem[Rossa et al.(2007)]{ros07}
Rossa, J., van der Marel, R. P., Hibbard, J. E., Mihos, J. C., Laine, S., 
B\"oker, T., \& Zabludoff, A. I. 2007, in prep. [Paper~III]

\bibitem[Rothberg \& Joseph(2004)]{rot04}
Rothberg, B., \& Joseph, R. D. 2004, \aj, 128, 2098

\bibitem[Rothberg \& Joseph(2006)]{rot06}
Rothberg, B., \& Joseph, R. D. 2006, \aj, 131, 185

\bibitem[Sakamoto et al.(2006)]{sak06}
Sakamoto, K., Ho, P. T. P., \& Peck, A. B. 2006, \apj, 644, 862

\bibitem[Sanders \& Mirabel(1996)]{san96}
Sanders, D. B., \& Mirabel, I. F. 1996, \araa, 34, 749

\bibitem[Sanders et al.(1988)]{san88}
Sanders, D. B., Soifer, B. T., Elias, J. H., Madore, B. F., Matthews, K., 
Neugebauer, G., \& Scoville, N. Z. 1988, \apj, 325, 74

\bibitem[Saviane et al.(2004)]{sav04}
Saviane, I., Hibbard, J. E., \& Rich, R. M. 2004, \aj, 127, 660

\bibitem[Schade et al.(2002)]{sch02}
Schade, D., Micol, A., Durand, D., Pirenne, B., Simard, L., Dolensky, 
M., \& Stetson, P. B. 2002, The WFPC2 Associations Science Products 
Pipeline at CADC and ST-ECF (Victoria: Canadian Astronomy Data Centre)
[$\rm{http://archive.eso.org/archive/hst/wfpc2
\_asn/3sites/WFPC2\_Newsletter.pdf}$]

\bibitem[Schreier et al.(1998)]{schr98}
Schreier, E. J., et al. 1998, \apj, 499, L143

\bibitem[Schweizer(1982)]{sch82}
Schweizer, F. 1982, \apj, 252, 455

\bibitem[Schweizer(1996)]{sch96a}
Schweizer, F. 1996, \aj, 111, 109

\bibitem[Schweizer(1998)]{sch98}
Schweizer, F. 1998, in: ``Galaxies: Interactions and Induced Star
Formation'', Saas Fee Advanced Course 26, eds. D. Friedli, L. Martinet, 
and D. Pfenniger, (Heidelberg: Springer), p.~105

\bibitem[Schweizer et al.(1996)]{sch96}
Schweizer, F., Miller, B. W., Whitmore, B. C., \& Fall, S. M. 1996, 
\aj, 112, 1839

\bibitem[Schweizer et al.(2004)]{sch04}
Schweizer, F., Seitzer, P., \& Brodie, J. P. 2004, \aj, 128, 202

\bibitem[Scoville et al.(1998)]{sco98}
Scoville, N. Z., et al. 1998, \apj, 492, L107

\bibitem[Scoville et al.(2000)]{sco00}
Scoville, N. Z., et al. 2000, \aj, 119, 991

\bibitem[Seigar et al.(2002)]{sei02} Seigar, M., Carollo, C. M.,
Stiavelli, M., de Zeeuw, P. T., \& Dejonghe, H. 2002, \aj, 123, 184

\bibitem[Silva \& Bothun(1998)]{sil98}
Silva, D. R., \& Bothun, G. D. 1998, \aj, 116, 85

\bibitem[Springel(2000)]{spr00}
Springel, V. 2000, \mnras, 312, 859

\bibitem[Springel et al.(2005)]{spr05}
Springel, V., Di Matteo, T., \& Hernquist, L. 2005, \mnras, 361, 776

\bibitem[Stanford \& Bushouse(1991)]{sta91}
Stanford, S. A., \& Bushouse, H. A. 1991, \apj, 371, 92

\bibitem[Toomre(1977)]{too77}
Toomre, A. 1977, in: ``The Evolution of Galaxies and Stellar Populations,'' 
eds. B. M. Tinsley and R. B. Larson (New Haven: Yale University), p.~401

\bibitem[Toomre \& Toomre(1972)]{too72}
Toomre, A., \& Toomre, J. 1972, \apj, 178, 623

\bibitem[Verdoes Kleijn et al.(2002)]{ver02}
Verdoes Kleijn, G. A., Baum, S. A., de Zeeuw, P. T., \& O'Dea, C. P. 
2002, AJ, 123, 1334

\bibitem[Wang et al.(1992)]{wan92}
Wang, Z., Schweizer, F., \& Scoville, N. Z. 1992, \apj, 396, 570

\bibitem[Weilbacher et al.(2003)]{wei03}
Weilbacher, P. M., Duc, P.-A., \& Fritze-v. Alvensleben, U. 2003,
\aap, 397, 545

\bibitem[Whitmore \& Schweizer(1995)]{whi95}
Whitmore, B. C., \& Schweizer, F. 1995, \aj, 109, 960

\bibitem[Whitmore \& Zhang(2002)]{whi02}
Whitmore, B. C., \& Zhang, Q. 2002, \aj, 124, 1418

\bibitem[Whitmore et al.(1993)]{whi93}
Whitmore, B. C., Schweizer, F., Leitherer, C., Borne, K., \& Robert, C. 
1993, \aj, 106, 1354

\bibitem[Whitmore et al.(1999)]{whi99}
Whitmore, B. C., Zhang, Q., Leitherer, C., Fall, S. M., Schweizer, F., 
\& Miller, B. W. 1999, \aj, 118, 1551

\bibitem[Whitmore et al.(2005)]{whi05}
Whitmore, B. C., Gilmore, D., Leitherer, C., Fall, S. M., Chandar, R., 
Blair, W. P., Schweizer, F., Zhang, Q., \& Miller, B. W. 2005, \aj, 130, 2104

\bibitem[Wright et al.(1990)]{wri90}
Wright, G. S., James, P. A., Joseph, R. D., \& McLean, I. S. 1990, 
\nat, 344, 417

\bibitem[Yang et al.(2004)]{yan04}
Yang, Y., Zabludoff, A. I., Zaritsky, D., Lauer, T. R., \& Mihos, J. C. 
2004, \apj, 607, 258

\bibitem[Yun \& Hibbard(2001)]{yun01}
Yun, M. S., \& Hibbard, J. H. 2001, \apj, 550, 104

\bibitem[Zezas et al.(2002a)]{zez02a}
Zezas, A., Fabbiano, G., Rots, A. H., \& Murray, S. S. 2002a, \apj, 
577, 710

\bibitem[Zezas et al.(2002b)]{zez02b}
Zezas, A., Fabbiano, G., Rots, A. H., \& Murray, S. S. 2002b, \apjs, 
142, 239

\end{thebibliography}
\end{document}